\documentclass[11pt]{article}
\usepackage{appendix}

\usepackage[english]{babel}

\usepackage[authoryear]{natbib}
\usepackage[T1]{fontenc}
\usepackage[latin1]{inputenc}
\usepackage{textcomp}

\usepackage{amsthm}
\usepackage{amsmath}
\usepackage{amsfonts}
\usepackage{mathrsfs}
\usepackage{dsfont}

\usepackage{amssymb}
\usepackage{enumerate}
\usepackage{graphicx}
\usepackage{mathrsfs}
\usepackage{algorithm}
\usepackage{algorithmicx}
\usepackage{algpseudocode}
\usepackage{array}
\usepackage{caption}
\usepackage{booktabs}
\usepackage{url}

\usepackage{color}
\usepackage[normalem]{ulem}  %

\title{Sequential online subsampling for thinning experimental designs}

\author{Luc Pronzato\footnotemark[1]\ \ and HaiYing Wang\footnotemark[2]}

\DeclareMathOperator{\diag}{diag}
\DeclareMathOperator{\tr}{trace}

\numberwithin{equation}{section}

\def\rad{\stackrel{\rm d}{\ra}}

\newcommand{\bea}{\begin{eqnarray*}}
\newcommand{\eea}{\end{eqnarray*}}
\newcommand{\be}{\begin{eqnarray}}
\newcommand{\ee}{\end{eqnarray}}
\def\ra{\rightarrow}
\def\e1{\mathsf{e}}
\def\mp{\partial}
\def\0b{\boldsymbol{0}}
\def\1b{\boldsymbol{1}}
\def\ind{\mathbb{I}}

\def\TT{^\top}
\def\dd{\mbox{\rm d}}

\def\SA{\mathcal{A}}
\def\SB{{\mathscr B}}

\def\SF{{\mathscr F}}

\def\SI{{\mathcal I}}

\def\SL{{\mathscr L}}

\def\SM{{\mathscr M}}
\def\SMM{\mathbb{M}}
\def\SO{{\mathcal O}}
\def\SP{{\mathscr P}}

\def\SSp{{\mathscr S}}
\def\SN{{\mathscr N}}
\def\ST{{\mathscr T}}
\def\SU{{\mathscr U}}

\def\SX{{\mathscr X}}
\def\SY{{\mathscr Y}}
\def\SZ{{\mathscr Z}}
\def\ma{\alpha}

\def\mg{\gamma}
\def\ml{\lambda}
\def\me{\epsilon}
\def\mve{\varepsilon}

\def\ms{\sigma}
\def\mt{\theta}

\def\mtb{\boldsymbol{\theta}}
\def\bmtb{\overline{\boldsymbol{\theta}}}
\def\hmtb{\hat{\boldsymbol{\theta}}}

\def\mo{\omega}

\def\Sigmab{\boldsymbol{\Sigma}}

\def\Sigmab{\boldsymbol{\Sigma}}
\def\mOb{\boldsymbol{\Omega}}

\def\Ex{\mathsf{E}}

\def\IBOSS{\mathsf{IBOSS}}

\def\var{\mathsf{var}}

\def\Prob{\mathsf{Prob}}
\def\Ab{\mathbf{A}}

\def\fb{\mathbf{f}}

\def\Ib{\mathbf{I}}

\def\Mb{\mathbf{M}}

\def\ub{\mathbf{u}}

\def\vb{\mathbf{v}}
\def\Vb{\mathbf{V}}

\def\xb{\mathbf{x}}

\def\zb{\mathbf{z}}

\def\hMb{\widehat{\Mb}}
\def\hC{\widehat{C}}
\def\hf{\widehat{f}}

\theoremstyle{plain}

\newtheorem{theo}{Theorem}[section]

\newtheorem{lemma}{Lemma}[section]

\theoremstyle{definition}

\newtheorem{remark}{Remark}[section]

\newcommand{\carre} {\mbox{}~\hfill\rule{2mm}{2mm}}
\newcommand{\fin} {\mbox{}~\hfill{\lower-0.3ex\hbox{$\triangleleft$}}}

\newcommand{\vsp} {\vspace{0.3cm}}

\begin{document}
\maketitle

\renewcommand{\thefootnote}{\fnsymbol{footnote}}
\footnotetext[1]{CNRS, Universit\'e C\^ote d'Azur, I3S, France, Luc.Pronzato@cnrs.fr (corresponding author)}
\footnotetext[2]{Department of Statistics, University of Connecticut, USA, haiying.wang@uconn.edu}
\renewcommand{\thefootnote}{\arabic{footnote}}

\begin{abstract}
We consider a design problem where experimental conditions (design points $X_i$) are presented in the form of a sequence of i.i.d.\ random variables, generated with an unknown probability measure $\mu$, and only a given proportion $\alpha\in(0,1)$ can be selected. The objective is to select good candidates $X_i$ on the fly and maximize a concave function $\Phi$ of the corresponding information matrix. The optimal solution corresponds to the construction of an optimal bounded design measure $\xi_\alpha^*\leq \mu/\alpha$, with the difficulty that $\mu$ is unknown and $\xi_\alpha^*$ must be constructed online. The construction proposed relies on the definition of a threshold $\tau$ on the directional  derivative of $\Phi$ at the current information matrix, the value of $\tau$ being fixed by a certain quantile of the distribution of this directional derivative. Combination with recursive quantile estimation yields a nonlinear two-time-scale stochastic approximation method. It can be applied to very long design sequences since only the current information matrix and estimated quantile need to be stored. Convergence to an optimum design is proved. Various illustrative examples are presented.
\end{abstract}



\section{Introduction}\label{S:Intro}

Consider a rather general parameter estimation problem in a model with independent observations $Y_i=Y_i(x_i)$ conditionally on the experimental variables $x_i$, with $x_i$ in some set $\SX$. Suppose that for any $x\in\SX$ there exists a measurable set $\SY_x\in\mathds{R}$ and a $\ms$-finite measure $\mu_x$ on $\SY_x$ such that $Y(x)$ has the density $\varphi_{x,\bmtb}$ with respect to $\mu_x$, with $\bmtb$ the true value of the model parameters $\mtb$ to be estimated, $\mtb\in\mathds{R}^p$. In particular, this covers the case of regression models, with $\mu_x$ the Lebesgue measure on $\SY_x=\mathds{R}$ and $Y(x)=\eta(\bmtb,x)+\mve(x)$, where the $\mve(x_i)$ are independently distributed with zero mean and known variance $\ms_i^2$ (or unknown but constant variance $\ms^2$), and the case of generalized linear models, with $\varphi_{x,\bmtb}$ in the exponential family and logistic regression as a special case. Denoting by $\hmtb^n$ the estimated value of $\mtb$ from data $(X_i,Y_i)$, $i=1,\ldots,n$, under rather weak conditions on the $x_i$ and $\varphi_{x,\bmtb}$, see below, we have
\be \label{AN}
\sqrt{n} (\hmtb^n-\bmtb) \rad \SN(\0b,\Mb^{-1}(\xi,\bmtb)) \ \mbox{ as } n\ra\infty\,,
\ee
where $\Mb(\xi,\mtb)$ denotes the (normalized) Fisher information matrix for parameters $\mtb$ and (asymptotic) design $\xi$ (that is, a probability measure on $\SX$),
\bea
\Mb(\xi,\mtb) &=& \lim_{n\ra\infty} \frac1n\, \Ex_{x_1,\ldots,x_n,\mtb} \left\{ \sum_{i=1}^n \frac{\mp \log\varphi_{x,\mtb}(Y_i)}{\mp\mtb} \sum_{j=1}^n \frac{\mp \log\varphi_{x,\mtb}(Y_j)}{\mp\mtb\TT} \right\} \\
&=& \int_\SX \left[ \int_{\SY_x} \frac{\mp \log\varphi_{x,\mtb}(y)}{\mp\mtb} \frac{\mp \log\varphi_{x,\mtb}(y)}{\mp\mtb\TT}\, \varphi_{x,\mtb}(y)\, \mu_x(\dd y)\right] \xi(\dd x) \,.
\eea
This is true in particular for randomized designs such that the $x_i$ are independently sampled from $\xi$, and for asymptotically discrete designs, such that $\xi$ is a discrete measure on $\SX$ and the empirical design measure $\xi_n=\sum_{i=1}^n \delta_{x_i}$ converges strongly to $\xi$; see \citet{PP2013}. The former case corresponds to the situation considered here. The choice of $\mu_x$ is somewhat arbitrary, provided that $\int_{\SY_x} \varphi_{x,\bmtb}(y)\, \mu_x(\dd y)=1$ for all $x$, and we shall assume that $\mu_x(\dd y)\equiv 1$. We can then write
\bea
\Mb(\xi,\mtb) = \int_\SX \SM(x,\mtb) \, \xi(\dd x)\,, \mbox{ where } \SM(x,\mtb) = \int_{\SY_x} \frac{\mp \log\varphi_{x,\bmtb}(y)}{\mp\mtb} \frac{\mp \log\varphi_{x,\bmtb}(y)}{\mp\mtb\TT}\, \varphi_{x,\bmtb}(y) \,\dd y
\eea
denotes the elementary information matrix at $x$.

Taking motivation from \eqref{AN}, optimal experimental design (approximate theory) aims at choosing a measure $\xi\*$ that minimizes a scalar function of the asymptotic covariance matrix $\Mb^{-1}(\xi,\bmtb)$ of $\hmtb^n$, or equivalently, that maximizes a function $\Phi$ of $\Mb(\xi,\bmtb)$. For a nonlinear model $\SM(x,\mtb)$ and $\Mb(\xi,\mtb)$ depend on the model parameters $\mtb$. Since $\bmtb$ is unknown, the standard approach is local, and consists in constructing an optimal design for a nominal value $\mtb_0$ of $\mtb$. This is the point of view we shall adopt here --- although sequential estimation of $\mtb$ is possible, see Section~\ref{S:conclusions}. When $\mtb$ is fixed at some $\mtb_0$, there is fundamentally no difference with experimental design in a linear model for which $\SM(x,\mtb)$ and $\Mb(\xi,\mtb)$ do not depend on $\mtb$. For example, in the linear regression model
\bea
Y(X_i)= \fb\TT(X_i)\bmtb+\mve_i \,,
\eea
where the errors $\mve_i$ are independent and identically distributed (i.i.d.), with a density $\varphi_\mve$ with respect to the Lebesgue measure having finite Fisher information for location $I_\mve=\int_\mathds{R} \left\{[\varphi'_\mve(t)]^2/\varphi_\mve(t)\right\}\dd t < \infty$ ($I_\mve=1/\ms^2$ for normal errors $\SN(0,\ms^2)$), then $\SM(x)=I_\mve\, \fb(x)\fb\TT(x)$, $\Mb(\xi)=I_\mve\, \int_\SX \fb(x)\fb\TT(x)\, \xi(\dd x)$. Polynomial regression provides typical examples of such a situation and will be used for illustration in Section~\ref{S:Example}. The construction of an optimal design measure $\xi^*$ maximizing $\Phi[\Mb(\xi,\mt_0)]$ usually relies on the application of a specialized algorithm to a discretization of the design space $\SX$; see, e.g., \citet[Chap.~9]{PP2013}.

With the rapid development of connected sensors and the pervasive usage of computers, there exist more and more situations where extraordinary amounts of massive data $(X_i,Y_i)$, $i=1,\ldots,N$, are available to construct models.
When $N$ is very large, using all the data to construct $\hmtb^N$ is then unfeasible, and selecting the most informative subset through the construction of an $n$-point optimal design, $n\ll N$, over the discrete set $\SX_N=\{X_i, i=1,\ldots,N\}$ is also not feasible. The objective of this paper is to present a method to explore $\SX_N$ sequentially and select a proportion $n=\lfloor \ma N \rfloor$ of the $N$ data points to be used to estimate $\mtb$. Each candidate $X_i$ is considered only once, which allows very large datasets to be processed: when the $X_i$ are i.i.d.\ and are received sequentially, they can be selected on the fly which makes the method applicable to data streaming; when $N$ data points are available simultaneously, a random permutation allows $\SX_N$ to be processed as an i.i.d.\ sequence. When $N$ is too large for the storage capacity and the i.i.d.\ assumption is not tenable, interleaving or scrambling techniques can be used. Since de-scrambling is not necessary here (the objective is only to randomize the sequence), a simple random selection in a fixed size buffer may be sufficient; an example is presented in Section~\ref{S:non-iid}.

The method is based on the construction of an optimal bounded design measure and draws on the paper \citep{Pa05}. In that paper, the sequential selection of the $X_i$ relies on a threshold set on the directional derivative of the design criterion, given by the $(1-\ma)$-quantile of the distribution of this derivative. At stage $k$, all previous $X_i$, $i=1,\ldots,k$, are used for the estimation of the quantile $C_k$ that defines the threshold for the possible selection of the candidate $X_{k+1}$. In the present paper, we combine this approach with the recursive estimation of $C_k$, following \citep{Tierney83}: as a result, the construction is fully sequential and only requires to record the current value of the information matrix $\Mb_k$ and of the estimated quantile $\hC_k$ of the distribution of the directional derivative. It relies on a reinterpretation of the approach in \citep{Pa05} as a stochastic approximation method for the solution of the necessary and sufficient optimality conditions for a bounded design measure, which we combine with another stochastic approximation method for quantile estimation to obtain a two-time-scale stochastic approximation scheme.

The paper is organized as follows. Section~\ref{S:OptimalBounded} introduces the notation and assumptions and recalls main results on optimal bounded design measures. Section~\ref{S:SequantialOptimalBounded} presents our subsampling algorithm based on a two-time-scale stochastic approximation procedure and contains the main result of the paper. Several illustrative examples are presented in Section~\ref{S:Example}. We are not aware of any other method for thinning experimental designs that is applicable to data streaming; nevertheless, in Section~\ref{S:other-methods} we compare our algorithm with an exchange method and with the IBOSS algorithm of \citet{WangYS2019} in the case where the $N$ design points are available and can be processed simultaneously. Section~\ref{S:conclusions} concludes and suggests a few directions for further developments. A series of technical results are provided in the Appendix.

\section{Optimal bounded design measures}\label{S:OptimalBounded}
\subsection{Notation and assumptions}\label{S:notation}

Suppose that $X$ is distributed with the probability measure $\mu$ on $\SX\subseteq\mathds{R}^d$, a subset of $\mathds{R}^d$ with nonempty interior, with $d\geq 1$. For any $\xi\in\SP^+(\SX)$, the set of positive measure $\xi$ on $\SX$ (not necessarily of mass one), we denote $\Mb(\xi)=\int_\SX \SM(x)\,\xi(\dd x)$ where, for all $x$ in $\SX$, $\SM(x)\in\SMM^\geq$, the set (cone) of symmetric non-negative definite $p\times p$ matrices. We assume that $p>1$ in the rest of the paper (the optimal selection of information in the case $p=1$ forms a variant of the secretary problem for which an asymptotically optimal solution can be derived, see \citet{AlbrightD72, Pronzato01a_ieee}).

We denote by $\Phi: \SMM^\geq \ra \mathds{R} \cup \{-\infty\}$ the design criterion we wish to maximize, and by $\ml_{\min}(\Mb)$ and $\ml_{\max}(\Mb)$ the minimum and maximum eigenvalues of $\Mb$, respectively; we shall use the $\ell_2$ norm for vectors and Frobenius norm for matrices, $\|\Mb\|=\tr^{1/2}[\Mb\Mb\TT]$; all vectors are column vectors. For any $t\in\mathds{R}$, we denote $[t]^+=\max\{t,0\}$ and, for any $t\in\mathds{R}^+$, $\lfloor t \rfloor$ denotes the largest integer smaller than $t$. For $0\leq \ell\leq L$ we denote by $\SMM^\geq_{\ell,L}$ the (convex) set defined by
\bea
\SMM^\geq_{\ell,L} = \{\Mb\in\SMM^\geq: \ell < \ml_{\min}(\Mb) \mbox{ and } \ml_{\max}(\Mb) < L \}\,,
\eea
and by $\SMM^>$ the open cone of symmetric positive definite $p\times p$ matrices.
We make the following assumptions on $\Phi$.

\begin{description}
  \item[H$_\Phi$] $\Phi$ is strictly concave on $\SMM^>$, linearly differentiable and increasing for Loewner ordering; its gradient $\nabla_\Phi(\Mb)$ is well defined in $\SMM^\geq$ for any $\Mb\in\SMM^>$ and satisfies
      $\|\nabla_\Phi(\Mb)\|<A(\ell)$ and $\ml_{\min}[\nabla_\Phi(\Mb)]>a(L)$ for any $\Mb\in\SMM^\geq_{\ell,L}$, for some $a(L)>0$ and $A(\ell)<\infty$; moreover, $\nabla_\Phi$ satisfies the following Lipschitz condition: for all $\Mb_1$ and $\Mb_2$ in $\SMM^\geq$ such that $\ml_{\min}(\Mb_i)>\ell>0$, $i=1,2$, there exists $K_\ell<\infty$ such that $\|\nabla_\Phi(\Mb_2)-\nabla_\Phi(\Mb_1)\|<K_\ell\,\|\Mb_2-\Mb_1\|$.
\end{description}

The criterion $\Phi_0(\Mb)=\log\det(\Mb)$ and criteria $\Phi_q(\Mb)=-\tr(\Mb^{-q})$, $q\in(-1,\infty)$, $q\neq 0$, with $\Phi_q(\Mb)=-\infty$ if $\Mb$ is singular, which are often used in optimal design (in particular with $q$ a positive integer) satisfy H$_\Phi$; see, e.g., \citet[Chap.~6]{Pukelsheim93}. Their gradients are $\nabla_{\Phi_0}(\Mb)=\Mb^{-1}$ and $\nabla_{\Phi_q}(\Mb)=q\,\Mb^{-(q+1)}$, $q\neq 0$; the constants $a(L)$ and $A(\ell)$ are respectively given by $a(L)=1/L$, $A(\ell)=\sqrt{p}/\ell$ for $\Phi_0$ and $a(L)=q/L^{q+1}$, $A(\ell)=q\sqrt{p}/\ell^{q+1}$ for $\Phi_q$. The Lispchitz condition follows from the fact that the criteria are twice differentiable on $\SMM^>$. The positively homogeneous versions $\Phi_0^+(\Mb)=\det^{1/p}(\Mb)$ and $\Phi_q^+(\Mb)=[(1/p)\tr(\Mb^{-q})]^{-1/q}$, which satisfy $\Phi^+(a\Mb)=a\,\Phi^+(\Mb)$ for any $a>0$ and any $\Mb\in\SMM^\geq$, and $\Phi^+(\Ib_p)=1$, with $\Ib_p$ the $p\times p$ identity matrix, could be considered too; see \citet[Chaps.~5, 6]{Pukelsheim93}.
The strict concavity of $\Phi$ implies that, for any convex subset $\widehat \SMM$ of $\SMM^>$, there exists a unique matrix $\Mb^*$ maximizing $\Phi(\Mb)$ with respect to  $\Mb\in\widehat \SMM$.

We denote by $F_\Phi(\Mb,\Mb')$ the directional derivative of $\Phi$ at $\Mb$ in the direction $\Mb'$,
\bea
F_\Phi(\Mb,\Mb') = \lim_{\mg\ra 0^+} \frac{\Phi[(1-\mg)\Mb+\mg\Mb']-\Phi(\Mb)}{\mg} = \tr[\nabla_\Phi(\Mb)(\Mb'-\Mb)]\,,
\eea
and we make the following assumptions on $\mu$ and $\SM$.

\begin{description}
  \item[H$_{\mu}$] $\mu$ has a bounded positive density $\varphi$ with respect to the Lebesgue measure on every open subset of $\SX$. 
  
  \item[H$_{\SM}$] (\textit{i}) $\SM$ is continuous on $\SX$ and satisfies $\int_\SX \|\SM(x)\|^2\, \mu(\dd x)<B<\infty$;

  \noindent (\textit{ii}) for any $\SX_\me\subset\SX$ of measure $\mu(\SX_\me)=\me>0$,
 $\ml_{\min} \left\{\int_{\SX_\me} \SM(x)\,\mu(\dd x) \right\} > \ell_\me$ for some $\ell_\me>0$.
\end{description}

Since all the designs considered will be formed by points sampled from $\mu$, we shall confound $\SX$ with the support of $\mu$: $\SX=\{x\in\mathds{R}^d: \mu(\SB_d(x,\me))>0 \quad \forall\me>0\}$, with $\SB_d(x,\me)$ the open ball with center $x$ and radius $\me$. Notice that H$_{\SM}$-(\textit{i}) implies that $\ml_{\max}[\Mb(\mu)] < \sqrt{B}$ and $\|\Mb(\mu)\|< \sqrt{p\,B}$.

Our sequential selection procedure will rely on the estimation of the $(1-\ma)$-quantile $C_{1-\ma}(\Mb)$ of the distribution $F_\Mb(z)$ of the directional derivative $Z_\Mb(X)=F_\Phi[\Mb,\SM(X)]$ when $X\sim\mu$, and we shall assume that H$_{\mu,\SM}$ below is satisfied. It implies in particular that $C_{1-\ma}(\Mb)$ is uniquely defined by $F_\Mb(C_{\Mb,1-\ma})=1-\ma$.

\begin{description}
  \item[H$_{\mu,\SM}$] For all $\Mb\in\SMM^\geq_{\ell,L}$, $F_\Mb$ has a uniformly bounded density $\varphi_\Mb$; moreover, for any $\ma\in(0,1)$, there exists $\me_{\ell,L}>0$ such that $\varphi_\Mb[C_{1-\ma}(\Mb)]>\me_{\ell,L}$ and $\varphi_\Mb$ is continuous at $C_{1-\ma}(\Mb)$.
\end{description}

H$_{\mu,\SM}$ is overrestricting (we only need the existence and boundedness of $\varphi_\Mb$, and its positiveness and continuity at $C_{1-\ma}(\Mb)$), but is satisfied is many common situations; see Section~\ref{S:Example} for examples. Let us emphasize that H$_{\mu}$ and H$_{\SM}$ are not enough to guarantee the existence of a density $\varphi_\Mb$, since $\tr[\nabla_\Phi(\Mb)\SM(x)]$ may remain constant over subsets of $\SX$ having positive measure. Assuming the existence of $\varphi_\Mb$ and the continuity of $\varphi$ on $\SX$ is also insufficient, since $\varphi_\Mb$ is generally not continuous when $Z_\Mb(x)$ is not differentiable in $x$, and $\varphi_\Mb$ is not necessarily bounded.

\subsection{Optimal design}

As mentioned in introduction, when the cardinality of $\SX_N$ is very large, one may wish to select only $n$ candidates $X_i$ among the $N$ available, a fraction $n=\lfloor \ma N \rfloor$ say, with $\ma\in(0,1)$.
For any $n\leq N$, we denote by $\Mb_{n,N}^*$ a design matrix (non necessarily unique) obtained by selecting $n$ points optimally within $\SX_N$; that is, $\Mb_{n,N}^*$ gives the maximum of $\Phi(\Mb_n)$ with respect to $\Mb_n=(1/n)\,\sum_{j=1}^n \SM(X_{i_j})$, where the $X_{i_j}$ are $n$ distinct points in $\SX_N$. Note that this forms a difficult combinatorial problem, unfeasible for large $n$ and $N$.
If one assumes that the $X_i$ are i.i.d., with $\mu$ their probability measure on $\SX$,
for large $N$ the optimal selection of $n=\lfloor \ma N \rfloor$ points
amounts at constructing an optimal bounded design measure $\xi_\ma^*$, such that $\Phi[\Mb(\xi_\ma^*)]$ is maximum and $\xi_\ma\leq\mu/\ma$ (in the sense $\xi_\ma(\SA)\leq\mu(\SA)/\ma$ for any $\mu$-measurable set $\SA$, which makes $\xi_\ma$ absolutely continuous with respect to $\mu$). Indeed,
Lemma~\ref{L:opt-nonsequential} in Appendix~\ref{S:max-Phi} indicates that $\limsup_{N\ra\infty} \Phi(\Mb_{\lfloor \ma N \rfloor,N}^*)=\Phi[\Mb(\xi_\ma^*)]$. Also, under H$_\Phi$, $\Ex\{\Phi(\Mb_{n,N}^*)\} \leq \Phi[\Mb(\xi_{n/N}^*)]$ for all $N\geq n>0$; see \citet[Lemma~3]{Pa05}.

A key result is that, when all subsets of $\SX$ with constant $Z_\Mb(x)$ have zero measure, $Z_{\Mb(\xi_\ma^*)}(x)= F_\Phi[\Mb(\xi_\ma^*),\SM(x)]$ separates two sets $\SX_\ma^*$ and $\SX\setminus\SX_\ma^*$, with $F_\Phi[\Mb(\xi_\ma^*),\SM(x)]\geq C_{1-\ma}^*$ and $\xi_\ma^*=\mu/\ma$ on $\SX_\ma^*$, and $F_\Phi[\Mb(\xi_\ma^*),\SM(x)] \leq C_{1-\ma}^*$ and $\xi_\ma^*=0$ on $\SX\setminus\SX_\ma^*$, for some constant $C_{1-\ma}^*$; moreover, $\int_\SX F_\Phi[\Mb(\xi_\ma^*),\SM(x)]\, \xi_\ma^*(\dd x)=\int_{\SX_\ma^*} F_\Phi[\Mb(\xi_\ma^*),\SM(x)]\, \mu(\dd x)=0$; see \citet{Wynn82, Fedorov89} and \citet[Chap.~4]{FedorovH97}. (The condition mentioned in those references is that $\mu$ has no atoms, but the example in Section~\ref{S:ZX-discrete} will show that this is not sufficient; extension to arbitrary measures is considered in \citep{SahmS2000}.)

For $\ma\in(0,1)$, denote
\bea
\SMM(\ma)=\left\{\Mb(\xi_\ma)=\int_\SX \SM(x)\,\xi_\ma(\dd x): \, \xi_\ma\in\SP^+(\SX), \, \xi_\ma\leq\frac{\mu}{\ma},\, \int_\SX \xi_\ma(\dd x) = 1 \right\} \,.
\eea
In \citep{Pa05}, it is shown that, for any $\Mb\in\SMM^>$,
\be\label{E1}
\Mb^+(\Mb,\ma) = \arg\max_{\Mb'\in\SMM(\ma)} F_\Phi(\Mb,\Mb')= \frac1\ma\, \int_\SX \ind_{\{F_\Phi[\Mb,\SM(x)]\geq C_{1-\ma}\}}\, \SM(x)\,\mu(\dd x) \,,
\ee
where, for any proposition $\SA$, $\ind_{\{\SA\}}=1$ if $\SA$ is true and is zero otherwise, and $C_{1-\ma}=C_{1-\ma}(\Mb)$ is an $(1-\ma)$-quantile of $F_\Phi[\Mb,\SM(X)]$ when $X \sim \mu$ and satisfies
\be\label{C_ma}
\int_\SX \ind_{\{F_\Phi[\Mb,\SM(x)]\geq C_{1-\ma}(\Mb)\}} \,\mu(\dd x) = \ma\,.
\ee
Therefore, $\Mb_\ma^*=\Mb(\xi_\ma^*)$ is the optimum information matrix in $\SMM(\ma)$ (unique since $\Phi$ is strictly concave) if and only if it satisfies $\max_{\Mb'\in\SMM(\ma)} F_\Phi(\Mb_\ma^*,\Mb')=0$, or equivalently $\Mb_\ma^*=\Mb^+(\Mb_\ma^*,\ma)$, and the constant $C_{1-\ma}^*$ equals $C_{1-\ma}(\Mb_\ma^*)$; see \citep[Th.~5]{Pa05}; see also \citet{Pa04}. Note that $C_{1-\ma}^*\leq 0$ since $\int_\SX F_\Phi[\Mb(\xi_\ma^*),\SM(x)]\, \xi_\ma^*(\dd x)=0$ and $F_\Phi[\Mb(\xi_\ma^*),\SM(x)]\geq C_{1-\ma}^*$ on the support of $\xi_\ma^*$. 

\section{Sequential construction of an optimal bounded design measure}\label{S:SequantialOptimalBounded}
\subsection{A stochastic approximation problem}\label{S:Basis}

Suppose that the $X_i$ are i.i.d.\ with $\mu$.
The solution of $\Mb=\Mb^+(\Mb,\ma)$, $\ma\in(0,1)$, with respect to $\Mb$ by stochastic approximation yields the iterations
\be\label{iter-main}
\begin{array}{rcl}
n_{k+1} &=& n_k+ \ind_{\{F_\Phi[\Mb_{n_k},\SM(X_{k+1})] \geq C_{1-\ma}(\Mb_{n_k}) \}} \,, \\
\\
\Mb_{n_{k+1}} &=& \Mb_{n_k} + \frac{1}{n_k+1}\, \ind_{\{F_\Phi[\Mb_{n_k},\SM(X_{k+1})] \geq C_{1-\ma}(\Mb_{n_k}) \}} \, \left[ \SM(X_{k+1}) - \Mb_{n_k} \right] \,.
\end{array}
\ee
Note that $\Ex\left\{\ind_{\{F_\Phi[\Mb,\SM(X)] \geq C_{1-\ma}(\Mb) \}} \, \left[ \SM(X) - \Mb \right] \right\}= \ma\,[\Mb^+(\Mb,\ma)-\Mb]$.
The almost sure (a.s.) convergence of $\Mb_{n_k}$ in \eqref{iter-main} to $\Mb(\xi_\ma^*)$ that maximizes $\Phi(\Mb)$ with respect $\Mb\in\SMM(\ma)$ is proved in \citep{Pa05} under rather weak assumptions on $\Phi$, $\SM$ and $\mu$.

The construction \eqref{iter-main} requires the calculation of the $(1-\ma)$-quantile $C_{1-\ma}(\Mb_{n_k})$ for all $n_k$, see \eqref{C_ma}, which is not feasible when $\mu$ is unknown and has a prohibitive computational cost when we know $\mu$. For that reason, it is proposed in \citep{Pa05} to replace $C_{1-\ma}(\Mb_{n_k})$ by the empirical quantile $\widetilde C_{\ma,k}(\Mb_{n_k})$ that uses the empirical measure $\mu_k=(1/k)\,\sum_{i=1}^k \delta_{X_i}$ of the $X_i$ that have been observed up to stage $k$. This construction preserves the a.s.\ convergence of $\Mb_{n_k}$ to $\Mb(\xi_\ma^*)$ in \eqref{iter-main}, but its computational cost and storage requirement increase with $k$, which makes it unadapted to situations with very large $N$. The next section considers the recursive estimation of $C_{1-\ma}(\Mb_{n_k})$ and contains the main result of the paper.

\subsection{Recursive quantile estimation}\label{S:RecursiveQuantileEst}

The idea is to plug a recursive estimator of the $(1-\ma)$-quantile $C_{1-\ma}(\Mb_{n_k})$  in \eqref{iter-main}. Under mild assumptions, for random variables $Z_i$ that are i.i.d.\ with distribution function $F$ such that the solution of the equation $F(z)=1-\ma$ is unique, the recursion
\be\label{SAQuantile-basic}
\hC_{k+1} = \hC_k + \frac{\beta}{k+1}\, \left(\ind_{\{Z_{k+1} \geq \hC_k\}} - \ma\right)
\ee
with $\beta>0$ converges a.s.\ to the quantile $C_{1-\ma}$ such that $F(C_{1-\ma})=1-\ma$. Here, we shall use a construction based on \citep{Tierney83}. In that paper, a clever dynamical choice of $\beta=\beta_k$ is shown to provide the optimal asymptotic rate of convergence of $\hC_k$ towards $C_{1-\ma}$, with $\sqrt{k}(\hC_k-C_{1-\ma}) \rad \SN(0,\ma(1-\ma)/f^2(C_{1-\ma}))$ as $k\ra\infty$, where $f(z)=\dd F(z)/\dd z$ is the p.d.f.\ of the $Z_i$ --- note that it coincides with the asymptotic behavior of the sample (empirical) quantile. The only conditions on $F$ are that $f(z)$ exists for all $z$ and is uniformly bounded, and that $f$ is continuous and positive at the unique root $C_{1-\ma}$ of $F(z)=1-\ma$.

There is a noticeable difference, however, with the estimation of $C_{1-\ma}(\Mb_{n_k})$: in our case we need to estimate a quantile of $Z_k(X)=F_\Phi[\Mb_{n_k},\SM(X)]$ for $X \sim \mu$, with the distribution of $Z_k(X)$ evolving with $k$. For that reason, we shall impose a faster dynamic to the evolution of $\hC_k$, and replace \eqref{SAQuantile-basic} by
\be\label{SAQuantile}
\hC_{k+1} = \hC_k + \frac{\beta_k}{(k+1)^q}\, \left(\ind_{\{Z_k(X_{k+1}) \geq \hC_k\}} - \ma\right)
\ee
for some $q\in(0,1)$. The combination of \eqref{SAQuantile} with \eqref{iter-main} yields a particular nonlinear two-time-scale stochastic approximation scheme. There exist advanced results on the convergence of linear two-time-scale stochastic approximation, see \citet{KondaT2004, DalalSTM2018}. To the best of our knowledge, however, there are few results on convergence for nonlinear schemes. Convergence is shown in \citep{Borkar97} under the assumption of boundedness of the iterates using the ODE method of \citet{Ljung77}; sufficient conditions for stability are provided in \citep{LakshminarayananB2017}, also using the ODE approach.
In the proof of Theorem~\ref{P:main} we provide justifications for our construction, based on the analyses and results in the references mentioned above.

\vsp
The construction is summarized in Algorithm~1 below. %
The presence of the small number $\me_1$
is only due to technical reasons: setting $z_{k+1}=+\infty$ when $n_k/k <\me_1$ in \eqref{perturbation} has the effect of always selecting $X_{k+1}$ when less than $\me_1\,k$ points have been selected previously; it ensures that $n_{k+1}/k>\me_1$ for all $k$ and thus that $\Mb_{n_k}$ always belongs to $\SMM^\geq_{\ell,L}$ for some $\ell>0$ and $L<\infty$; see Lemma~\ref{L:min-eigenvalue} in Appendix.

\begin{samepage}
\begin{itemize}
\item[] {\bf Algorithm~1: sequential selection ($\ma$ given).}
  \item[0)] Choose $k_0\geq p$, $q\in(1/2,1)$, $\mg\in(0,q-1/2)$, and $0<\me_1\ll \ma$.
  \item[1)] Initialization: select $X_1,\ldots,X_{k_0}$, compute $\Mb_{n_{k_0}}=(1/k_0)\,\sum_{i=1}^{k_0} \SM(X_i)$. If $\Mb_{n_{k_0}}$ is singular, increase $k_0$ and select the next points until $\Mb_{n_{k_0}}$ has full rank. Set $k=n_{k}=k_0$, the number of points selected.

  Compute $\zeta_i=Z_{k_0}(X_i)$, for $i=1,\ldots,k_0$ and order the $\zeta_i$ as $\zeta_{1:k_0} \leq \zeta_{2:k_0} \leq \cdots \leq \zeta_{k_0:k_0}$; denote $k_0^+=\lceil (1-\ma/2)\,k_0 \rceil$ and $k_0^-=\max\{\lfloor (1-3\,\ma/2)\,k_0 \rfloor, 1\}$.

      Initialize $\hC_{k_0}$ at $\zeta_{\lceil (1-\ma)\,k_0\rceil:k_0}$; set
      $\beta_0=k_0/(k_0^+-k_0^-)$,
      $h=(\zeta_{k_0^+:k_0}-\zeta_{k_0^-:k_0})$, $h_{k_0}=h/k_0^\mg$ and
      $\hf_{k_0}= \left[\sum_{i=1}^{k_0} \ind_{\{|\zeta_i-\hC_{k_0}|\leq h_{k_0}\}} \right]/(2\,k_0\,h_{k_0})$.

  \item[2)] Iteration $k+1$: collect $X_{k+1}$ and compute $Z_k(X_{k+1})=F_\Phi[\Mb_{n_k},\SM(X_{k+1})]$.
  \be\label{perturbation}
  \begin{array}{ll}
  \mbox{If } n_k/k <\me_1 & \mbox{ set } z_{k+1}=+\infty \,; \\
  \mbox{otherwise } & \mbox{ set } z_{k+1}=Z_k(X_{k+1}) \,.
  \end{array}
  \ee
  If $z_{k+1} \geq \hC_k$, update $n_k$ into $n_{k+1}=n_k+1$ and $\Mb_{n_k}$ into
  \be\label{iter-basic-k2}
  \Mb_{n_{k+1}}=\Mb_{n_k} + \frac{1}{n_k+1}\, \left[ \SM(X_{k+1}) - \Mb_{n_k} \right] \,;
  \ee
  otherwise, set $n_{k+1}=n_k$.

  \item[3)] Compute
  $\beta_k= \min\{1/\hf_k,\beta_0\,k^\mg\}$;
  update $\hC_k$ using \eqref{SAQuantile}.

  Set $h_{k+1}=h/(k+1)^\mg$ and update $\hf_k$ into
  \bea
  \hf_{k+1}= \hf_k + \frac{1}{(k+1)^q}\, \left[ \frac{1}{2\,h_{k+1}}\, \ind_{\{|Z_k(X_{k+1})-\hC_k|\leq h_{k+1}\}} - \hf_k \right] \,.
  \eea

  \item[4)] $k \leftarrow k+1$, return to Step 2.

\end{itemize}
\end{samepage}

Note that $\hC_k$ is updated whatever the value of $Z_k(X_{k+1})$. Recursive quantile estimation by \eqref{SAQuantile} follows \citep{Tierney83}. To ensure a faster dynamic for the evolution of $\hC_k$ than for $\Mb_{n_k}$, we take $q<1$ instead of $q=1$ in \citep{Tierney83}, and the construction of $\hf_k$ and the choices of $\beta_k$ and $h_k$ are modified accordingly. Following the same arguments as in the proof of Proposition~1 of \citep{Tierney83}, the a.s.\ convergence of $\hC_k$ to $C_{1-\ma}$ in the modified version of \eqref{SAQuantile-basic} is proved in Theorem~\ref{P:CV-of-Ck} (Appendix~\ref{S:fk}).

The next theorem establishes the convergence of the combined stochastic approximation schemes with two time-scales.

\begin{theo}\label{P:main} Under H$_\Phi$, H$_\mu$, H$_\SM$ and H$_{\mu,\SM}$, the normalized information matrix $\Mb_{n_k}$ corresponding to the $n_k$ candidates selected after $k$ iterations of Algorithm~1 converges a.s.\ to the optimal matrix $\Mb_\ma^*$ in $\SMM(\ma)$ as $k\ra\infty$.
\end{theo}

\noindent{\emph{Proof}.}
Our analysis is based on \citep{Borkar97}. We denote by $\SF_n$ the increasing sequence of $\ms$-fields generated by the $X_i$. According to \eqref{SAQuantile}, we can write $\hC_{k+1} = \hC_k + [\beta_k/(k+1)^q]\, V_{k+1}$ with $V_{k+1}=\ind_{\{Z_k(X_{k+1}) \geq \hC_k\}}-\ma$. Therefore, $\Ex\{V_{k+1}|\SF_k\}=\int_\SX [\ind_{\{Z_k(x) \geq \hC_k\}}-\ma]\,\mu(\dd x)$ and $\var\{V_{k+1}|\SF_k\}=F_k(\hC_k)[1-F_k(\hC_k)]$, with
$F_k$ the distribution function of $Z_k(X)$. From Lemma~\ref{L:min-eigenvalue} (Appendix~\ref{S:min-eigenvalue}) and H$_{\mu,\SM}$, $F_k$ has a well defined density $f_k$ for all $k$, with $f_k(t)>0$ for all $t$ and $f_k$ bounded.
The first part of the proof of Theorem~\ref{P:CV-of-Ck} applies (see Appendix~\ref{S:fk}): $\hf_k$ is a.s.\ bounded and $\beta_k$ is bounded away from zero a.s.
Therefore,
$\sum_k \beta_k/(k+1)^q \ra\infty$ a.s.\ and $(k+1)^q/[\beta_k\,(k+1)] \ra 0$ a.s.; also, $\sum_k [\beta_k/(k+1)^q]^2 <\infty$ since $q-\mg>1/2$.

The o.d.e.\ associated with \eqref{SAQuantile}, for a fixed matrix $\Mb$ and thus a fixed $Z(\cdot)$, such that $Z(X)=F_\Phi[\Mb,\SM(X)]$ has the distribution function $F$ and density $f$, is
\bea
\frac{\dd C(t)}{\dd t } = 1-F[C(t)] - \ma = F(C_{1-\ma})-F[C(t)]\,,
\eea
where $C_{1-\ma}=C_{1-\ma}(\Mb)$ satisfies $F(C_{1-\ma})=1-\ma$. Consider the Lyapunov function $L(C)=[F(C)-F(C_{1-\ma})]^2$. It satisfies $\dd L[C(t)]/\dd t=-2\,f[C(t)]\,L[C(t)] \leq 0$, with $\dd L[C(t)]/\dd t=0$ if and only if $C=C_{1-\ma}$.
Moreover, $C_{1-\ma}$ is Lipschitz continuous in $\Mb$; see Lemma~\ref{L:Lipschitz} in Appendix~\ref{S:Lipschitz}.
The conditions for Theorem~1.1 in \citep{Borkar97} are thus satisfied concerning the iterations for $\hC_k$.

\vsp
Denote $\hMb_k=\Mb_{n_k}$ and $\rho_k=n_k/k$, so that \eqref{perturbation} implies $k\,\rho_k\geq \me_1\,(k-1)$ for all $k$; see Lemma~\ref{L:min-eigenvalue} in Appendix. They satisfy
\be\label{iter-rho}
\rho_{k+1} = \rho_k + \frac{R_{k+1}}{k+1} \mbox{ and } \hMb_{k+1} = \hMb_k + \frac{\mOb_{k+1}}{k+1} \,,
\ee
where $R_{k+1}=\ind_{\{Z_k(X_{k+1}) \geq \hC_k \}} - \rho_k$, and
$\mOb_{k+1}=(1/\rho_{k+1})\, \ind_{\{Z_k(X_{k+1}) \geq \hC_k \}}\,\left[\SM(X_{k+1})-\hMb_k \right]$.
We have $\Ex\{R_{k+1}|\SF_k\} = \int_\SX \ind_{\{Z_k(x) \geq \hC_k\}}\,\mu(\dd x) - \rho_k$ and $\var\{R_{k+1}|\SF_k\}=F_k(\hC_k)[1-F_k(\hC_k)]$, with
$F_k$ the distribution function of $Z_k(X)$, which, from H$_{\mu,\SM}$, has a well defined density $f_k$ for all $k$.
Also,
\bea
\Ex\{\mOb_{k+1}|\SF_k\} &=& \frac{\SI_k}{\rho_k+ \frac{1-\rho_k}{k+1}} = \frac{1+1/k}{\rho_k+ 1/k}\, \SI_k \,,
\eea
where $\SI_k=\int_\SX \ind_{\{Z_k(x) \geq \hC_k\}}\,\left[\SM(x)-\hMb_k \right]\, \mu(\dd x)$. Denote $\Delta_{k+1}=\mOb_{k+1}-\SI_k/\rho_k$, so that
\be\label{iter-hMb}
\hMb_{k+1} = \hMb_k + \frac{1}{k+1} \, \frac{\SI_k}{\rho_k} + \frac{\Delta_{k+1}}{k+1} \,.
\ee
We get
$\Ex\{\Delta_{k+1}|\SF_k\} = (\rho_k-1)/[\rho_k\,(k\,\rho_k+1)]\, \SI_k$ and
\bea
\var\{\{\Delta_{k+1}\}_{i,j}|\SF_k\} &=& \var\{\{\mOb_{k+1}\}_{i,j}|\SF_k\} \\
&=& \frac{(k+1)^2}{(k\,\rho_k+1)^2}\, \left[\int_\SX \ind_{\{Z_k(x) \geq \hC_k\}}\, \{\SM(x)-\hMb_k\}_{i,j}^2\,\mu(\dd x)- \{\SI_k\}_{i,j}^2 \right]\,,
\eea
where \eqref{perturbation} implies that $\rho_k>\me_1/2$, and therefore $(k+1)\,(1-\rho_k)/[\rho_k\,(k\,\rho_k+1)]<4/\me_1^2$, and $\var\{\{\Delta_{k+1}\}_{i,j}|\SF_k\}$ is a.s.\ bounded from H$_\SM$-(\textit{i}). This implies that $\sum_k \Delta_{k+1}/(k+1) <\infty$ a.s. The limiting o.d.e.\ associated with \eqref{iter-rho} and \eqref{iter-hMb} are
\bea
\frac{\dd \rho(t)}{\dd t} &=& \int_\SX \ind_{\{F_\Phi[\hMb(t),\SM(x)] \geq C_{1-\ma}[\hMb(t)] \}}\, \mu(\dd x) - \rho(t) = \ma - \rho(t) \,, \\
\frac{\dd \hMb(t)}{\dd t} &=& \frac{1}{\rho(t)}\, \int_\SX \ind_{\{F_\Phi[\hMb(t),\SM(x)] \geq C_{1-\ma}[\hMb(t)] \}}\, \left[\SM(x)-\hMb(t) \right]\, \mu(\dd x) \\
&=& \frac{\ma}{\rho(t)}\, \left\{ \Mb^+[\hMb(t),\ma]-\hMb(t) \right\}\,,
\eea
where $\Mb^+[\hMb(t),\ma]$ is defined by \eqref{E1}. The first equation implies that $\rho(t)$ converges exponentially fast to $\ma$, with $\rho(t)=\ma+[\rho(0)-\ma]\,\exp(-t)$; the second equation gives
\bea
\frac{\dd \Phi[\hMb(t)]}{\dd t} = \tr\left[\nabla_\Phi[\hMb(t)]\, \frac{\dd \hMb(t)}{\dd t} \right] = \frac{\ma}{\rho(t)}\, \max_{\Mb'\in\SMM(\ma)} F_\Phi[\hMb(t),\Mb'] \geq 0\,,
\eea
with a strict inequality if $\hMb(t)\neq\Mb_\ma^*$, the optimal matrix in $\SMM(\ma)$. The conditions of Theorem~1.1 in \citep{Borkar97} are thus satisfied, and $\hMb_k$ converges to $\Mb_\ma^*$ a.s.
\carre

\begin{remark}\label{R:R1}\mbox{}
\begin{description}
  \item[(\textit{i})] Algorithm~1 does not require the knowledge of $\mu$ and has minimum storage requirements: apart for the current matrix $\Mb_{n_k}$, we only need to update the scalar variables $\hC_k$ and $f_k$. Its complexity is $\SO(d^3\,N)$ in general, considering that the complexity of the calculation of $F_\Phi[\Mb,\SM(X)]$ is $\SO(d^3)$. It can be reduced to $\SO(d^2\,N)$ when $\SM(X)$ has rank one and $\Mb_{n_k}^{-1}$ is updated instead of $\Mb_{n_k}$ (see remark (\textit{iii}) below), for D-optimality and $\Phi_q$-optimality with $q$ integer; see Section~\ref{S:notation}. Very long sequences $(X_i)$ can thus be processed.

  \item[(\textit{ii})] Numerical simulations indicate that we do not need to take $q<1$ in Algorithm~1: \eqref{SAQuantile} with $q=1$ yields satisfactory performance, provided the step-size obeys Kersten's rule \nocite{Kesten58} and does not decrease at each iteration. 

  \item[(\textit{iii})] 
    The substitution of $\tr[\nabla_\Phi(\Mb)\SM(X)]$ for $F_\Phi[\Mb,\SM(X)]=\tr\{\nabla_\Phi(\Mb)[\SM(X)-\Mb]\}$ everywhere does not change the behavior of the algorithm. When $\nabla_\Phi(\Mb)$ only depends on $\Mb^{-1}$ (which is often the case for classical design criteria, see the discussion following the presentation of H$_\Phi$), and if $\SM(X)$ is a low rank matrix, it may be preferable to update $\Mb_{n_k}^{-1}$ instead of $\Mb_{n_k}$, thereby avoiding matrix inversions.
    For example, if $\SM(X_{k+1})=I_\mve\,\fb(X_{k+1})\fb\TT(X_{k+1})$,
    then, instead of updating \eqref{iter-basic-k2}, it is preferable to update the following
\bea
  \Mb_{n_{k+1}}^{-1}
  =\left(1+\frac{1}{n_k}\right) \left[\Mb_{n_{k}}^{-1}
    - \frac{I_\mve\, \Mb_{n_{k}}^{-1}\fb(X_{k+1})\fb\TT(X_{k+1})\Mb_{n_{k}}^{-1}}
    {n_k+I_\mve\,\fb\TT(X_{k+1})\Mb_{n_{k}}^{-1}\fb(X_{k+1})} \right]\,.
\eea
    Low-rank updates of the Cholesky decomposition of the matrix can be considered too.

  \item[(\textit{iv})] Algorithm~1 can be adapted to the case where the number of iterations is fixed (equal to the size $N$ of the candidate set $\SX_N$) and the number of candidates $n$ to be selected is imposed. A straightforward modification is to introduce truncation and forced selection: we run the algorithm with $\ma=n/N$ and, at Step 2, we set $z_{k+1}=-\infty$ (reject $X_{k+1}$) if $n_k \geq n$ and set $z_{k+1}=+\infty$ (select $X_{k+1}$) if $n-n_k \geq N-k$. However, this may induce the selection of points $X_i$ carrying little information when $k$ approaches $N$ in case $n_k$ is excessively small. For that reason, adaptation of $\ma$ to $n_k$, obtained by substituting $\ma_k=(n-n_k)/(N-k)$ for the constant $\ma$ everywhere, seems preferable. This is illustrated by an example in Section~\ref{S:Ex2}.

\item[(\textit{v})]
    The case when $\mu$ has discrete components (atoms), or more precisely when there exist subsets of $\SX$ of positive measure where $Z_\Mb(x)$ is constant (see Section~\ref{S:ZX-discrete}), requires additional technical developments which we do not detail here.

    A first difficulty is that H$_\SM$-(\textit{ii}) may not be satisfied when the matrices $\SM(x)$ do not have full rank, unless we only consider large enough $\me$. Unless $\me_1$ in \eqref{perturbation} is large enough, Lemma~\ref{L:min-eigenvalue} is not valid, and other arguments are required in the proof of Theorem~\ref{P:main}. Possible remedies may consist (a) in adding a regularization matrix $\mg\Ib_p$ with a small $\mg$ to all matrices $\SM(x)$ (which amounts at considering optimal design for Bayesian estimation with a vague prior; see, e.g., \citet{Pilz83}), or (b) in replacing the condition in \eqref{perturbation} by [If $\ml_{\min}(\Mb_{n_k}) <\me_1$, set $z_{k+1}=+\infty$].

    A second difficulty is that $C_{1-\ma}(\Mb_\ma^*)$ may correspond to a point of discontinuity of the distribution function of $F_\Phi[\Mb_\ma^*,\SM(X)]$. The estimated value $f_k$ of the density of $F_\Phi[\Mb_{n_k},\SM(X)]$ at $\hC_k$ (Step~3 of Algorithm~1) may then increase to infinity and $\beta_k$ tend to zero in \eqref{SAQuantile}. This can be avoided by setting $\beta_k= \max\{\me_2,\min(1/\hf_k,\beta_0\,k^\mg)\}$ for some $\me_2>0$.

    In \citep{Pa05}, where empirical quantiles are used, measures needed to be taken to avoid the acceptance of too many points, for instance based on the adaptation of $\ma$ through $\ma_k=(n-n_k)/(N-k)$, see remark (\textit{iv}) above, or via the addition of the extra condition [if $n_k/k>\ma$, set $z_{k+1}=-\infty$] to \eqref{perturbation} in case $n$ is not specified. Such measures do not appear to be necessary when quantiles are estimated by \eqref{SAQuantile}; see the examples in Section~\ref{S:Ex3}.
          \fin
\end{description}
\end{remark}

\section{Examples}\label{S:Example}
We always take $k_0=5\,p$, $q=5/8$, $\mg=1/10$ in Algorithm~1 (our simulations indicate that these choices are not critical); we also set
$\me_1=0$.

\subsection{Example 1: quadratic regression with normal independent variables}\label{S:Ex1}

Take $\SM(x)=\fb(x)\fb\TT(x)$, with $\fb(x)=(1,\ x,\ x^2)\TT$ and $\Phi(\Mb)=\log\det(\Mb)$, and let the $X_i$ be i.i.d.\ standard normal variables $\SN(0,1)$. The D-optimal design for $x$ in an interval $[t,t']$ corresponds to $\xi^*=(1/3)\, \left(\delta_t + \delta_{(t+t')/2} + \delta_{t'} \right)$. In the data thinning problem, the optimal solution corresponds to the selection of $X_i$ in the union of three intervals; that is, with the notation of Section~\ref{S:OptimalBounded}, $\SX_\ma^*=(-\infty,-a]\cup [-b,b]\cup[a,\infty)$. The values of $a$ and $b$ are obtained by solving the pair of equations
$\int_0^b \varphi(x)\,\dd x + \int_a^\infty \varphi(x)\,\dd x =\ma/2$ and $\tr[\Mb^{-1}(\xi)\SM(a)]=\tr[\Mb^{-1}(\xi)\SM(b)]$, with $\varphi$ the standard normal density and $\Mb(\xi)=[\int_{-\infty}^{-a} \SM(x)\, \varphi(x)\,\dd x + \int_{-b}^{b} \SM(x)\, \varphi(x)\,\dd x + \int_{a}^{\infty} \SM(x)\, \varphi(x)\,\dd x]/\ma$.

We set the horizon $N$ at $100\,000$ and consider the two cases $\ma=1/2$ and $\ma=1/10$. In each case we keep $\ma$ constant but apply the rule of Remark~\ref{R:R1}-(\textit{iv}) (truncation/forced selection) to select exactly $n=50\,000$ and $n=10\,000$ design points, respectively. For $\ma=1/2$, we have $a \simeq 1.0280$, $b\simeq 0.2482$, and $\Phi_\ma^*=\Phi(\Mb_\ma^*)=\max_{\Mb\in\SMM(\ma)}\Phi(\Mb) \simeq 1.6354$, $C_{1-\ma}(\Mb_\ma^*) \simeq -1.2470$; when $\ma=1/10$, we have $a \simeq 1.8842$, $b\simeq 0.0507$, and $\Phi_\ma^* \simeq 3.2963$, $C_{1-\ma}(\Mb_\ma^*) \simeq -0.8513$.
The figures below present results obtained for one simulation (i.e., one random set $\SX_N$), but they are rather typical in the sense that different $\SX_N$ yield similar behaviors.

\begin{figure}[ht!]
\begin{center}
\includegraphics[width=.49\linewidth]{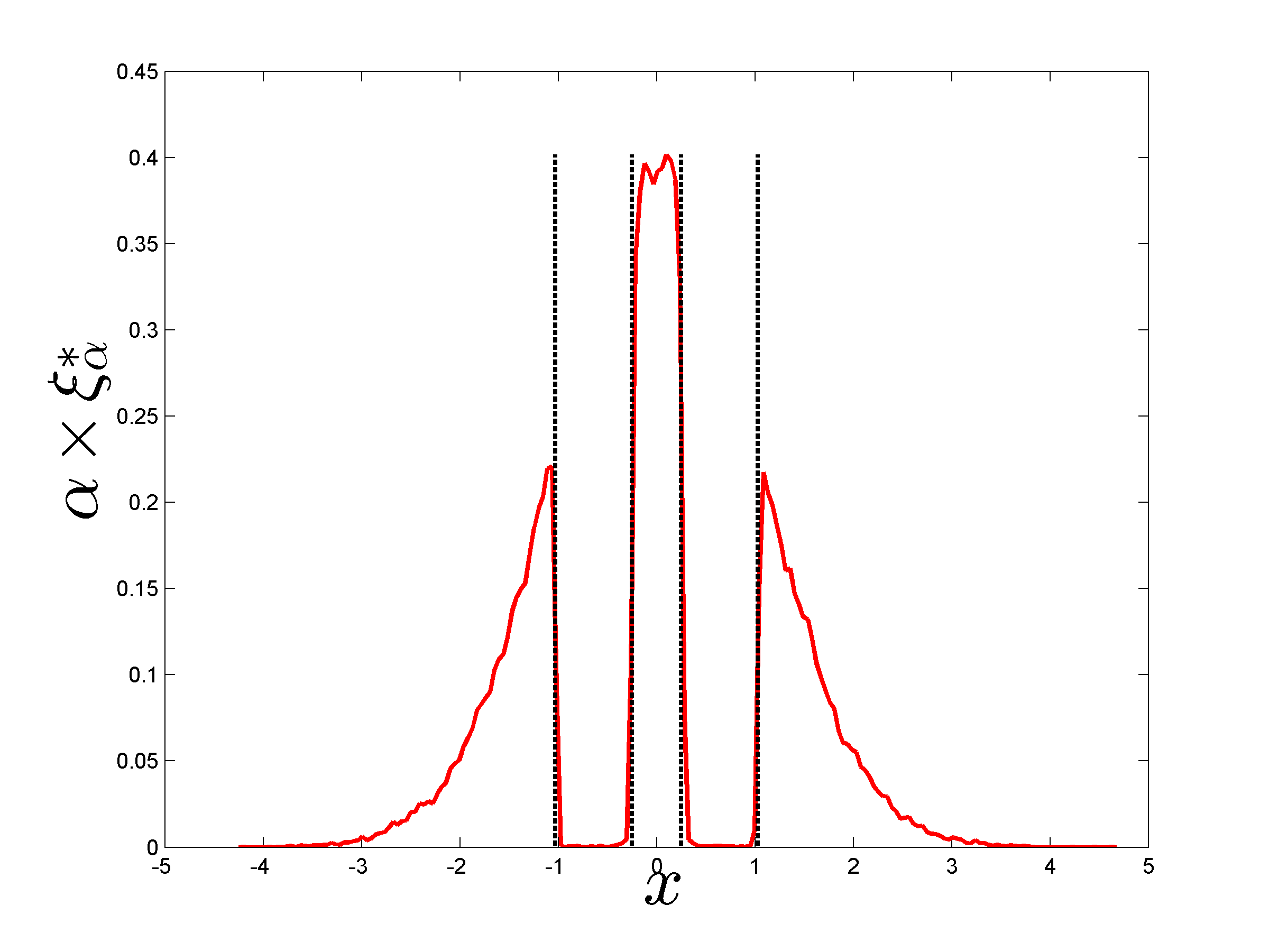} \includegraphics[width=.49\linewidth]{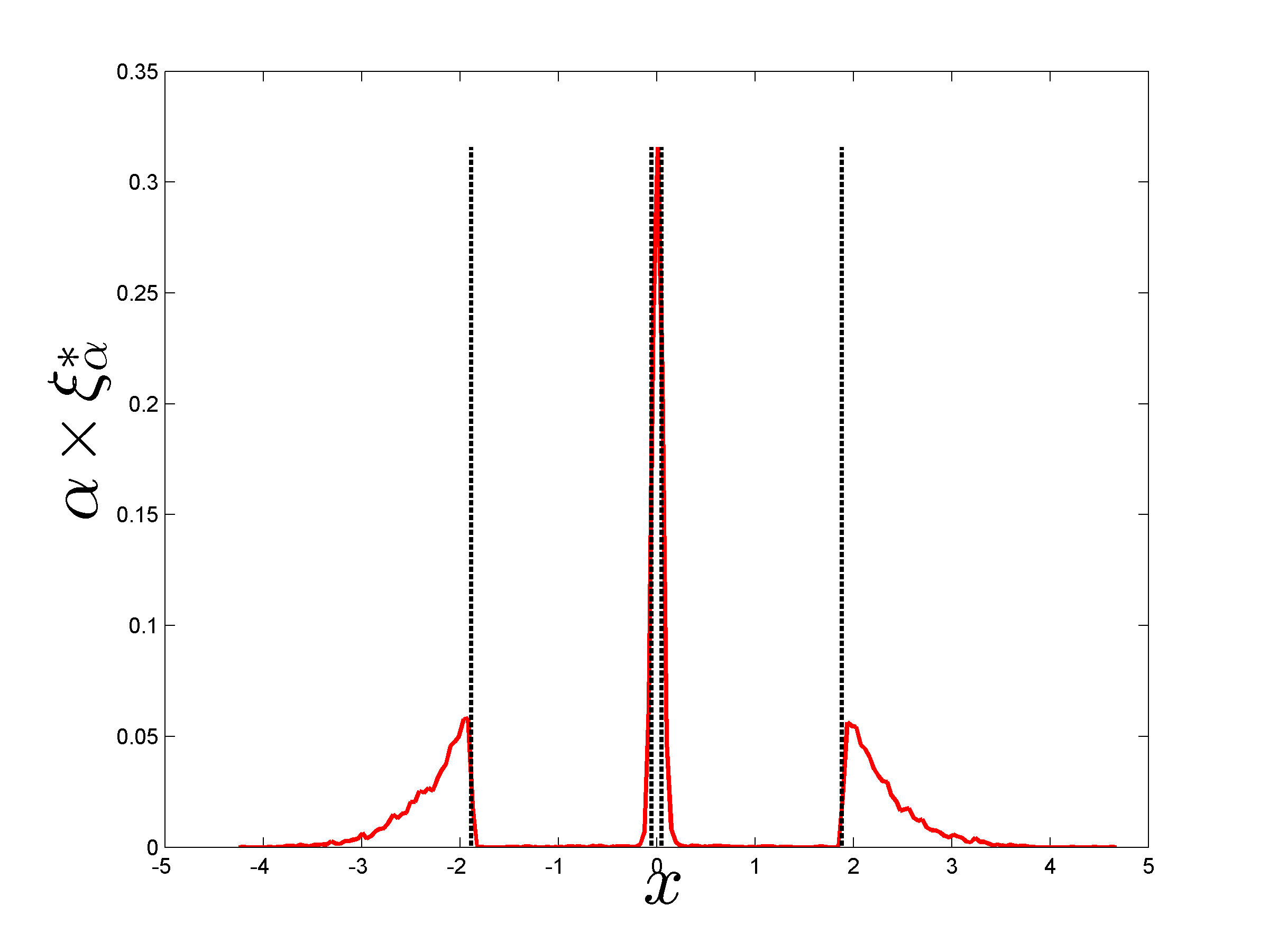}
\end{center}
\caption{\small Smoothed histogram of the $X_i$ selected by Algorithm~1; the vertical dotted lines indicate the positions of $-a,-b,b,a$ that define the set $\SX_\ma^*=(-\infty,-a]\cup [-b,b]\cup[a,\infty)$ where $\xi_\ma^*=\mu/\ma$; $N=100\,000$; Left: $\ma=1/2$ ($n=50\,000$); Right: $\ma=1/10$ ($n=10\,000$).}
\label{F:xiopt_quadratic_normal}
\end{figure}

\vsp
Figure~\ref{F:xiopt_quadratic_normal} shows a smoothed histogram (Epanechnikov kernel, bandwidth equal to $1/1000$ of the range of the $X_i$ in $\SX_N$) of the design points selected by Algorithm~1, for $\ma=1/2$ (left) and $\ma=1/10$ (right). There is good adequation with the theoretical optimal density, which corresponds to a truncation of the normal density at values indicated by the vertical dotted lines.

\begin{figure}[ht!]
\begin{center}
\includegraphics[width=.49\linewidth]{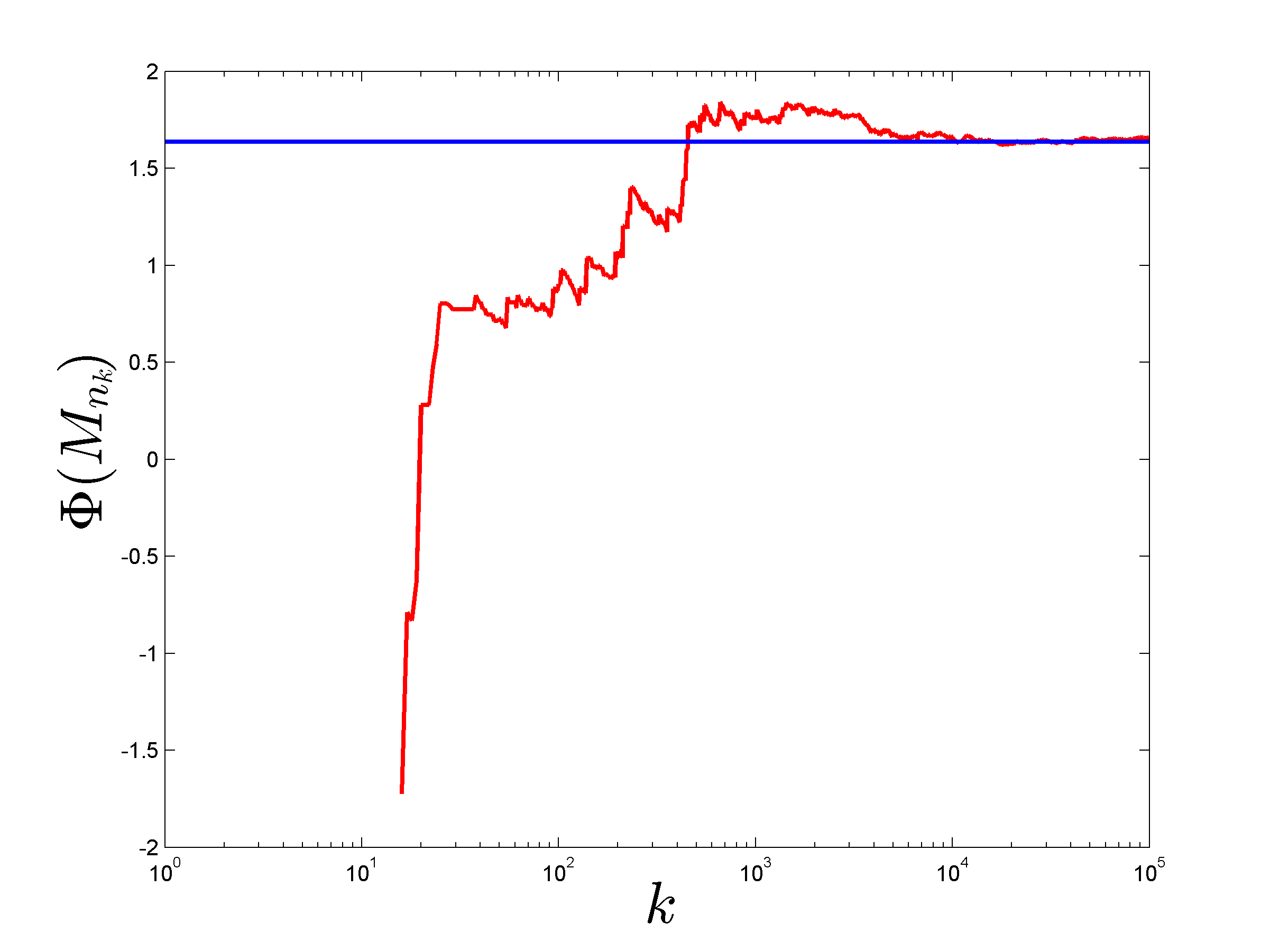} \includegraphics[width=.49\linewidth]{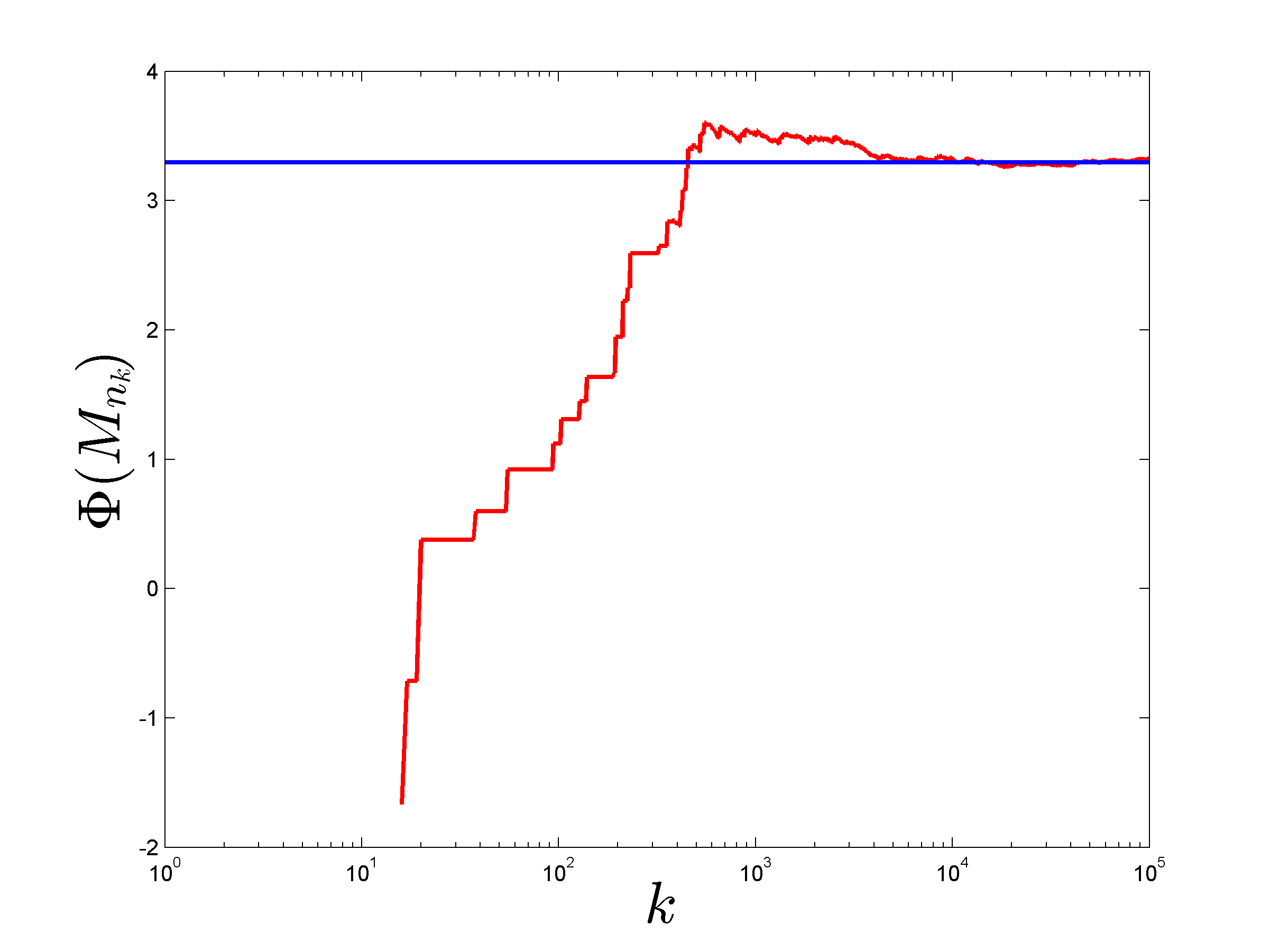}
\end{center}
\caption{\small Evolution of $\Phi(\Mb_{n_k})$ obtained with Algorithm~1 as a function of $k$ (log scale); the horizontal line indicates the optimal value $\Phi_\ma^*$; $N=100\,000$; Left: $\ma=1/2$ ($n=50\,000$); Right: $\ma=1/10$ ($n=10\,000$).}
\label{F:Phink_quadratic_normal}
\end{figure}

Figure~\ref{F:Phink_quadratic_normal} presents the evolution of $\Phi(\Mb_{n_k})$ as a function of $k$, together with the optimal value $\Phi_\ma^*$ (horizontal line), for the two choices of $\ma$ considered (the figures show some similarity on the two panels since the same set $\SX_N$ is used for both). Convergence of $\Phi(\Mb_{n_k})$ to $\Phi_\ma^*$ is fast in both cases; the presence of steps on the evolution of $\Phi(\Mb_{n_k})$, more visible on the right panel, is due to long subsequences of samples consecutively rejected.

\begin{figure}[ht!]
\begin{center}
\includegraphics[width=.49\linewidth]{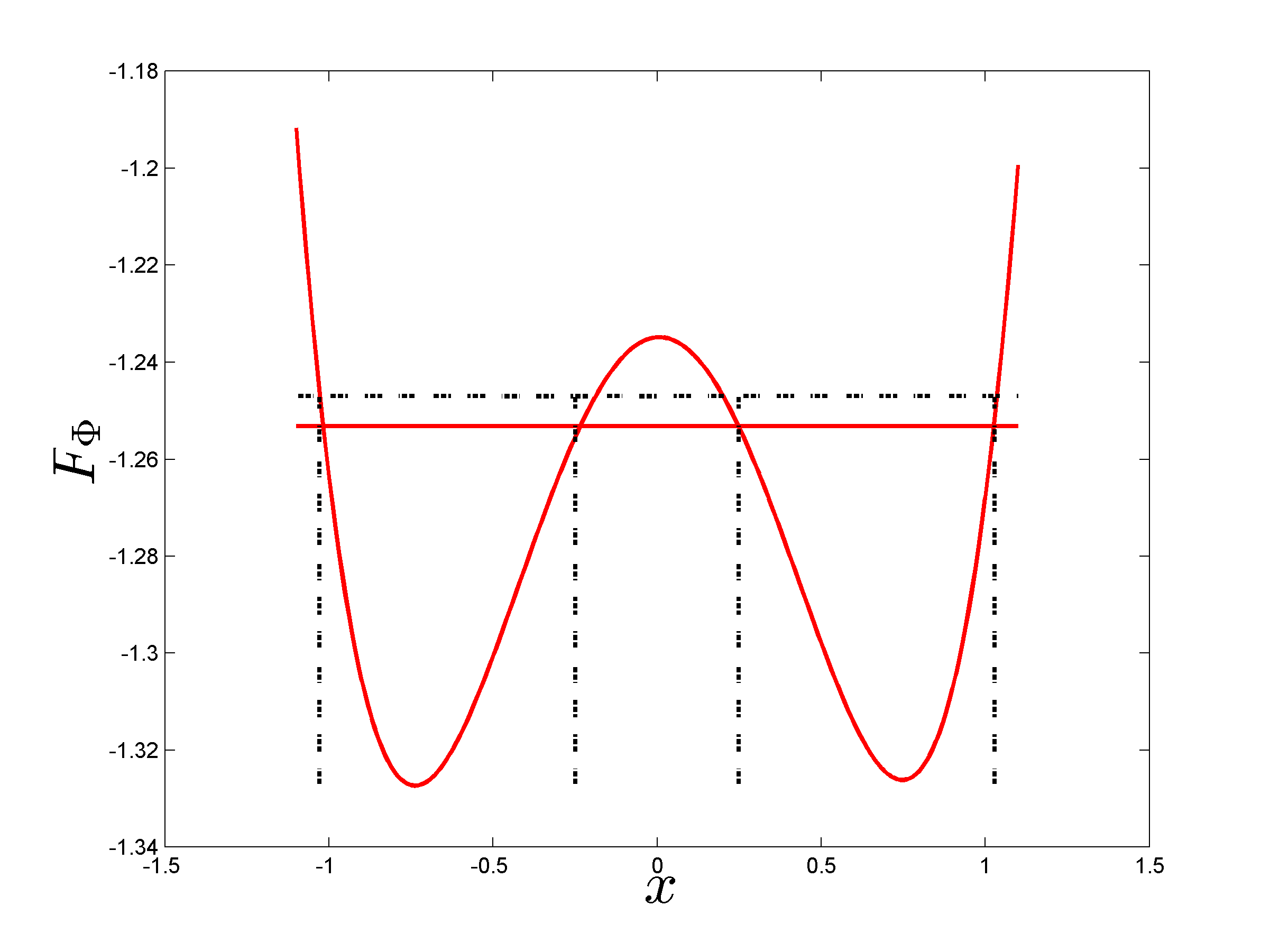} \includegraphics[width=.49\linewidth]{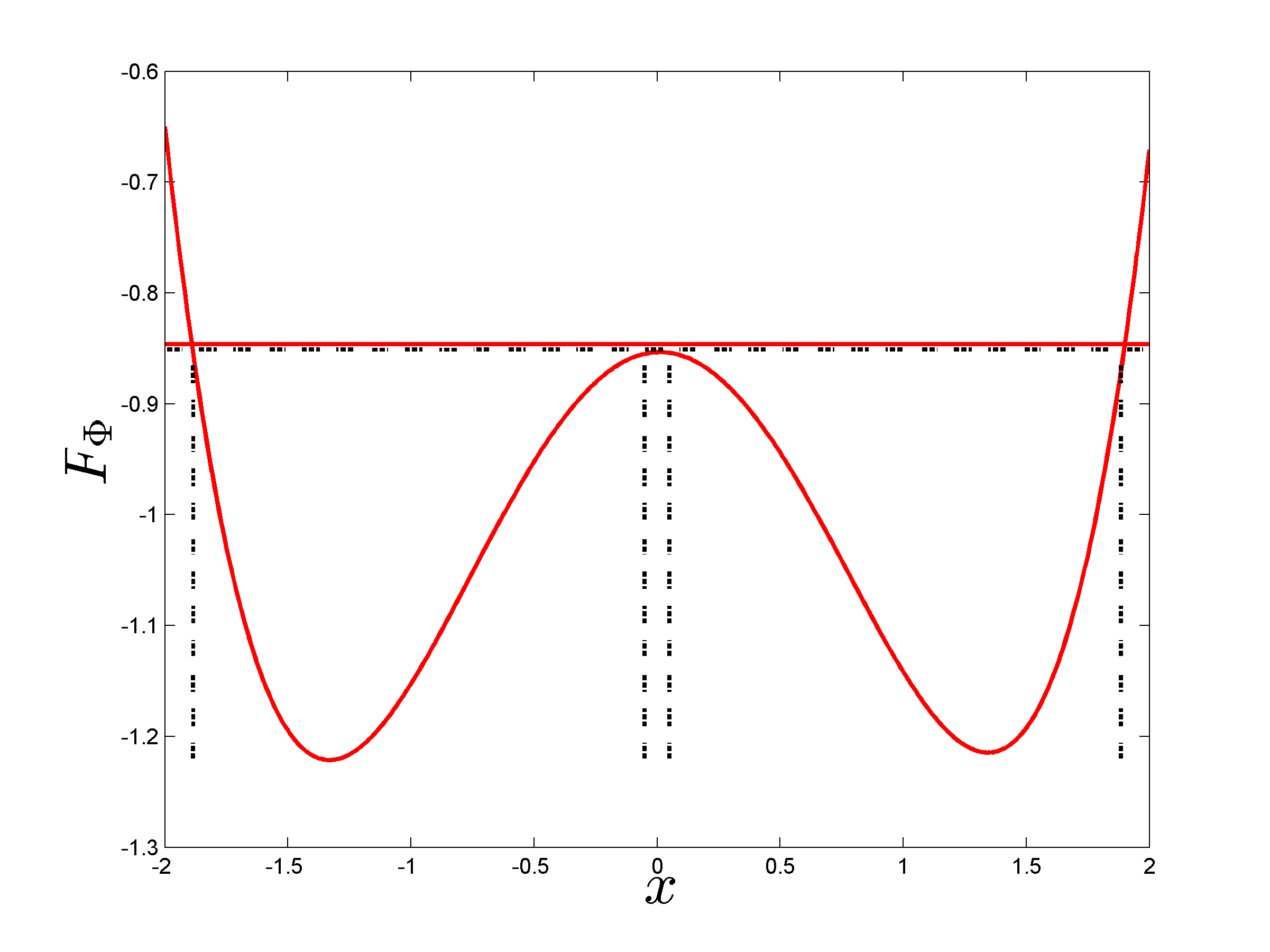}
\end{center}
\caption{\small $F_\Phi[\Mb_{n_N},\SM(x)]=\tr[\Mb_{n_N}^{-1}\,\SM(x)]-3$ as a function of $x$ (solid line); the horizontal solid (respectively, dashed) line indicates the value of $\hC_N$ (respectively, $C_{1-\ma}(\Mb_\ma^*)$), the vertical lines indicate the positions of $-a,-b,b,a$ where $F_\Phi[\Mb_\ma^*,\SM(x)]=C_{1-\ma}(\Mb_\ma^*)$; $N=100\,000$; Left: $\ma=1/2$ ($n=50\,000$); Right: $\ma=1/10$ ($n=10\,000$).}
\label{F:Dirder_quadratic_normal}
\end{figure}

Figure~\ref{F:Dirder_quadratic_normal} shows the behavior of the final directional derivative $F_\Phi[\Mb_{n_N},\SM(x)]$, after observation of all $X_i$ in $\SX_N$, together with the value of its estimated quantile $\hC_N$ (horizontal solid line). The theoretical values $C_{1-\ma}(\Mb_\ma^*)$ (horizontal dashed line) and the values $-a,-b,b,a$ where $F_\Phi[\Mb_\ma^*,\SM(x)]=C_{1-\ma}(\Mb_\ma^*)$ (vertical dashed lines) are also shown ($\hC_N$ and $C_{1-\ma}(\Mb_\ma^*)$ are indistinguishable on the right panel). Although the figure indicates that $F_\Phi[\Mb_{n_N},\SM(x)]$ differs significantly from $F_\Phi[\Mb_\ma^*,\SM(x)]$, they are close enough to allow selection of the most informative $X_i$, as illustrated by Figures~\ref{F:xiopt_quadratic_normal} and \ref{F:Phink_quadratic_normal}.

\begin{figure}[ht!]
\begin{center}
\includegraphics[width=.49\linewidth]{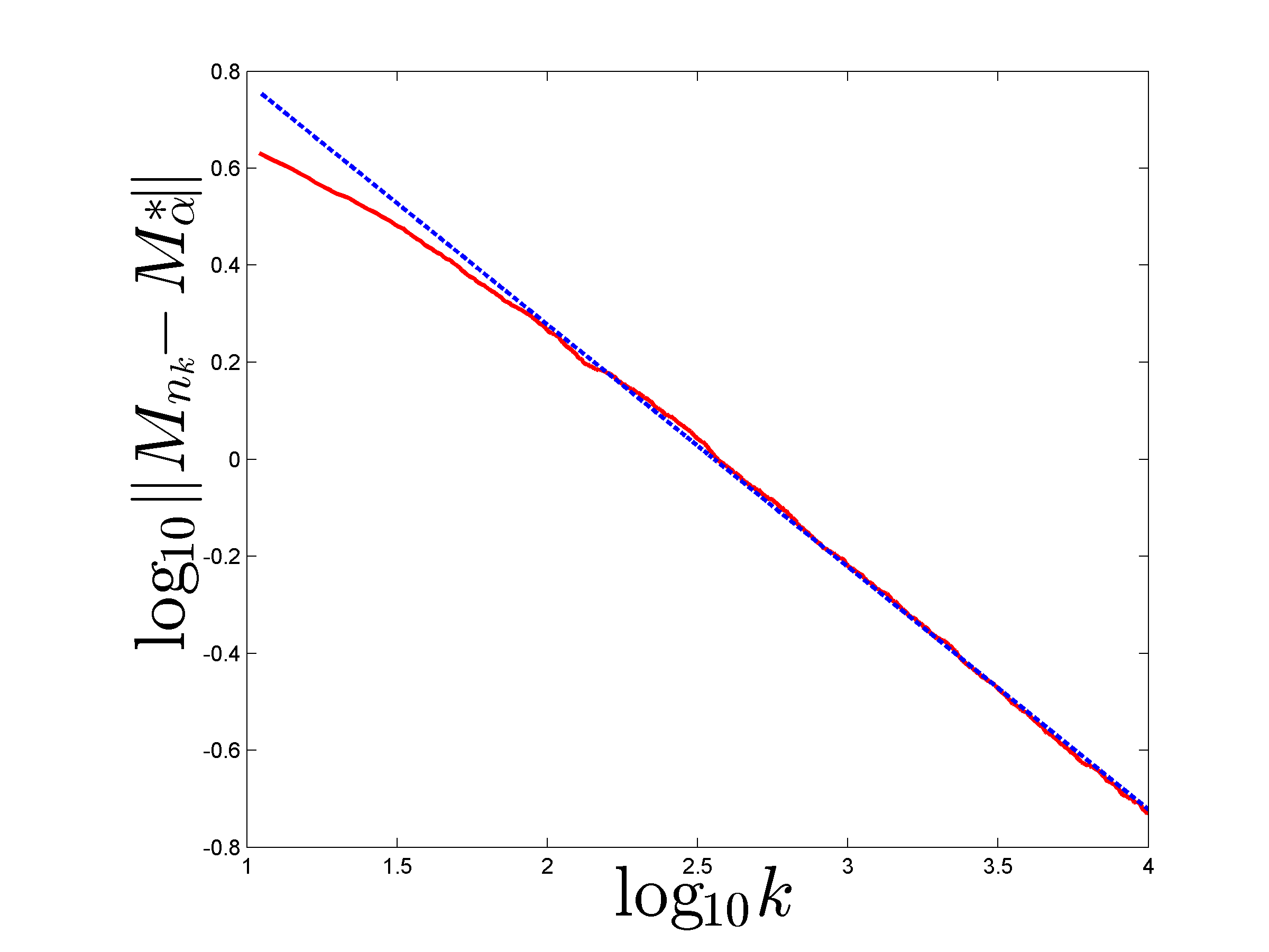} %
\end{center}
\caption{\small Evolution of $\log_{10}\,\|\Mb_{n_k}-\Mb_\ma^*\|$ as a function of $\log_{10}\,k$ (values averaged over 1\,000 random samples $\SX_N$); the dashed line has slope $-1/2$ ($\ma=1/2$: $n=5\,000$, $N=10\,000$).}
\label{F:DeltaM_quadraticnormal}
\end{figure}

Figure~\ref{F:DeltaM_quadraticnormal} shows $\|\Mb_{n_k}-\Mb_\ma^*\|$ (Frobenius norm) as a function of $k$ (log scale), averaged over $1\,000$ independent repetitions with random samples $\SX_N$ of size $N=10\,000$, for $\ma=1/2$. It suggests that $\|\Mb_{n_k}-\Mb_\ma^*\|=\SO(1/\sqrt{k})$ for large $k$, although the conditions in \citep{KondaT2004} are not satisfied since the scheme we consider is nonlinear. This convergence rate is significantly faster than what is suggested by \citet{DalalSTM2018}. These investigations require further developments and will be pursued elsewhere.

\subsection{Example 2: multilinear regression with normal independent variables}\label{S:Ex2}

Take $\SM(X)=XX\TT$, with $X=(x_1,\ x_2,\ldots,x_d)\TT$, $d>1$, and $\Phi(\Mb)=\log\det(\Mb)$, the vectors $X_i$ being i.i.d.\ $\SN(\0b,\Ib_d)$ (so that $p=d$). Denote by $\varphi(\xb)=(2\,\pi)^{-d/2}\,\exp(- \|\xb\|^2/2)$ the probability density of $X$.
For symmetry reasons, for any $\ma\in(0,1)$ the optimal (normalized) information matrix is $\Mb_\ma^*=\rho_\ma\, \Ib_d$, with $\Phi(\Mb_\ma^*)=d\, \log \rho_\ma$, where
\bea
\rho_\ma = \frac{1}{\ma}\, \int_{\|\xb\|\geq R_\ma} x_1^2\, \varphi(\xb)\, \dd\xb &=& \frac{1}{d\,\ma}\, \int_{\|\xb\|\geq R_\ma} \|\xb\|^2\, \varphi(\xb)\, \dd\xb \\
&=& \frac{1}{d\,\ma}\, \int_{r \geq R_\ma} r^2\, \frac{1}{(2\,\pi)^{d/2}}\,\exp(-r^2/2)\,S_d(1)\,r^{d-1}\, \dd r \,,
\eea
with $S_d(1)=2\,\pi^{d/2}/\Gamma(d/2)$, the surface area of the $d$-dimensional unit sphere, and $R_\ma$ the solution of
\bea
\ma = \int_{\|\xb\|\geq R_\ma} \varphi(\xb)\, \dd\xb = \int_{r \geq R_\ma} \frac{1}{(2\,\pi)^{d/2}}\,\exp(-r^2/2)\,S_d(1)\,r^{d-1}\, \dd r \,.
\eea
Since $F_\Phi[\Mb,\SM(X)]=\tr[\Mb^{-1}\SM(X)]-d$, we get $F_\Phi[\Mb_\ma^*,\SM(X)]=\|\xb\|^2/\rho_\ma-d$, $C_{1-\ma}(\Mb_\ma^*)=R_\ma^2/\rho_\ma-d \leq 0$, and $\Phi_\ma^*=\Phi(\Mb_\ma^*)$ is differentiable with respect to $\ma$, with $\dd \Phi_\ma^*/\dd\ma=C_{1-\ma}(\Mb_\ma^*)/\ma$; see \citet[Th.~4]{Pa04}.
Closed-form expressions are available for $d=2$, with $R_\ma=\sqrt{-2\,\log \ma}$ and $\rho_\ma = 1-\log \ma$; $R_\ma$ and $\rho_\ma$ can easily be computed numerically for any $d>2$ and $\ma\in(0,1)$. One may notice that, from a result by \citet{Harman2004-MODA}, the design matrix $\Mb_\ma^*$ is optimal for any other orthogonally invariant criterion $\Phi$.

For the linear model with intercept, such that $\SM'(X)=\fb(X)\fb\TT(X)$ with $\fb(X)=[1,\ X\TT]\TT$, the optimal matrix is
\bea %
{\Mb'}_\ma^* = \left(
                 \begin{array}{cc}
                   1 & \0b\TT \\
                   \0b & \Mb_\ma^* \\
                 \end{array}
               \right)
\eea
with $\Mb_\ma^*=\rho_\ma\, \Ib_d$ the optimal matrix for the model without intercept. The same design is thus optimal for both models. Also, when the $X_i$ are i.i.d.\ $\SN(0,\Sigmab)$, the optimal matrix $\Mb_{\Sigma,\ma}^*$ for $\Phi(\cdot)=\log\det(\cdot)$ simply equals $\Sigmab^{1/2}\,\Mb_\ma^*\,\Sigmab^{1/2}$.

\begin{figure}[ht!]
\begin{center}
\includegraphics[width=.49\linewidth]{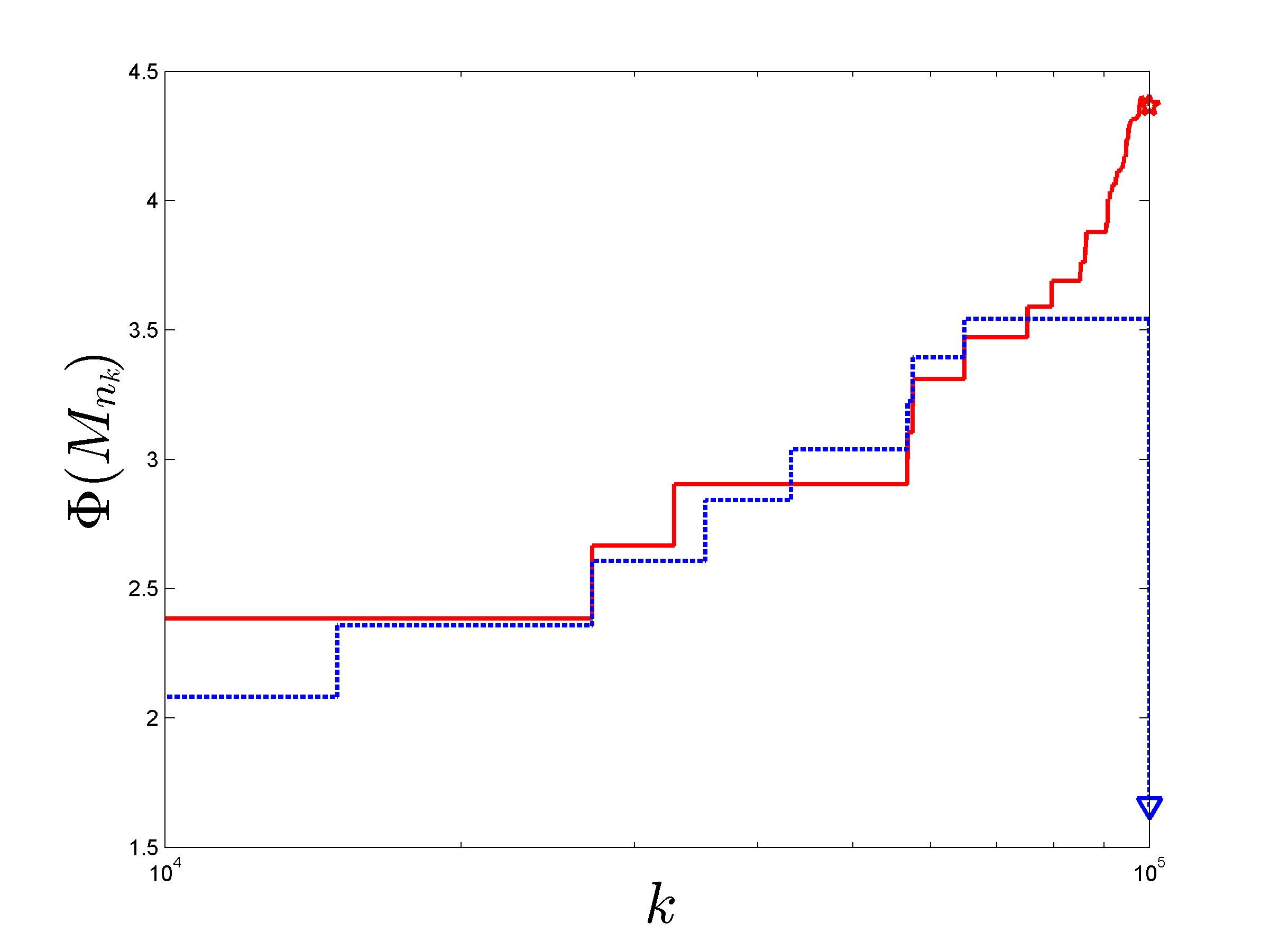}
\end{center}
\caption{\small Evolution of $\Phi(\Mb_{n_k})$ obtained with Algorithm~1 as a function of $k$ (log scale):
$d=3$, $N=100\,000$, 
exactly $n=100$ points are collected using truncation/forced selection (blue dashed line and $\triangledown$) or adaptation of $\ma$ (red solid line and $\bigstar$); see Remark~\ref{R:R1}-(\textit{iv}).}
\label{F:Phink_linearnormal}
\end{figure}

\vsp
Again, we present results obtained for one random set $\SX_N$.
Figure~\ref{F:Phink_linearnormal} shows the evolution of $\Phi(\Mb_{n_k})$ as a function of $k$ for $d=3$ with $\ma=1/1\,000$ and $N=100\,000$ when we want we select exactly $100$ points: the blue dashed-line is when we combine truncation and forced selection; the red solid line is when we adapt $\ma$ according to $\ma_k=(n-n_k)/(N-k)$; see Remark~\ref{R:R1}-(\textit{iv}) --- the final values, for $k=N$, are indicated by a triangle and a star, respectively; we only show the evolution of $\Phi(\Mb_{n_k})$ for $k$ between 10\,000 and 100\,000 since the curves are confounded for smaller $k$ (they are based on the same $\SX_N$). In the first case, the late forced selection of unimportant $X_i$ yields a significant decrease of $\Phi(\Mb_{n_k})$, whereas adaptation of $\ma$ anticipates the need of being less selective to reach the target number $n$ of selected points.

\begin{figure}[ht!]
\begin{center}
\includegraphics[width=.49\linewidth]{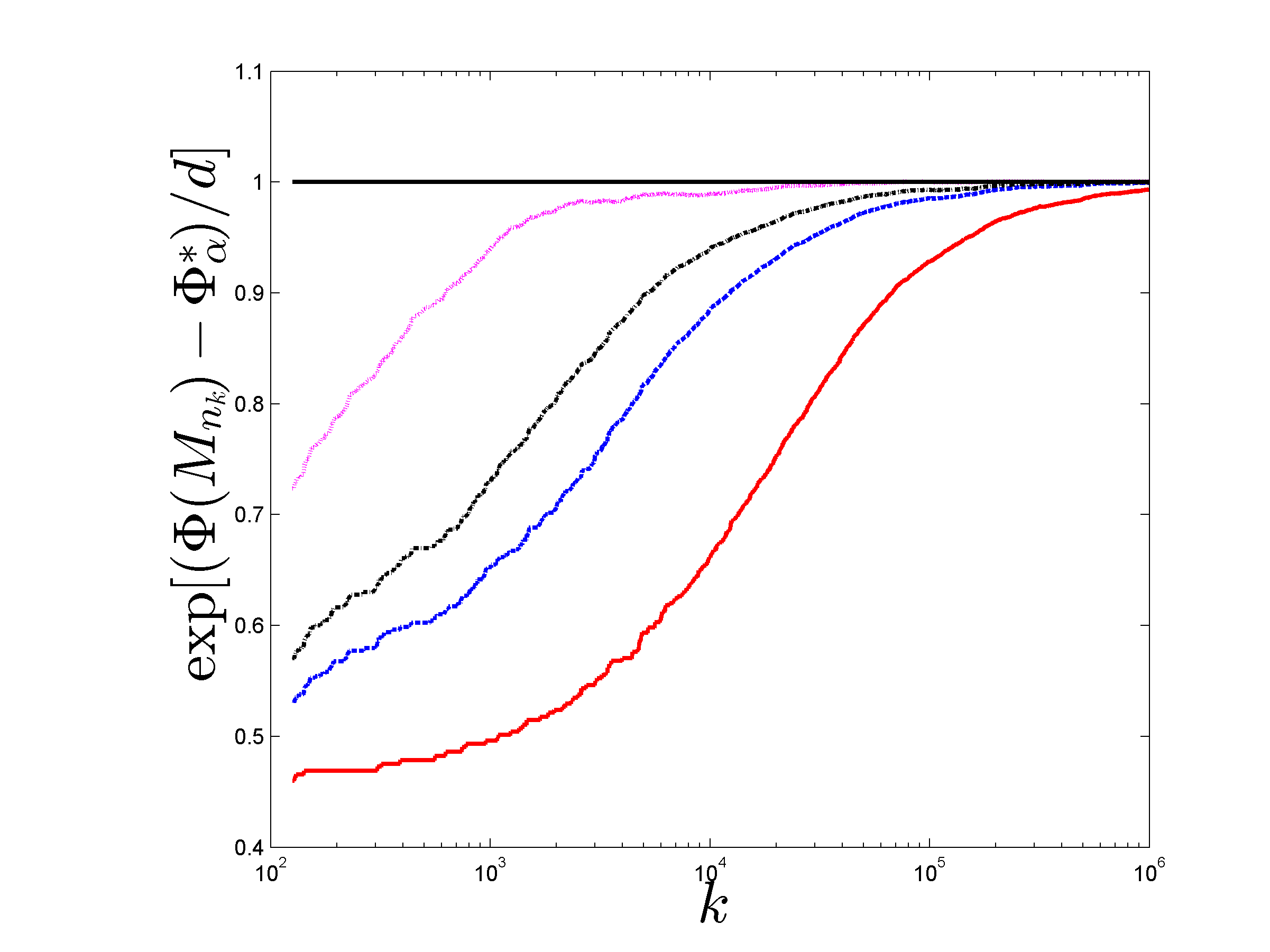} \includegraphics[width=.49\linewidth]{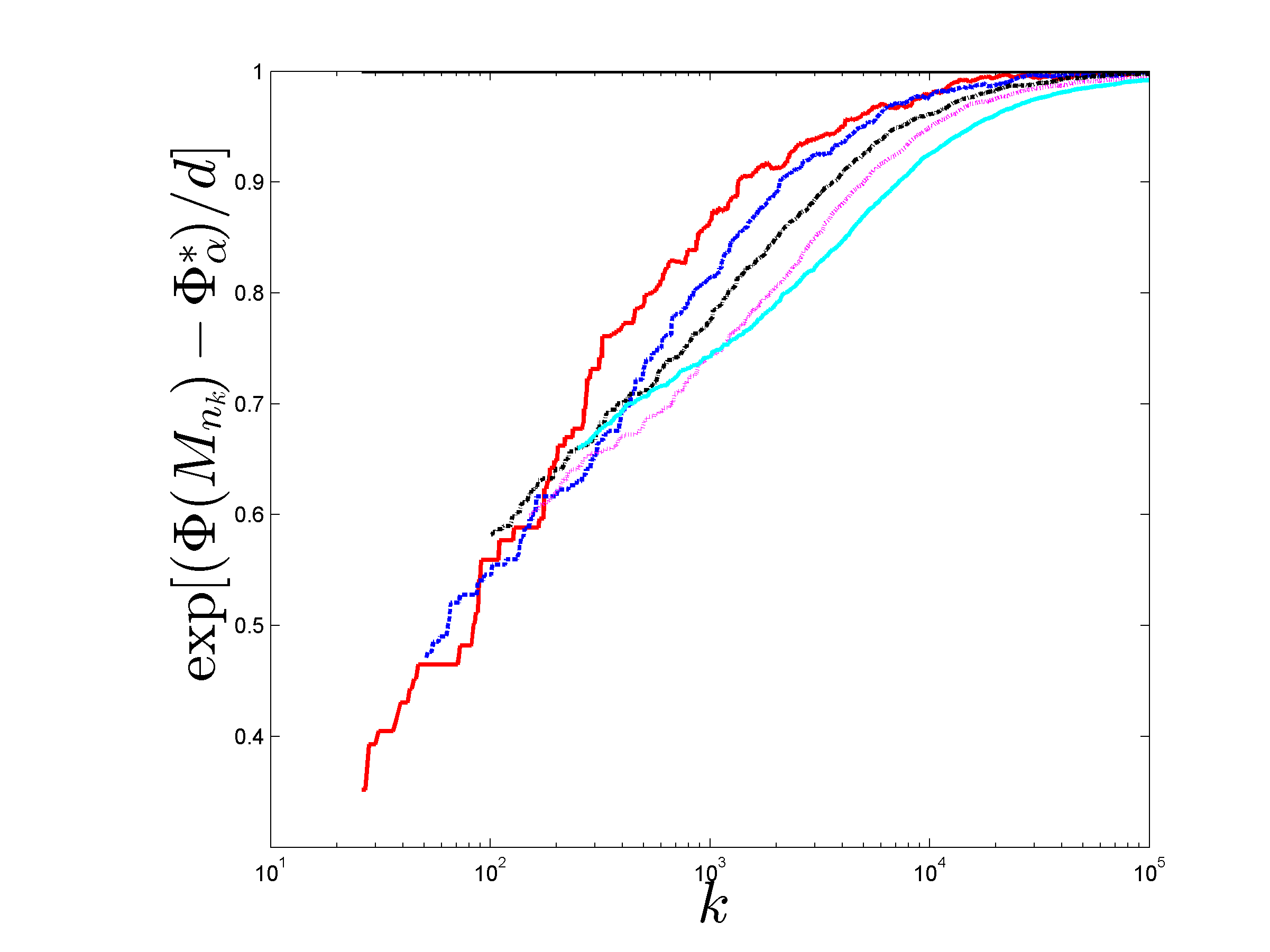}
\end{center}
\caption{\small Evolution of D-efficiency of $\Mb_{n_k}$ obtained with Algorithm~1 as a function of $k$ (log scale); the horizontal line indicates the optimal value 1. Left: $d=25$ and $\ma= 0.5$ (magenta dotted line), 0.1 (black), 0.05 (blue) and 0.01 (red solid line).
Right: $\ma=0.1$ and $d=5$ (red solid line), 10 (blue), 20 (black), 30 (magenta) and 50 (cyan solid line).}
\label{F:Phink_linearnormal2}
\end{figure}

Figure~\ref{F:Phink_quadratic_normal} has illustrated
the convergence of $\Phi(\Mb_{n_k})$ to $\Phi_\ma^*$ for a fixed $\ma$ as $k\ra\infty$, but in fact what really matters is that $n_k$ tends to infinity: indeed, $\Phi(\Mb_{n_k})$ does not converge to $\Phi_\ma^*$ if we fix $n_k=n$ and let $k$ tend to infinity, so that $\ma=n/k$ tends to zero (see also Section~\ref{S:exchange}). This is illustrated on the left panel of Figure~\ref{F:Phink_linearnormal2}, where $d=25$ and, from left to right, $\ma$ equals 0.5 (magenta dotted line), 0.1, 0.05 and 0.01 (red solid line). Since the optimal value $\Phi_\ma^*$ depends on $\ma$, here we present the evolution with $k$ of the D-efficiency $[\det(\Mb_{n_k})/\det(\Mb_\ma^*)]^{1/p}=\exp[(\Phi(\Mb_{n_k})-\Phi_\ma^*)/d]$.
The right panel is for fixed $\ma=0.1$ and varying $d$, with, from left to right, $d=5$ (red solid line), 10, 20, 30 and 50 (cyan solid line). As one may expect, performance (slightly) deteriorates as $d$ increases due to the increasing variability of $Z_\Mb(X)$, with $\var[Z_\Mb(X)]=\var[X\TT\Mb X]=2\,\tr(\Mb^2)$.

\subsection{Example 3: processing a non i.i.d.\ sequence}\label{S:non-iid}

When the design points $X_i$ are not from an i.i.d.\ sequence, Algorithm~1 cannot be used directly and some preprocessing is required. When storage of the whole sequence $\SX_N$ is possible, a random permutation can be applied to $\SX_N$ before using Algorithm~1. When $N$ is too large for that, for instance in the context of data streaming, and the sequence possesses a structured time-dependence, one may try to identify the dependence model through time series analysis and use forecasting to decide which design points should be selected. The data thinning mechanism is then totally dependent on the model of the sequence, and the investigation of the techniques to be used is beyond the scope of this paper. Examples of the application of a simple scrambling method to the sequence prior to selection by Algorithm~1 are presented below. The method corresponds to Algorithm~2 below; its output sequence $\widetilde X_k$ is used as input for Algorithm~1. We do not study the properties of the method in conjunction with the convergence properties of Algorithm~1, which would require further developments.

\begin{samepage}
\begin{itemize}
\item[] {\bf Algorithm~2: random scrambling in a buffer.}
  \item[1)] Initialization: choose the buffer size $B$, set $k=1$ and $\SX^{(1)}=\{X_1,\ldots,X_B\}$.

  \item[2)] Draw $\widetilde X_k$ by uniform sampling within $\SX^{(k)}$.

  \item[3)] Set $\SX^{(k+1)}=\SX^{(k)}\setminus\{\widetilde X_k\} \cup \{X_{B+k}\}$, $k \leftarrow k+1$, return to Step 2.
\end{itemize}
\end{samepage}

Direct calculation shows that the probability that $\widetilde X_k$ equals $X_i$ is
\bea
\Prob\{\widetilde X_k=X_i\} = \left\{
\begin{array}{ll}
\frac1B\, \left(1-\frac1B\right)^{k-1} & \mbox{ for } 1\leq i \leq B \\
\frac1B\, \left(1-\frac1B\right)^{k-1+B-i} & \mbox{ for } B+1 \leq i \leq B+k-1 \\
0 & \mbox{ for } B+k-1 < i
\end{array}
\right.
\eea
showing the limits of randomization via Algorithm~2 (in particular, the first points of the sequence $X_i$ will tend to appear first among the $\widetilde X_k$). However, the method will give satisfactory results if the size $B$ of the buffer is large enough, as its performance improves as $B$ increases.

As an illustration, we consider the same quadratic regression model as in Example~1, with $\Phi(\Mb)=\log \det(\Mb)$, in the extreme case where $X_i=x(t_i)$ with $x(t)$ a simple function of $t$.

\begin{figure}[ht!]
\begin{center}
\includegraphics[width=.49\linewidth]{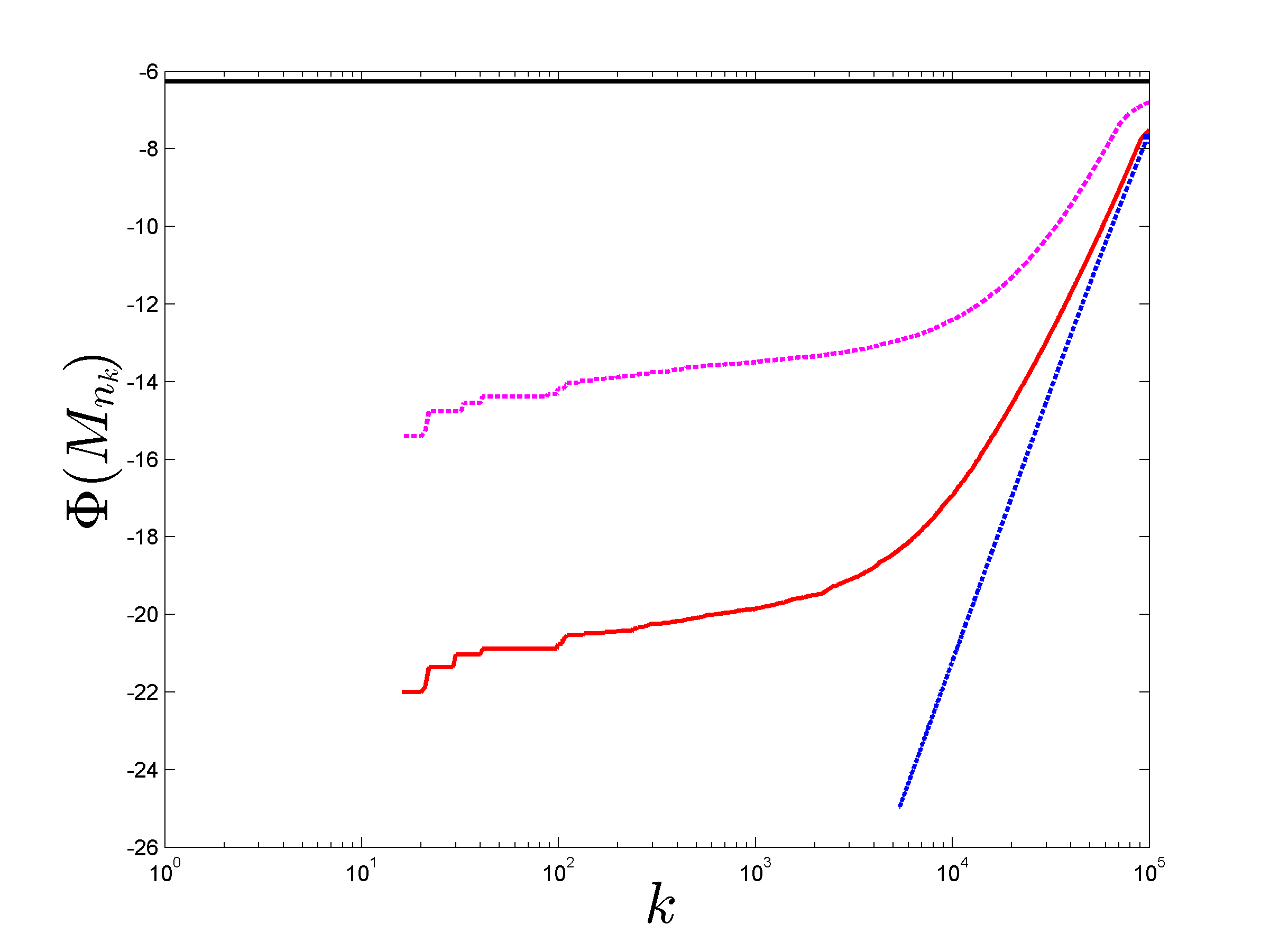} \end{center}
\caption{\small Evolution of $\Phi(\Mb_{n_k})$ obtained with Algorithm~1 as a function of $k$ (log scale), the horizontal line indicates the optimal value $\Phi_\ma^*$; $N=100\,000$, $\ma=1/10$; direct processing of the points $X_i=i/N$ (dotted blue line), preprocessing with Algorithm~2 with $B=\ma N$ (red solid line) and $B=3\,\ma N$ (magenta dotted line).}
\label{F:quadratic-t-buffer}
\end{figure}

First, we consider the extremely unfavorable situation where $x(t)=t$. When $t$ is uniformly distributed on $\ST=[0,T]$, the optimal design $\xi_\ma^*$ selects all points associated with $t$ in $[0,t_a]\cup[t_b,T-t_b]\cup[T-t_a,T]$, for some $t_a<t_b$ in $\ST$ satisfying $2\,t_a+T-2\,t_b=\ma\,T$. For $\ma=1/10$, we get $t_a/T \simeq 0.03227$ and $t_b/T\simeq 0.48227$.
The horizontal black line in Figure~\ref{F:quadratic-t-buffer} indicates the optimal value $\Phi_\ma^*$ when $T=1$. The blue dotted line shows $\Phi(\Mb_{n_k})$ when Algorithm~1 is directly applied to the points $X_i=i/N$, $i=1,\ldots,N=100\,000$. The red line is when randomization via Algorithm~2, with buffer size $B=\ma N=10\,000$, is applied first; the dotted curve in magenta is for $B=3\,\ma N$. The positive effects of randomization through Algorithm~2 and the influence of the buffer size are visible on the figure.
Here, the monotonicity of the $X_i$, which inhibits early exploration of their range of variation, prevents convergence to the optimum.

\begin{figure}[ht!]
\begin{center}
\includegraphics[width=.49\linewidth]{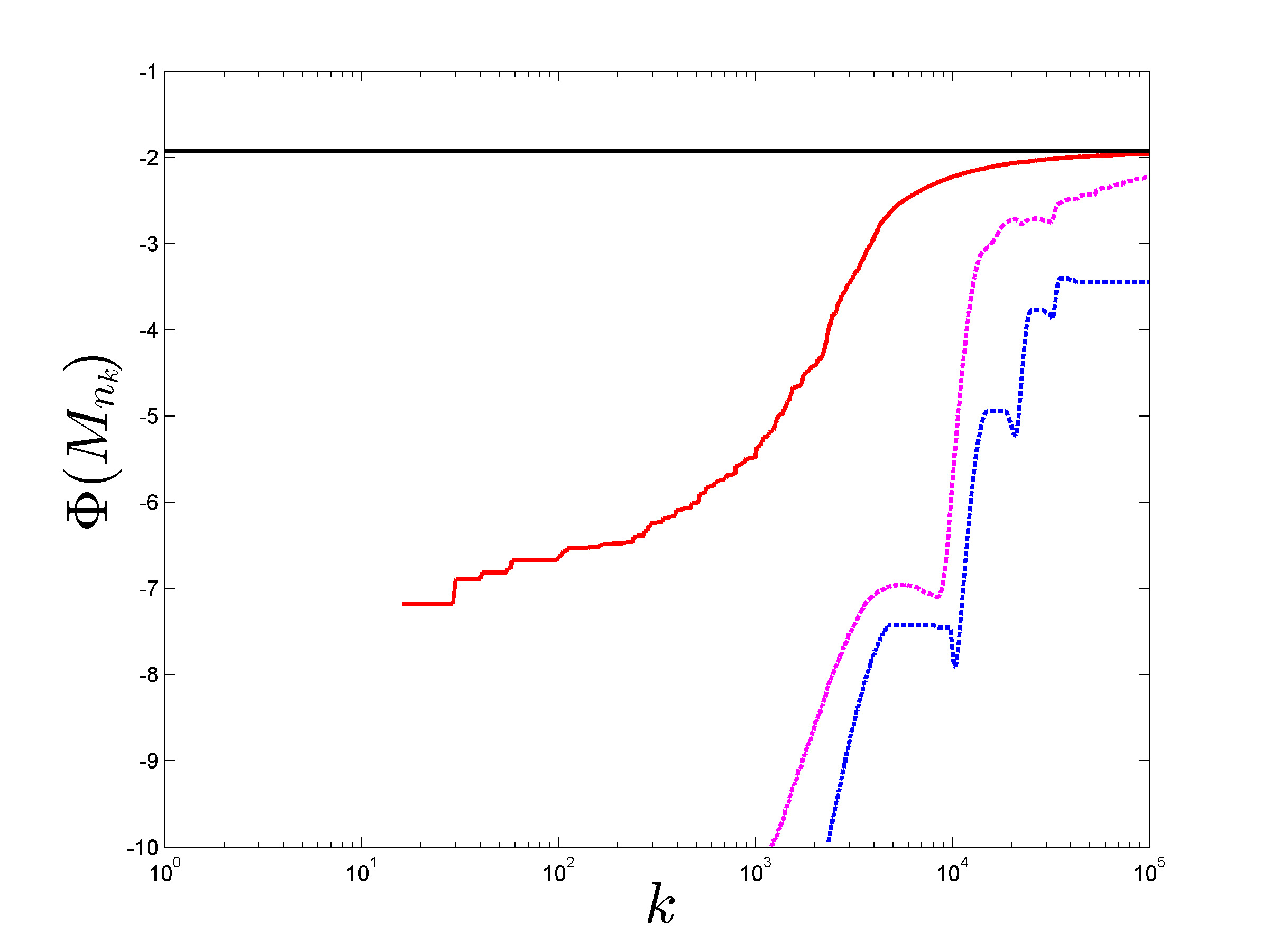}
\includegraphics[width=.49\linewidth]{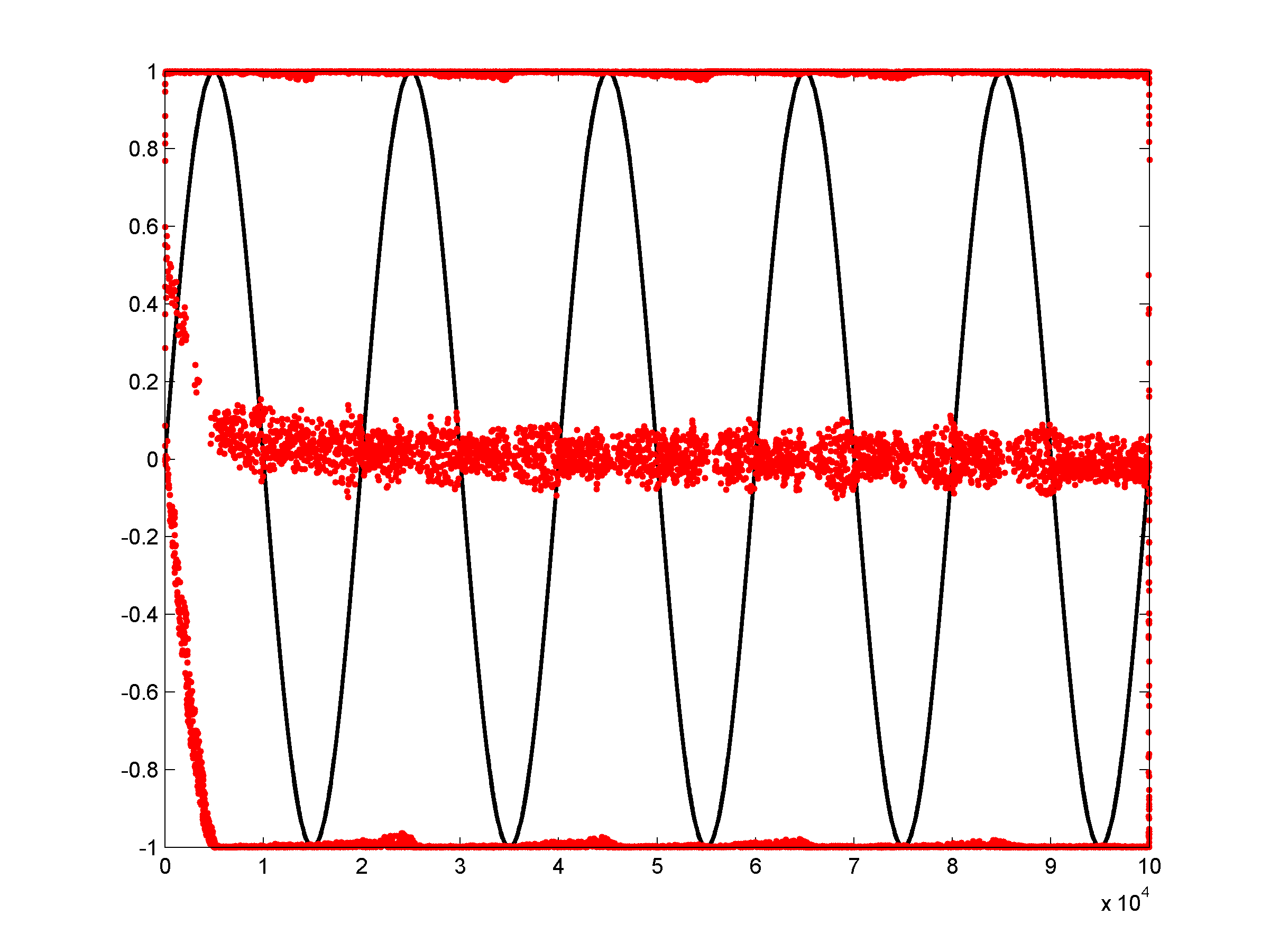}
\end{center}
\caption{\small Left: Evolution of $\Phi(\Mb_{n_k})$ obtained with Algorithm~1 as a function of $k$ (log scale), the horizontal line indicates the optimal value $\Phi_\ma^*$; $N=100\,000$, $\ma=1/10$; direct processing of the points $X_i=\sin(2\pi \nu i/N)$ (dotted blue line), preprocessing with Algorithm~2 with $B=\ma N$ (red solid line) and $B=\ma N/10$ (magenta dotted line). Right: $X_i$, $i=1,\ldots,N=100\,000$ (black) and points $\widetilde X_k$ produced by Algorithm~2 with $B=\ma N$ which are selected by Algorithm~1 (red).}
\label{F:quadratic-sin-buffer}
\end{figure}

We now consider the more favorable case where $X_i=\sin(2\pi \nu i/N)$, $i=1,\ldots,N=100\,000$, with $\nu=5$.
The left panel of Figure~\ref{F:quadratic-sin-buffer} shows the same information as Figure~\ref{F:quadratic-t-buffer}, when Algorithm~1 is applied directly to the $X_i$ (blue dotted line) and after preprocessing with Algorithm~2 with $B=\ma N$ (red line) and $B=\ma N/10$ (magenta dotted line). The early exploration of the range of variability of the $X_i$, possible here thanks to the periodicity of $x(t)$, makes the randomization through Algorithm~2 efficient enough to allow Algorithm~1 to behave correctly when $B=\ma N$ (red line). The situation improves when $B$ is increased, but naturally deteriorates if $B$ is too small (magenta dotted line). The right panel shows the points $\widetilde X_k$ produced by Algorithm~2 (with $B=\ma N= 10\,000$) which are selected by Algorithm~1. The effect of randomization is visible. For $k<5\,000$ all points in the buffer are in the interval $[\sin(2\pi \nu (k+B)/N),1] \subset [-1,1]$ and the points selected by Algorithm~1 are near the end points or the center of this moving interval. For larger $k$, randomization is strong enough to maintain the presence of suitable candidates in the buffer for selection by Algorithm~1.

\subsection{Examples with $Z_\Mb(x)$ constant on subsets of positive measure}\label{S:Ex3}

Here we consider situations where H$_{\mu,\SM}$ is violated due to the existence of subsets of $\SX$ of positive measure on which $Z_\Mb(x)$ is constant. The model is the same as in Section~\ref{S:Ex2}, with $X=(x_1,\ x_2,\ldots,x_d)\TT$, $\SM(X)=XX\TT$ and $\Phi(\Mb)=\log\det(\Mb)$.

\subsubsection{Example 4: $\mu$ has discrete components}

This is Example~11 in \citep{Pa05}, where $d=2$, $\mu=(1/2)\,\mu_\SN+(1/2)\,\mu_d$, with $\mu_\SN$ corresponding to the normal distribution $\SN(0,1)$ and $\mu_d$ the discrete measure that puts weight $1/4$ at each one of the points $(\pm 1,\pm1)$. Denote by $\SB(r)$ the closed ball centered at the origin $\0b$ with radius $r$, by $\mu[r]$ the measure equal to $\mu$ on its complement $\overline{\SB(r)}$, and let $\e1=\exp(1)$. The optimal matrix is $\Mb_\ma^*=\Mb(\xi_\ma^*)$, with $\xi_\ma^*$ the probability measure defined by:
\bea
\xi_\ma^* = \left\{
\begin{array}{ll}
\frac{1}{\ma}\, \mu[\sqrt{-2\,\log(2\,\ma)}]  &  \mbox{ if } 0<\ma\leq \frac{1}{2\e1} \,, \\
\frac{1}{\ma}\, \mu[\sqrt{2}]+ \frac{1}{\ma}\,[\ma-1/(2\,\e1)]\mu_d  &  \mbox{ if } \frac{1}{2\e1} <\ma\leq \frac{1}{2\e1} + \frac12 \,,\\
\frac{1}{\ma}\, \mu[\sqrt{-2\,\log(2\,\ma-1)}] &  \mbox{ if } \frac{1}{2\e1} + \frac12 <\ma <1 \,,
\end{array}
\right.
\eea
with associated $\Phi$-values
\bea
\Phi(\Mb_\ma^*) = \left\{
\begin{array}{ll}
2\, \log[1-\log(2\,\ma)] &  \mbox{ if } 0<\ma\leq \frac{1}{2\e1} \,,\\
2\, \log\left(1+\frac{1}{2\e1\,\ma}\right) &  \mbox{ if } \frac{1}{2\e1} <\ma\leq \frac{1}{2\e1} + \frac12 \,,\\
2\, \log\left(1 - \frac{(2\,\ma-1)\,\log(2\,\ma-1)}{2\,\ma} \right) &  \mbox{ if } \frac{1}{2\e1} + \frac12 <\ma \leq 1 \,.
\end{array}
\right.
\eea

\begin{figure}[ht!]
\begin{center}
\includegraphics[width=.49\linewidth]{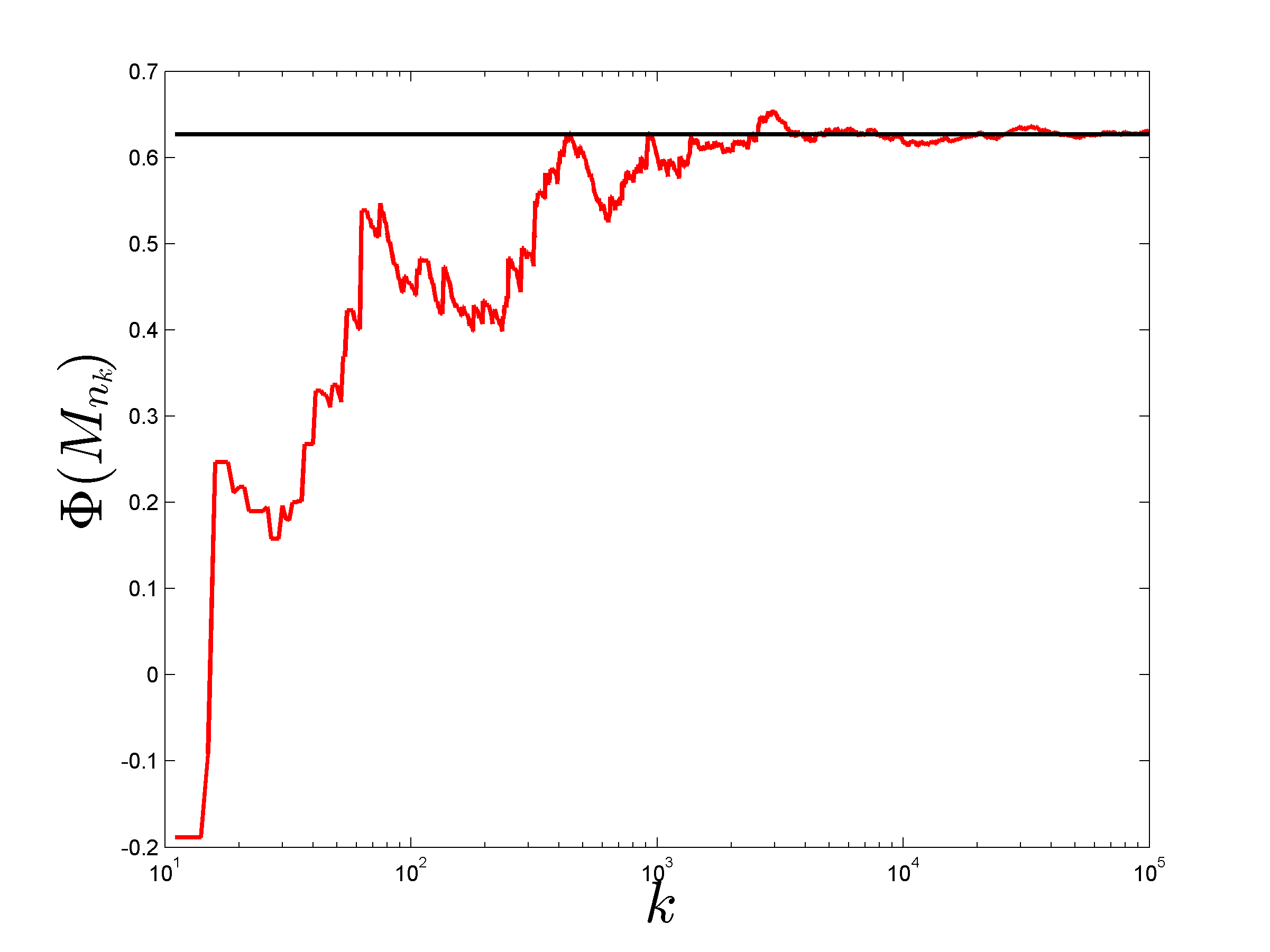}  \includegraphics[width=.49\linewidth]{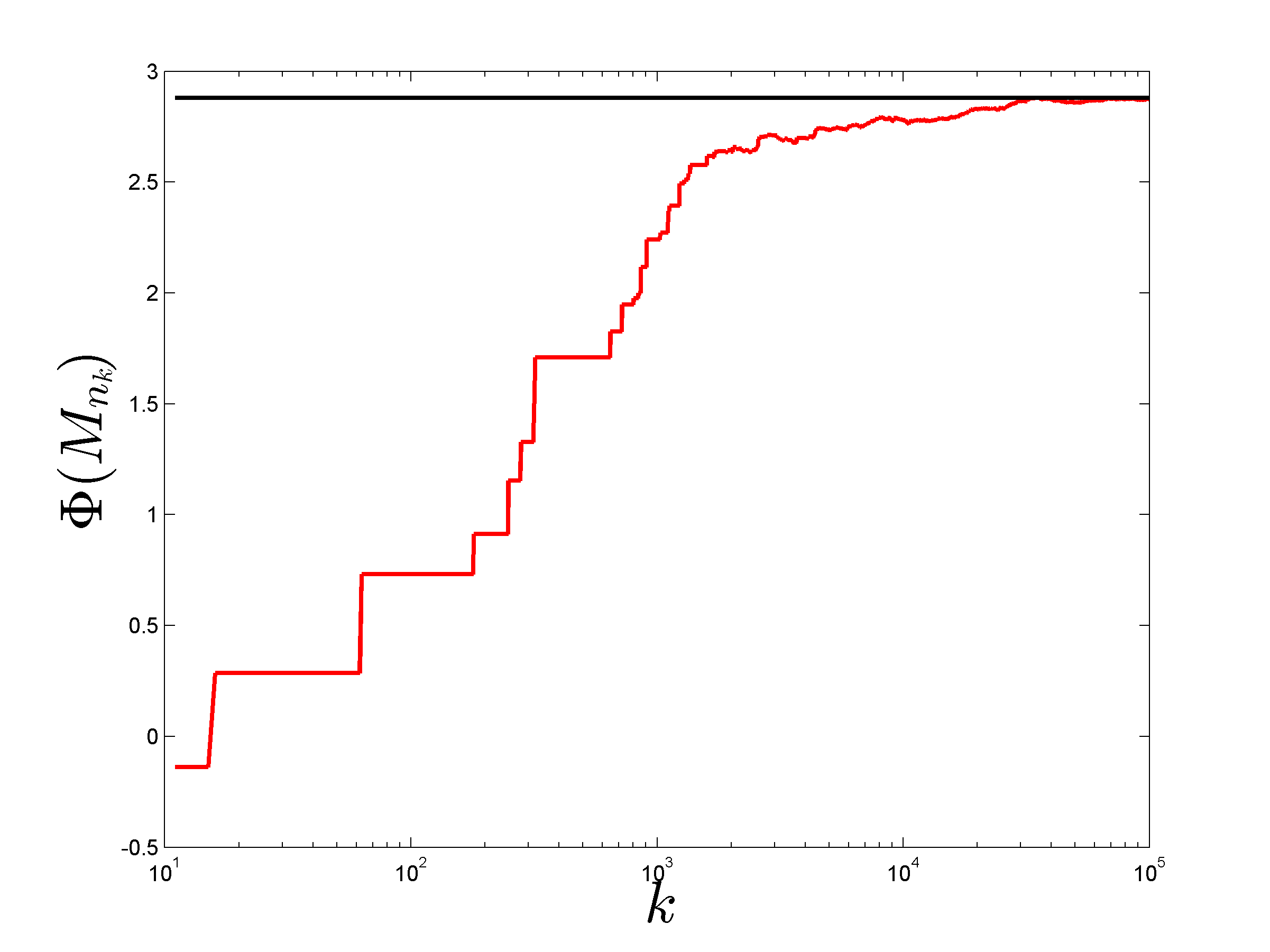}
\end{center}
\caption{\small Evolution of $\Phi(\Mb_{n_k})$ obtained with Algorithm~1 as a function of $k$ (log scale) when $d=2$ and $\mu=(1/2)\,\mu_\SN+(1/2)\,\mu_d$ with $\ma=0.5$ (left) and $\ma=0.02$ (right); the horizontal line indicates the optimal value $\Phi_\ma^*$.}
\label{F:Phink_linearnormal_mixture}
\end{figure}

Figure~\ref{F:Phink_linearnormal_mixture} shows the evolution of $\Phi(\Mb_{n_k})$ as a function of $k$ for $\ma=0.5$ (left) and $\ma=0.02$ (right). Note that $\ma < 1/(2\e1)$ in the second case, but $1/(2\e1) <\ma \leq 1/(2\e1) + 1/2$ in the first one, so that $\xi_\ma^*$ is neither zero nor $\mu/\ma$ on the four points $(\pm 1,\pm1)$. Figure~\ref{F:Phink_linearnormal_mixture} shows that Algorithm~1 nevertheless behaves satisfactorily in both cases.

\subsubsection{Example 5: the distribution of $Z_{\Mb_\ma^*}(X)$ has discrete components}\label{S:ZX-discrete}

Let $U[\SSp_d(\0b,r)]$ denote the uniform probability measure on the $d$-dimensional sphere $\SSp_d(\0b,r)$ with center $\0b$ and radius $r$. The probability measure of the $X_i$ is $\mu=(1/3) \sum_{i=1}^3 U[\SSp_d(\0b,r_i)]$, the mixture of distributions on three nested spheres with radii $r_1>r_2>r_3>0$. The optimal bounded measure is
\bea
\xi_\ma^* = \left\{ \begin{array}{ll}
U[\SSp_d(\0b,r_1)] & \mbox{if } 0< \ma\leq \frac13 \,,\\
\frac{1}{3\ma}\,U[\SSp_d(\0b,r_1)]+\frac{\ma-1/3}{\ma} U[\SSp_d(\0b,r_2)] & \mbox{if } \frac13 < \ma\leq \frac23 \,,\\
\frac{1}{3\ma}\,\left\{U[\SSp_d(\0b,r_1)]+U[\SSp_d(\0b,r_2)]\right\} + \frac{\ma-2/3}{\ma} U[\SSp_d(\0b,r_3)] & \mbox{if } \frac23 <\ma <1 \,,
\end{array} \right.
\eea
with associated $\Phi$-values
\bea
\Phi(\Mb_\ma^*) = \left\{ \begin{array}{ll}
d\,\log(r_1^2/d) & \mbox{if } 0< \ma\leq \frac13 \,,\\
d\,\log\left(\frac{r_1^2/3+(\ma-1/3)r_2^2}{\ma d}\right) & \mbox{if } \frac13 < \ma\leq \frac23 \,,\\
d\,\log\left(\frac{(r_1^2+r_2^2)/3+(\ma-2/3)r_3^2}{\ma d}\right)  & \mbox{if } \frac23 <\ma <1 \,.
\end{array} \right.
\eea
Notice that for $\ma\in(0,1/3)$ (respectively, $\ma\in(1/3,2/3)$) $\xi_\ma^*\neq 0$ and $\xi_\ma^*\neq \mu/\ma$ on $\SSp_d(\0b,r_1)$ (respectively, on $\SSp_d(\0b,r_2)$) although $\mu$ is atomless.

\begin{figure}[ht!]
\begin{center}
\includegraphics[width=.49\linewidth]{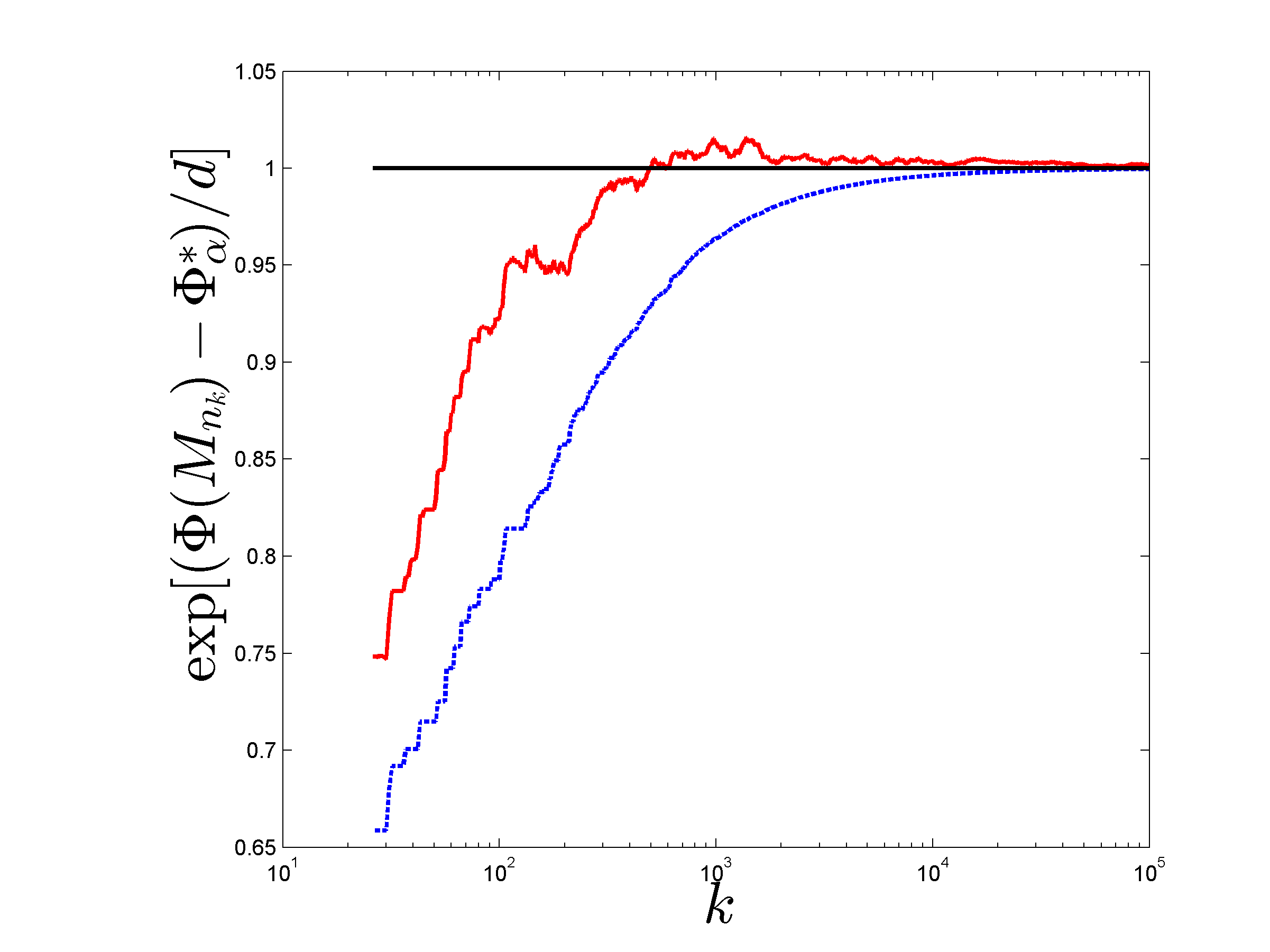}  \includegraphics[width=.49\linewidth]{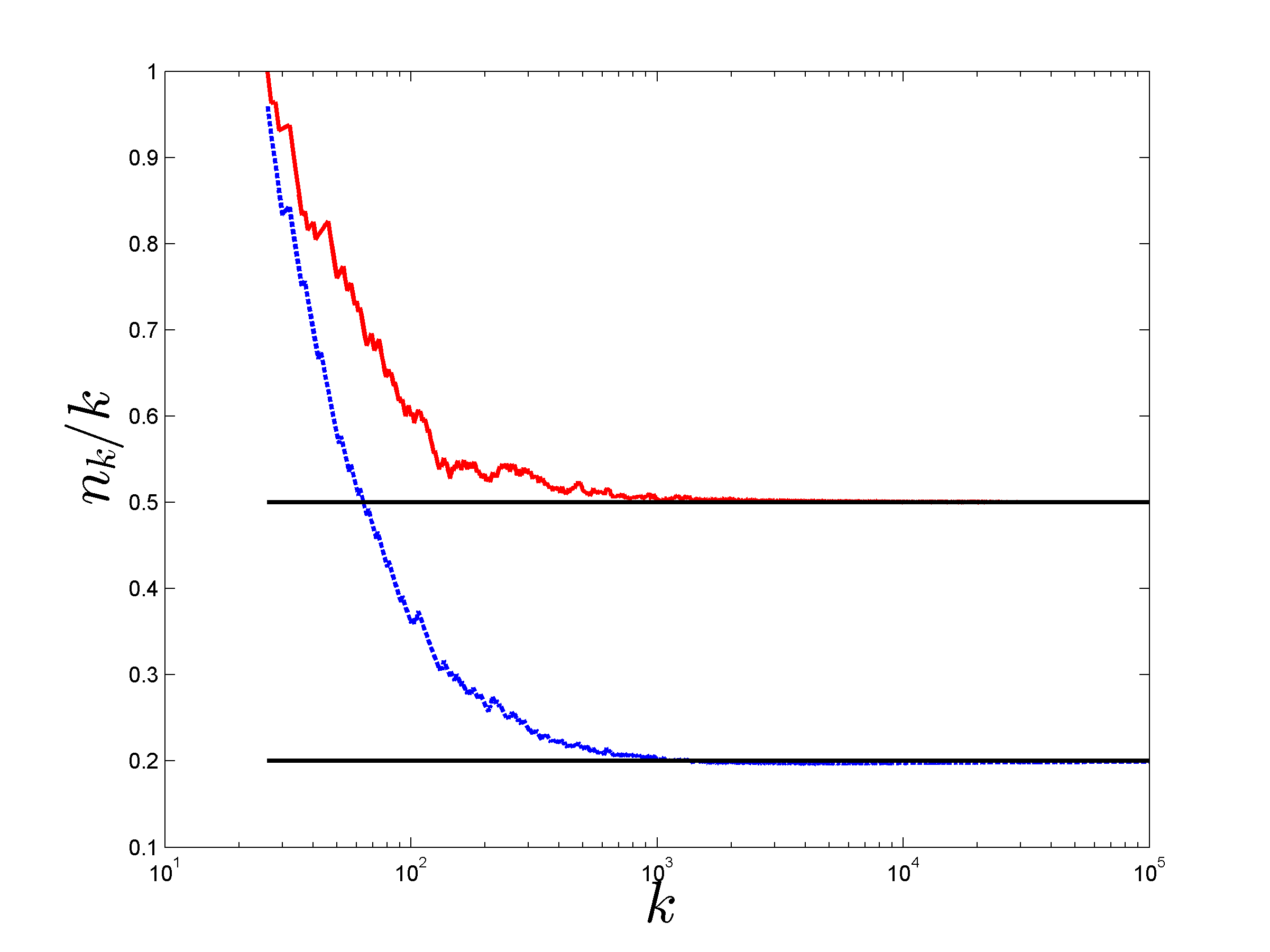}
\end{center}
\caption{\small Left: Evolution of D-efficiency of $\Mb_{n_k}$ obtained with Algorithm~1 as a function of $k$ (log scale) for $\ma=0.5$ (red solid line) and $\ma=0.2$ (blue dashed line); the horizontal line indicates the optimal value 1; $d=5$. Right: evolution of the ratio $n_k/k$ in the same simulations.}
\label{F:linearnormal_Sphere_mixture_d5_a05_a02}
\end{figure}

The left panel of Figure~\ref{F:linearnormal_Sphere_mixture_d5_a05_a02} gives the evolution with $k$ of the D-efficiency $[\det(\Mb_{n_k})/\det(\Mb_\ma^*)]^{1/p}=\exp[(\Phi(\Mb_{n_k})-\Phi_\ma^*)/d]$, for $\ma=0.5$ (red solid line) and $\ma=0.2$ (blue dashed line) when $d=5$. The right panel shows the evolution of the ratio $n_k/k$ for those two situations, with the limiting value $\ma$ indicated by a horizontal line. Although assumption H$_{\mu,\SM}$ is violated, Algorithm~1 continues to perform satisfactorily.

\section{Comparison with other methods}\label{S:other-methods}

\subsection{Case $n$ fixed with large $N$: comparison with an exchange method}\label{S:exchange}

The convergence of $\Phi(\Mb_{n_k})$ to $\Phi_\ma^*$ in Algorithm~1 relies on the fact that $n_k$ grows like $\SO(\ma\, k)$ for some $\ma>0$; see Theorem~\ref{P:main}. If the number $n$ of points to be selected is fixed, Algorithm~1 does not provide any performance guarantee when applied to a sequence of length $N\ra\infty$ (the situation is different when $p=1$ where an asymptotically optimal construction is available; see \citet{Pronzato01a_ieee}). In that case, a method of the exchange type may look more promising, although  large values of $N$ entail serious difficulties. Typically, the algorithm is initialized by a $n$ point design chosen within $\SX_N$, and at each iteration a temporarily selected $X_i$ is replaced by a better point in $\SX_N$. Fedorov's (1972)\nocite{Fedorov72} algorithm considers all $n\times(N-n)$ possible replacements at each iteration ($(N-n)$ instead of $N$ since we do not allow repetitions in the present context); its computational cost is prohibitive for large $N$. The variants suggested by \citet{CookN80}, or the DETMAX algorithm of \citet{Mitchell74}, still require the maximization of a function $g(X_j)$ with respect to $X_j\in\SX_N$ at each iteration, which remains unfeasible for very large $N$. Below, we consider a simplified version where all $N$ points are examined successively, and replacement is accepted when it improves the current criterion value.

\begin{samepage}
\begin{itemize}
\item[] {\bf Algorithm~3: sequential exchange ($n$ fixed).}
    \item[1)] Initialization: select $X_1,\ldots,X_{n}$, set $k=n$ and $\SX_k^*=\{X_1,\ldots,X_k\}$, compute $\Mb_{n,k}=(1/k)\,\sum_{i=1}^k \SM(X_i)$ and $\Phi(\Mb_{n,k})$.

  \item[2)] Iteration $k+1$: collect $X_{k+1}$. If $X_{k+1}\in\SX_k^*$, set $\Delta^{(k)}(X_{i^*},X_{k+1})=0$; otherwise compute
  \bea
  \Delta^{(k)}(X_{i^*},X_{k+1})=\max_{X_i\in\SX_k^*} \left[ \Phi\{\Mb_{n,k}+(1/n)[\SM(X_{k+1})-\SM(X_{i})]\}-\Phi(\Mb_{n,k}) \right] \,.
  \eea
  If $\Delta^{(k)}(X_{i^*},X_{k+1})>0$, set $\SX_{k+1}^*=\SX_k^*\setminus\{X_{i^*}\}\cup X_{k+1}$, update $\Mb_{n,k}$ into $\Mb_{n,k+1}=\Mb_{n,k}+(1/n)[\SM(X_{k+1})-\SM(X_{i^*})]$, compute $\Phi(\Mb_{n,k+1})$; 

  otherwise, set $\SX_{k+1}^*=\SX_k^*$, $\Mb_{n,k+1}=\Mb_{n,k}$.

  \item[3)] If $k+1=N$ stop; otherwise, $k \leftarrow k+1$, return to Step 2.

\end{itemize}
\end{samepage}

\begin{remark}\label{R:R2} %
When $\SM(x)$ has rank one, with $\SM(x)=\fb(x)\fb\TT(x)$ and $\Phi(\Mb)=\log\det(\Mb)$ or $\Phi(\Mb)=\det^{1/p}(\Mb)$ ($D$-optimal design), $\Delta^{(k)}(X_{i^*},X_{k+1})>0$ is equivalent to
\be\label{exchange-increase}
 \fb\TT(X_{k+1})\Mb_{n,k}^{-1}\fb\TT(X_{k+1})-\fb\TT(X_{i^*})\Mb_{n,k}^{-1}\fb\TT(X_{i^*}) + \delta^{(k)}(X_{i^*},X_{k+1}) >0 \,,
\ee
where
\be
\delta^{(k)}(X,X_{k+1}) = \frac{[\fb\TT(X_{k+1})\Mb_{n,k}^{-1}\fb\TT(X)]^2-[\fb\TT(X_{k+1})\Mb_{n,k}^{-1}\fb\TT(X_{k+1})][\fb\TT(X)\Mb_{n,k}^{-1}\fb\TT(X)]}{n} \,, \label{delta_ik+1}
\ee
see \citet[p.~164]{Fedorov72}. As for Algorithm~1 (see Remark~\ref{R:R1}-(\textit{iii})), we may update $\Mb_{n,k}^{-1}$ instead of $\Mb_{n,k}$ to avoid matrix inversions. For large enough $n$, the term \eqref{delta_ik+1} is negligible and the condition is almost $\fb\TT(X_{k+1})\Mb_{n,k}^{-1}\fb\TT(X_{k+1})>\fb\TT(X_{i^*})\Mb_{n,k}^{-1}\fb\TT(X_{i^*})$; that is,
$F_\Phi[\Mb_{n,k},\SM(X_{k+1})]>F_\Phi[\Mb_{n,k},\SM(X_{i^*})]$. This is the condition we use in the example below.
It does not guarantee in general that $\Phi(\Mb_{n,k+1})>\Phi(\Mb_{n,k})$ (since $\delta^{(k)}(X_i,X_{k+1})\leq 0$ from Cauchy-Schwartz inequality), but no significant difference was observed compared with the use of the exact condition \eqref{exchange-increase}.
Algorithm~3 has complexity $\SO(nd^3\,N)$ in general (the additional factor $n$ compared with Algorithm~1 is due to the calculation of the maximum over all $X_i$ in $\SX_k^*$ at Step~2).
\fin
\end{remark}

Neither Algorithm~1 with $\ma=n/N$ and $\Mb_{n,k}=\Mb_{n_k}$ nor Algorithm~3 ensures that $\Phi(\Mb_{n,N})$ tends to $\Phi_\ma^*=\Phi(\Mb_\ma^*)$ as $N\ra\infty$. Also, we can expect to have $\Phi(\Mb_{n,k}) \lesssim \Phi_{n/k}^*$ for all $k$ with Algorithm~3, since under H$_\Phi$ the matrix $\Mb_{n,k}^*$ corresponding to the optimal selection of $n$ distinct points among $\SX_k$ satisfies $\Ex\{\Phi(\Mb_{n,k}^*)\}\leq \Phi_{n/k}^*$ for all $k\geq n >0$; see \citet[Lemma~3]{Pa05}.

\paragraph{Example 6: $n$ fixed and $N$ large}

We consider the same situation as in Example~2 (Section~\ref{S:Ex2}), with $X=(x_1,\ x_2,\ldots,x_d)\TT$, $\SM(X)=XX\TT$, $\Phi(\Mb)=\log\det(\Mb)$; the $X_i$ are i.i.d.\ $\SN(\0b,\Ib_d)$, with $p=d=3$. We still take $k_0=5\,p$, $q=5/8$, $\mg=1/10$ in Algorithm~1. We have $\Ex\{\Mb_{n_{k_0}}\} = \Mb(\mu)$ in Algorithm~1, and, when $n$ is large enough, $\Mb_{n,n} \simeq \Mb(\mu)$ at Step~1 of Algorithm~3, with $\Mb(\mu)=\Ib_d$ and therefore $\Phi(\Mb_{n,n}) \simeq 0$.

\begin{figure}[ht!]
\begin{center}
\includegraphics[width=.49\linewidth]{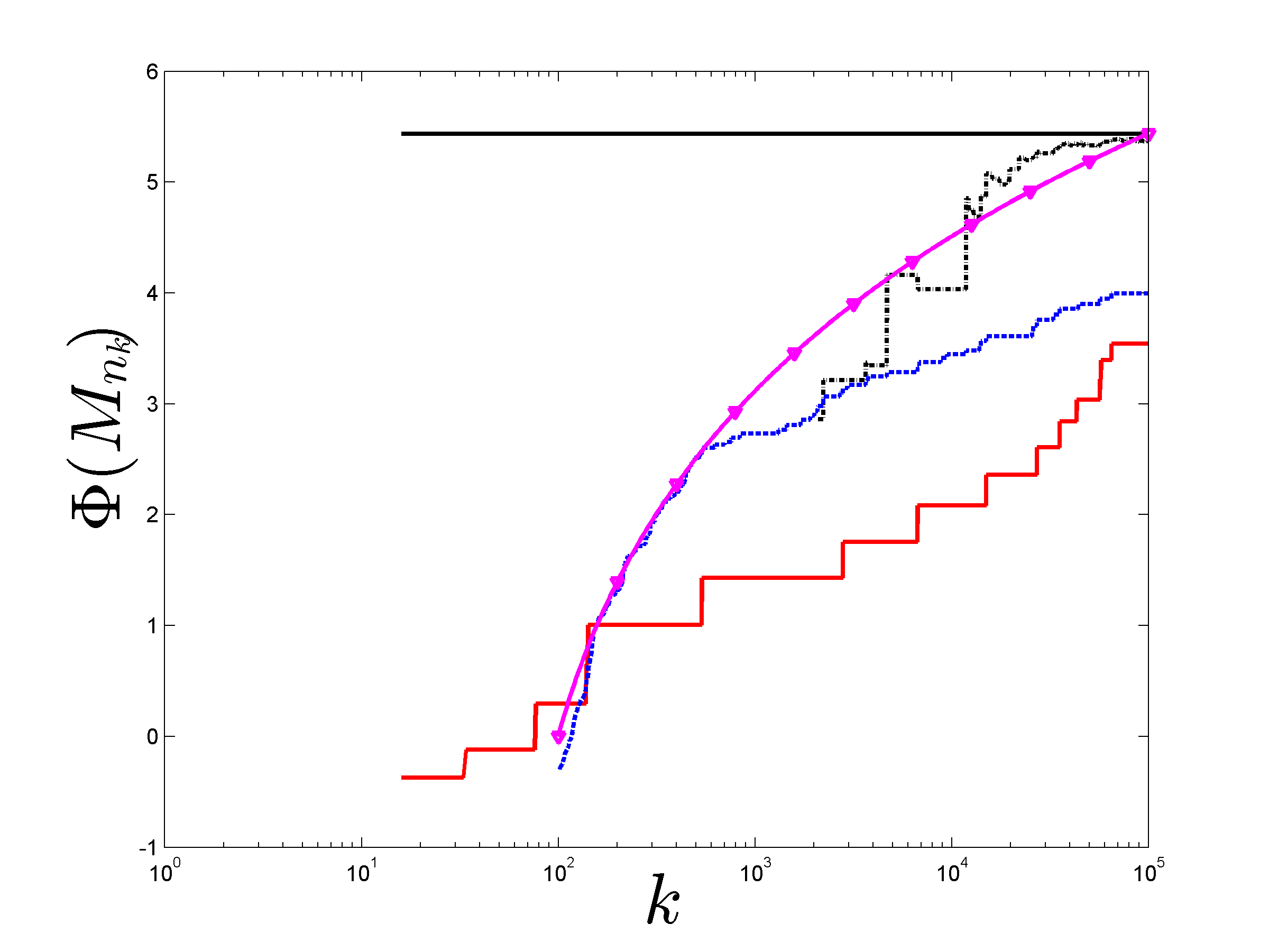} \includegraphics[width=.49\linewidth]{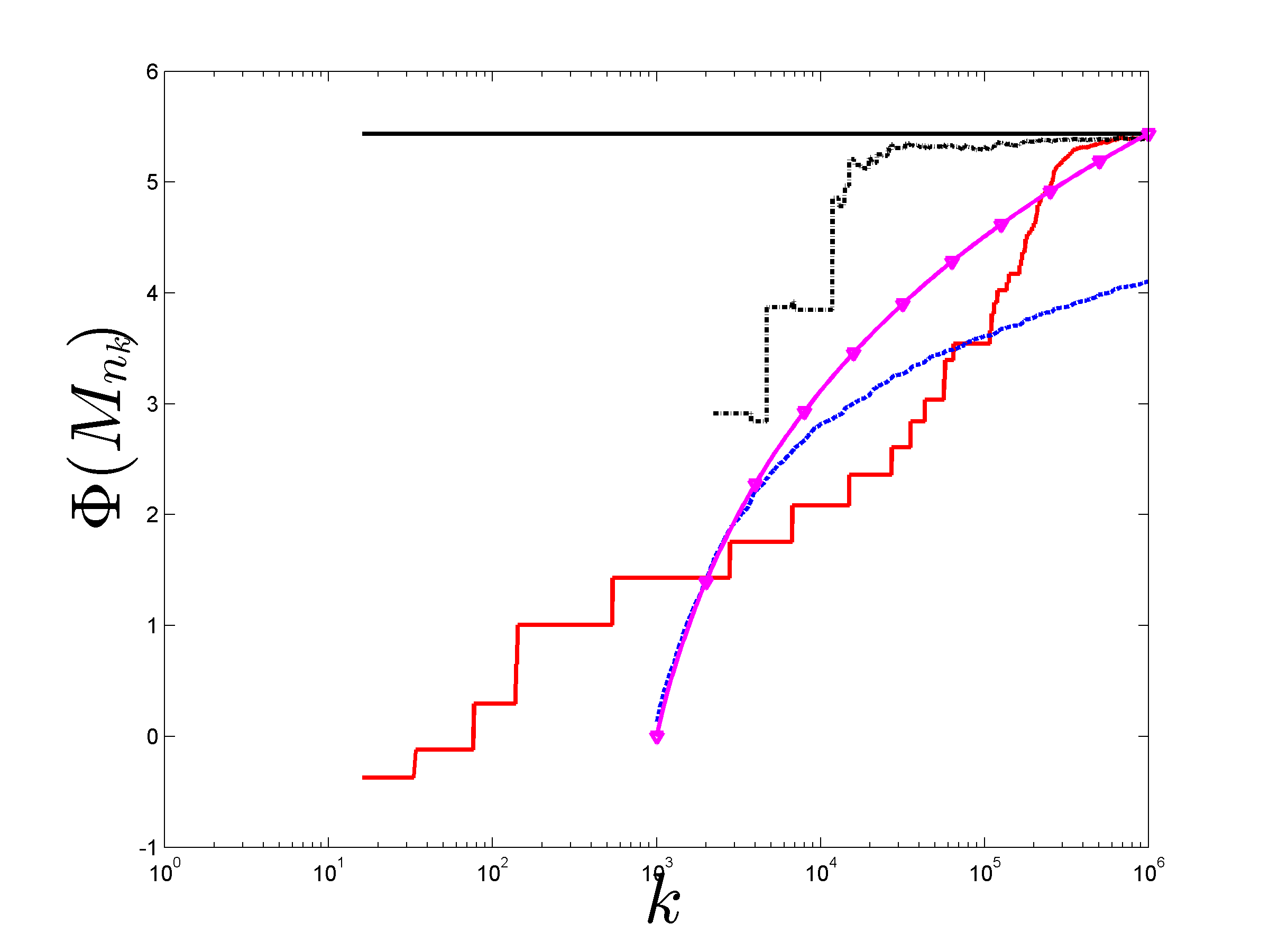}
\end{center}
\caption{\small Evolutions of $\Phi(\Mb_{n_k})$ ($k\geq k_0$, Algorithm~1, red solid line) $\Phi(\Mb_{n,k})$ ($k\geq n$, Algorithm~3, blue dashed line), and $\Phi_{n/k}^*$ (magenta curve with triangles) as functions of $k$; the horizontal black line corresponds to $\Phi_\ma^*$; the black dotted curve shows the evolution of $\Phi(\Mb_{n_k})$ as a function of $k$ when the selection is based on $F_\Phi[\Mb_{n_{mN}},\SM(X_{k+1})]>\hC_{mN}$, with $\Mb_{n_{mN}}$ and $\hC_{mN}$ obtained with Algorithm~1 applied to $\SX_{mN}=\SX_N\cup\SX_N^{(2)}\cup \cdots \cup \SX_N^{(m)}$. Left: $n=100$, $N=100\,000$, $m=9$; Right: $n=1\,000$, $N=1\,000\,000$, $m=1$.}
\label{F:Phik_linearnormal_n1e3_N1e6_d3_compare}
\end{figure}

We consider two values of $n$, $n=100$ and $n=1\,000$, with $\ma=10^{-3}$ (that is, with $N=100\,000$ and $N=1\,000\,000$, respectively). Figure~\ref{F:Phik_linearnormal_n1e3_N1e6_d3_compare} shows the evolutions of $\Phi(\Mb_{n_k})$ ($k\geq k_0$, Algorithm~1, red solid line) and $\Phi(\Mb_{n,k})$ ($k\geq n$, Algorithm~3, blue dashed line) as functions of $k$ in those two cases ($n=100$, left; $n=1\,000$, right). In order to select $n$ points exactly, adaptation of $\ma$ is used in Algorithm~1, see Remark~\ref{R:R1}-(\textit{iv}). The value of $n$ is too small for $\Phi(\Mb_{n_k})$ to approach $\Phi_\ma^*$ (indicated by the horizontal black line) in the first case, whereas $n=1\,000$ is large enough on the right panel; Algorithm~3 performs similarly in both cases and is superior to Algorithm~1 for $n=100$; the magenta curve with triangles shows $\Phi_{n/k}^*$, $k\geq n$, with $\Phi_{n/k}^* \gtrsim \Phi(\Mb_{n,k})$ for all $k$, as expected.
\fin

\vsp
In case it is possible to store the $N$ points $X_i$, we can replay both algorithms on the same data set in order to increase the final value of $\Phi$ for the sample selected.
For Algorithm~3, we can simply run the algorithm again on a set $\SX_N^{(2)}$ --- starting with $k=1$ at Step~1 since $n$ points have already been selected --- with $\SX_N^{(2)}=\SX_N$ or corresponding to a random permutation of it.
Series of runs on sets $\SX_N^{(2)}, \SX_N^{(3)},\ldots$ can be concatenated: the fact that $\Phi$ can only increase implies convergence for an infinite sequence of runs, but generally to a local maximum only; see the discussion in \citep[Sect.~2.4]{CookN80}. When applied to Example~6, this method was not able to improve the design obtained in the first run of Algorithm~3, 
with a similar behavior with or without permutations in the construction of the $\SX_N^{(i)}$.

Algorithm~1 requires a more subtle modification since points are selected without replacement. First, we run Algorithm~1 with $\ma$ fixed at $n/N$ on a set $\SX_{mN}=\SX_N\cup\SX_N^{(2)}\cup \cdots \cup \SX_N^{(m)}$, where the replications $\SX_N^{(i)}$ are all identical to $\SX_N$ or correspond to random permutations of it. The values of $\Mb_{n_{mN}}$ and $\hC_{mN}$ are then used in a second stage, where the $N$ points $X_1,\ldots,X_N$ in $\SX_N$ are inspected sequentially: starting at $k=0$ and $n_k=0$, a new point $X_{k+1}$ is selected if $n_k<n$ and $F_\Phi[\Mb_{n_{mN}},\SM(X_{k+1})]>\hC_{mN}$ (or if $n-n_k \geq N-k+1$, see Remark~\ref{R:R1}-(\textit{iv})). The set $\SX_N$ is thus used $m+1$ times in total.  The idea is that for $m$ large enough, we can expect $\Mb_{n_{mN}}$ to be close to $\Mb_\ma^*$ and $\hC_{mN}$ to be close to the true quantile $C_{1-\ma}(\Mb_\ma^*)$, whereas the optimal rule for selection is $F_\Phi[\Mb_\ma^*,\SM(X_{k+1})]>C_{1-\ma}(\Mb_\ma^*)$. Note that the quantile of the directional derivative is not estimated in this second phase, and updating of $\Mb_{n_k}$ is only used to follow the evolution of $\Phi(\Mb_{n_k})$ on plots.

\begin{figure}[ht!]
\begin{center}
\includegraphics[width=.49\linewidth]{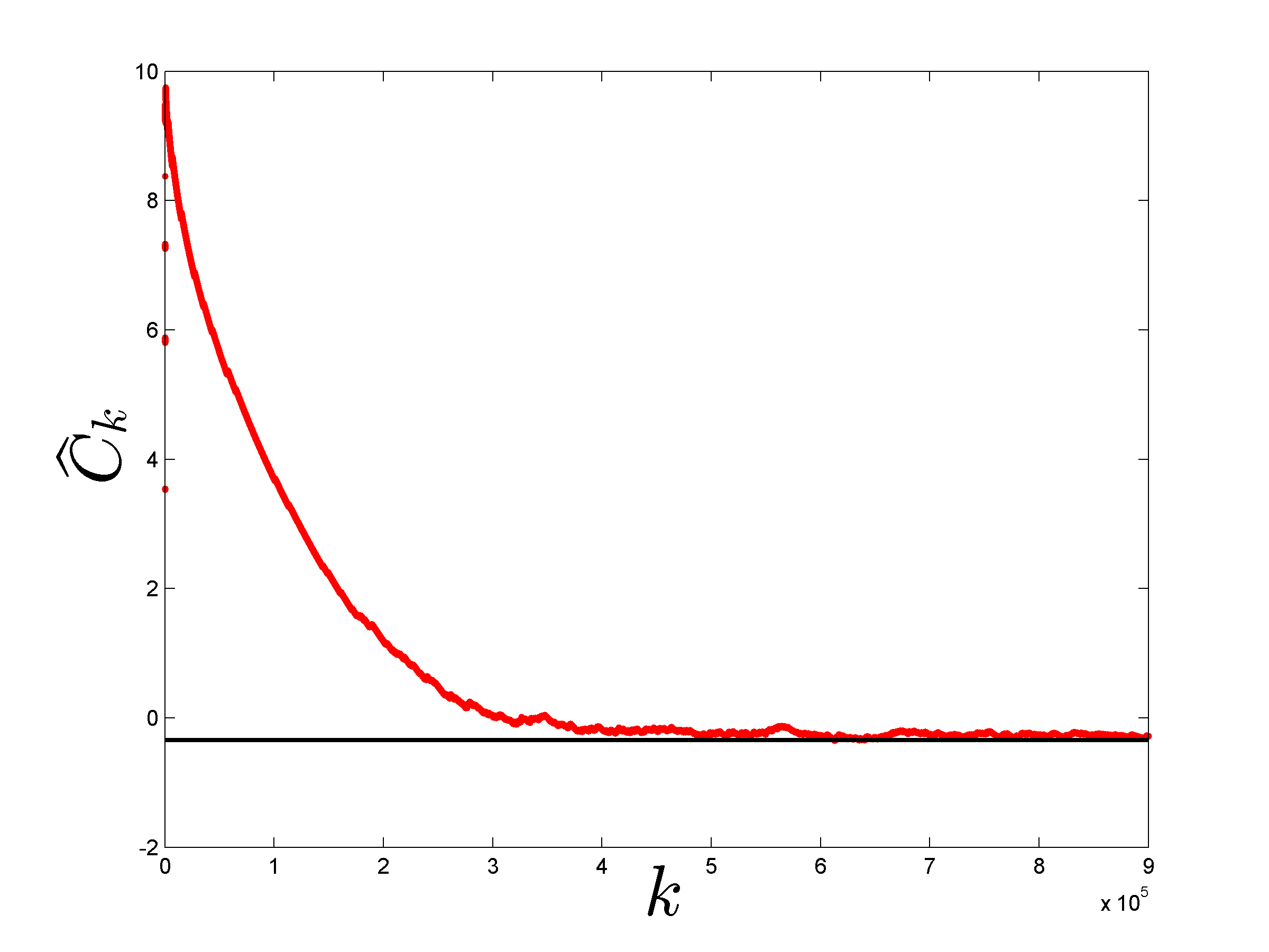} \includegraphics[width=.49\linewidth]{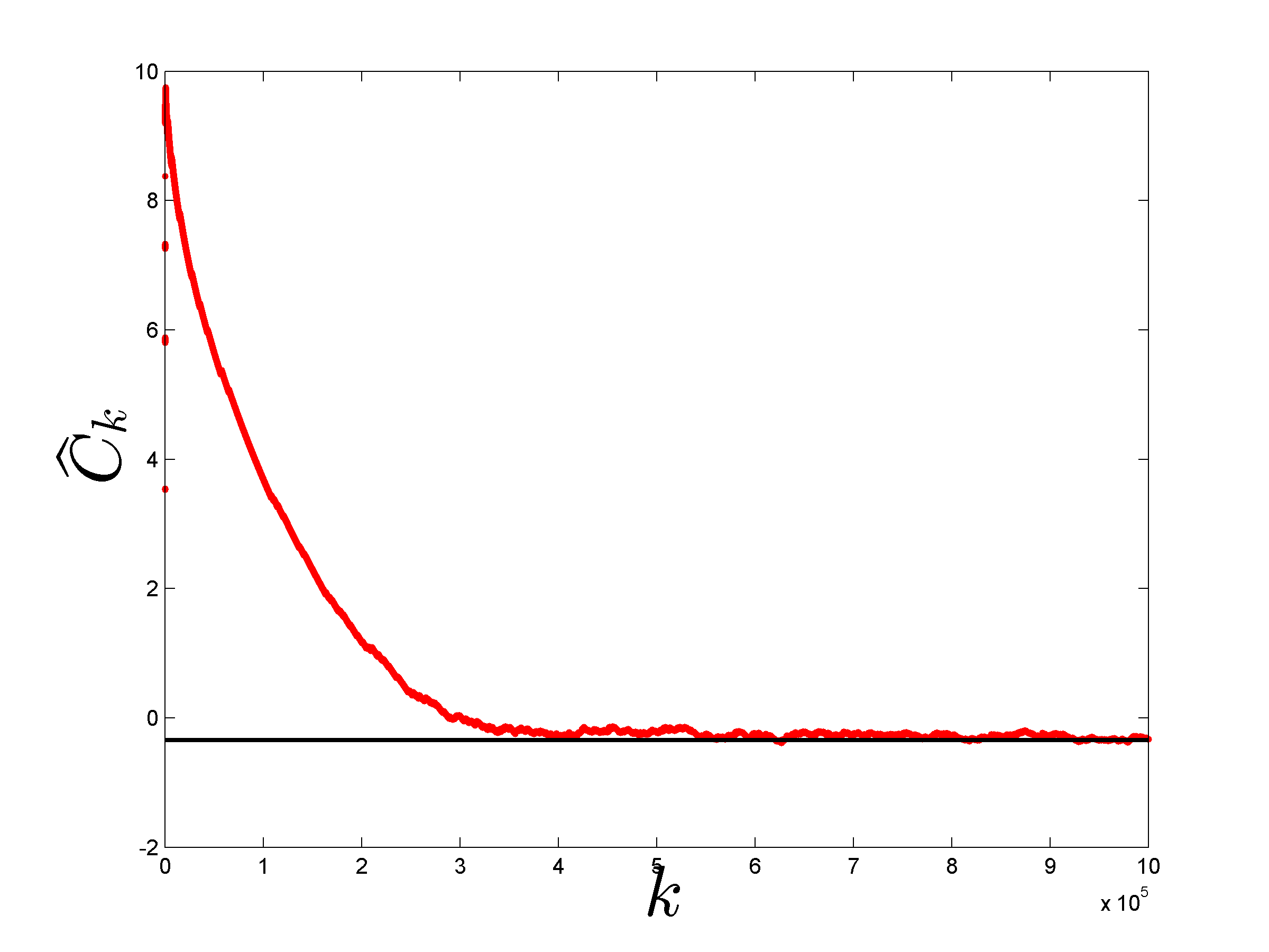}
\end{center}
\caption{\small Evolution of $\hC_k$ in Algorithm~1 when applied to $\SX_{mN}=\SX_N\cup\SX_N^{(2)}\cup \cdots \cup \SX_N^{(m)}$, the horizontal black line corresponds to $C_{1-\ma}(\Mb_\ma^*)$. Left: $n=100$, $N=100\,000$, $m=9$; Right: $n=1\,000$, $N=1\,000\,000$, $m=1$.}
\label{F:Phik_linearnormal_n1e2_N1e5_d3_Ck_repeat}
\end{figure}

\paragraph{Example~6 (continued)}
The black-dotted line in Figure~\ref{F:Phik_linearnormal_n1e3_N1e6_d3_compare} shows the evolution of $\Phi(\Mb_{n_k})$ as a function of $k$ in the second phase (for $k$ large enough to have $\Phi(\Mb_{n_k})>0$): we have taken $m=9$ for $n=100$ (left), so that $(m+1)N=1\,000\,000$ points are used in total (but 10 times the same), and $m=1$ for $n=1\,000$ (right), with $2\,000\,000$ points used (twice the same). Figure~\ref{F:Phik_linearnormal_n1e2_N1e5_d3_Ck_repeat} shows the evolution of $\hC_k$ for $k=1,\ldots,mN$, for $n=100$, $N=100\,000$, $m=9$ (left), and $n=1\,000$, $N=1\,000\,000$, $m=1$ (right); the horizontal black line indicates the value of $C_{1-\ma}(\Mb_\ma^*)$. The left panel indicates that $n=100$ is too small to estimate $C_{1-\ma}(\Mb_\ma^*)$ correctly with Algorithm~1 (note that $m=4$ would have been enough), which is consistent with the behavior of $\Phi(\Mb_{n_k})$ observed in Figure~\ref{F:Phik_linearnormal_n1e3_N1e6_d3_compare}-left (red solid line). The right panel of Figure~\ref{F:Phik_linearnormal_n1e2_N1e5_d3_Ck_repeat} shows that $\hC_k$ has converged before inspection of the $1\,000\,000$ points in $\SX_N$, which explains the satisfactory behavior of Algorithm~1 in Figure~\ref{F:Phik_linearnormal_n1e3_N1e6_d3_compare}-right. Notice the similarity between the left and right panels of Figure~\ref{F:Phik_linearnormal_n1e2_N1e5_d3_Ck_repeat} due to the fact that the same value $\ma=10^{-3}$ is used in both. Here the $\SX_N^{(i)}$ are constructed by random permutations of the points in $\SX_N$, but the behavior is similar without.

\subsection{Comparison with IBOSS}

IBOSS (Information-Based Optimal Subdata Selection, \citet{WangYS2019}) is a selection procedure motivated by D-optimality developed in the context of multilinear regression with intercept, where $\SM(X)=\fb(X)\fb\TT(X)$ with $\fb(X)=[1,\ X\TT]\TT$. All points $X_i$ in $\SX_N$ are processed simultaneously: the $d$ coordinates of the $X_i$ are examined successively; for each $k=1,\ldots,d$, the $r$ points with largest $k$-th coordinate and the $r$ points having smallest $k$-th coordinate are selected (and removed from $\SX_N$), where $r=n/(2d)$, possibly with suitable rounding, when exactly $n$ points have to be selected. The design selected is sensitive to the order in which coordinates are inspected. The necessity to find the largest or smallest coordinate values yields a complexity of $\SO(d\,N)$; parallelization with simultaneous sorting of each coordinate is possible. Like for any design selection algorithm, the matrix $\Mb_{n,N}$ obtained with IBOSS satisfies $\Ex\{\Phi(\Mb_{n,N})\} \leq \Phi_{n/N}^*$ for all $N\geq n>0$ \citep[Lemma~3]{Pa05}. The asymptotic performance of IBOSS (the behavior of $\Mb_{n,N}$ and $\Mb_{n,N}^{-1}$) for $n$ fixed and $N$ tending to infinity is investigated in \citep{WangYS2019} for $X$ following a multivariate normal or lognormal distribution. Next property concerns the situation where $n$ is a fraction of $N$, with $N\ra\infty$ and the components of $X$ are independent.

\begin{theo}\label{P:IBOSS} Suppose that the $X_i$ are i.i.d.\ with $\mu$ satisfying H$_\mu$ and, moreover, that their components $\{X_i\}_k$ are independent, with $\varphi_k$ the p.d.f.\ of $\{X_1\}_k$ for $k=1,\ldots,d$. Suppose, without any loss of generality, that coordinates are inspected in the order $1,\ldots,d$. Then, for any $\ma\in(0,1]$, the matrix $\Vb_{n,N}=(1/n)\sum_{j=1}^n X_{i_j}X_{i_j}\TT$ corresponding to the $n$ points $X_{i_j}$ selected by IBOSS satisfies $\Vb_{n,N} \ra \Vb_\ma^{\IBOSS}$ a.s.\ when $n=\lfloor \ma N \rfloor$ and $N\ra \infty$, with
\be
\{\Vb_\ma^{\IBOSS}\}_{k,k} &=& \frac{1}{\ma}\,\left[\Ex[\{X\}_k^2]-\pi_k\,s_k(\ma)\right] \,, \ k=1\ldots,d\,, \label{V_IBOSS_kk}\\
\{\Vb_\ma^{\IBOSS}\}_{k,k'} &=& \frac{1}{\ma}\,\left[\Ex[\{X\}_k]\,\Ex[\{X\}_k']- \frac{\pi_k\,\pi_{k'}}{1-\ma} \,m_k(\ma)\, m_{k'}(\ma)\right] \,, \  k\neq k' \,, \label{V_IBOSS_kk'}
\ee
where $\Ex[\{X\}_k]=\int_{-\infty}^\infty x\, \varphi_k(x)\,\dd x$, $\Ex[\{X\}_k^2]=\int_{-\infty}^\infty x^2\, \varphi_k(x)\,\dd x$,
$\pi_k=(1-\ma)[d-(k-1)\ma]/(d-k\ma)$,
\bea
s_k(\ma)=\int_{q_k\left(\frac{\ma}{2[d-(k-1)\ma]}\right)}^{q_k\left(1-\frac{\ma}{2[d-(k-1)\ma]}\right)} x^2\, \varphi_k(x)\,\dd x  \ \mbox{ and } \ m_k(\ma)=\int_{q_k\left(\frac{\ma}{2[d-(k-1)\ma]}\right)}^{q_k\left(1-\frac{\ma}{2[d-(k-1)\ma]}\right)} x\, \varphi_k(x)\,\dd x \,,
\eea
with $q_k(\cdot)$ the quantile function for $\varphi_k$, satisfying $\int_{-\infty}^{q_k(t)} \varphi_k(u)\,\dd u=t$ for any $t\in(0,1]$.
\end{theo}

\noindent{\emph{Proof}.}
By construction, IBOSS asymptotically first selects all points such that $\{X\}_1$ does not belong to $\SI_1=(q_1[\ma/(2d)], q_1[1-\ma/(2d)])$, then, among remaining points, all those such that
$\{X\}_2\not\in \SI_2=(q_2[\ma/(2d(1-\ma/d))], q_1[1-\ma/(2d(1-\ma/d))])$. By induction, all points such that $\{X\}_k\not\in\SI_k=(q_k[\ma/(2[d-(k-1)\ma])], q_k[1-\ma/(2[d-(k-1)\ma])])$ are selected at stage $k\in\{3,\ldots,d\}$.
Denote $\xb=(x_1,\ldots,x_d)\TT$. We have
\bea
\{\Vb_\ma^{\IBOSS}\}_{k,k} &=& \frac{1}{\ma} \int_{\SX\setminus \prod_{\ell=1}^d \SI_\ell } x_k^2\, \varphi(\xb)\,\dd\xb =
\frac{1}{\ma} \left[ \int_{\SX} x_k^2\, \varphi(\xb)\,\dd\xb - \int_{\prod_{\ell=1}^d \SI_\ell} x_k^2\, \varphi(\xb)\,\dd\xb \right] \\
&=&
\frac{1}{\ma} \left[ \Ex[\{X\}_k^2] - \left(\prod_{\ell\neq k} \Pr\{\{X\}_{\ell}\in\SI_\ell\}\right) \, \int_{\SI_k} x^2\, \varphi_k(x)\,\dd x \right]\,.
\eea
Direct calculation gives $\Pr\{X\in\prod_{k=1}^d \SI_k \}=1-\ma$ and
\bea
\prod_{\ell\neq k} \Pr\{\{X\}_{\ell}\in\SI_\ell\}= \frac{1-\ma}{\Pr\{\{X\}_k\in\SI_k\}}=\frac{1-\ma}{1-\frac{\ma}{d-(k-1)\ma}} = \pi_k \,,
\eea
which proves \eqref{V_IBOSS_kk}. Similarly,
\bea
\{\Vb_\ma^{\IBOSS}\}_{k,k'} &=& \frac{1}{\ma} \int_{\SX\setminus \prod_{\ell=1}^d \SI_\ell } x_k \, x_{k'}\, \varphi(\xb)\,\dd\xb =
\frac{1}{\ma} \left[ \int_{\SX} x_k\,x_{k'}\, \varphi(\xb)\,\dd\xb - \int_{\prod_{\ell=1}^d \SI_\ell} x_k\,x_{k'}\, \varphi(\xb)\,\dd\xb \right] \\
&=&
\frac{1}{\ma} \left[ \Ex[\{X\}_k]\,\Ex[\{X\}_k'] - \left(\prod_{\ell\neq k,\, \ell\neq k'} \Pr\{\{X\}_{\ell}\in\SI_\ell\}\right) \,m_k(\ma)\, m_{k'}(\ma) \right] \,,
\eea
with
\bea
\prod_{\ell\neq k,\, \ell\neq k'} \Pr\{\{X\}_{\ell}\in\SI_\ell\} = \frac{1-\ma}{\Pr\{\{X\}_k\in\SI_k\}\,\Pr\{\{X\}_{k'}\in\SI_{k'}\}} = \frac{\pi_k\,\pi_{k'}}{1-\ma} \,,
\eea
which proves \eqref{V_IBOSS_kk'} and concludes the proof.
\carre

\vsp
A key difference between IBOSS and Algorithm~1 is that IBOSS is nonsequential and therefore cannot be used in the streaming setting. Also, IBOSS is motivated by D-optimal design and may not perform well for other criteria, whereas Algorithm~1 converges to the optimal solution when $n=\lfloor\ma N\rfloor$ and $N\ra\infty$ for any criterion satisfying H$_\Phi$. Moreover, IBOSS strongly relies on the assumption that $\SM(X)=\fb(X)\fb\TT(X)$ with $\fb(X)=[1,\ X\TT]\TT$ and, as the next example illustrates, it  can perform poorly in other situations, in particular when the $X_i$ are functionally dependent.

\begin{figure}[ht!]
\begin{center}
\includegraphics[width=.49\linewidth]{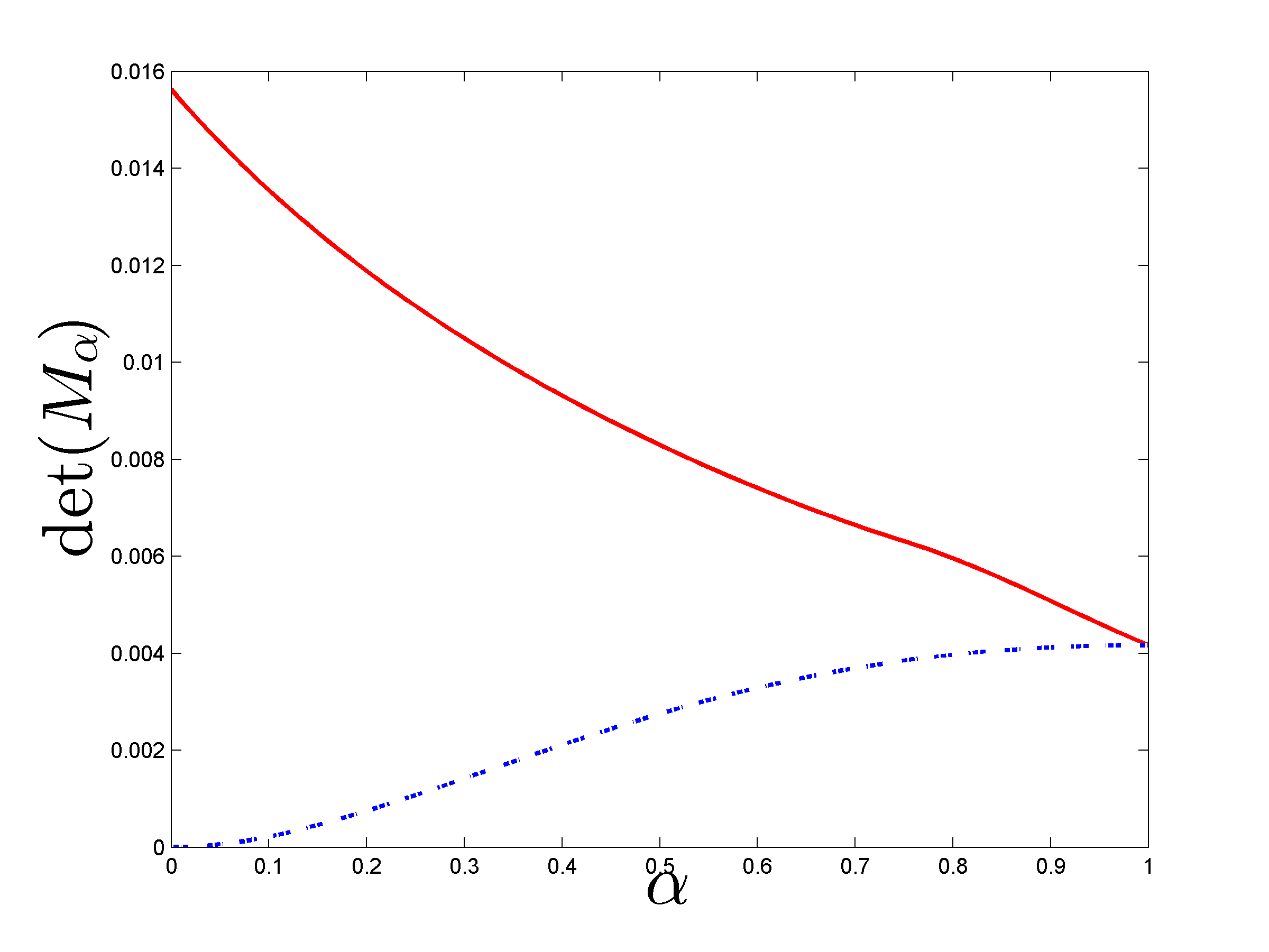} \includegraphics[width=.49\linewidth]{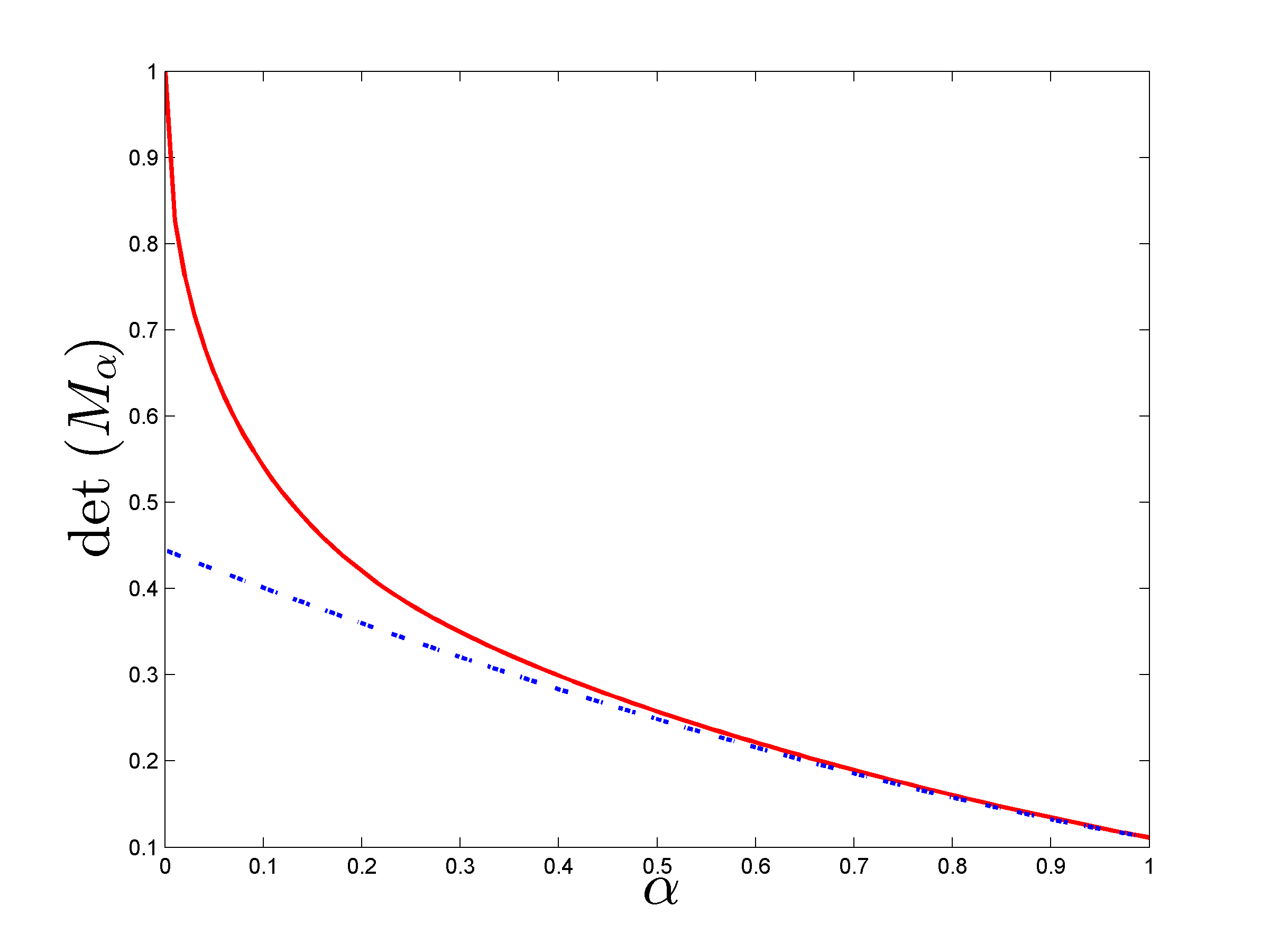}
\end{center}
\caption{\small $\det(\Mb_\ma^*)$ (red solid line) and $\det(\Mb_\ma^{\IBOSS})$ (blue dotted line) as functions of $\ma\in[0,1]$. Left: quadratic regression on $[0,1]$; Right: multilinear regression with intercept on $[-1,1]^2$.}
\label{F:thinning-IBOSS}
\end{figure}

\paragraph{Example 7: quadratic regression on $[0,1]$}
Take $\fb(X)=[X,\ X^2]\TT$, with $X$ uniformly distributed in $[0,1]$ and $\Phi(\Mb)=\log\det(\Mb)$.
For $\ma \leq \ma_* \simeq 0.754160$, the optimal measure $\xi_\ma^*$ equals $\mu/\ma$ on $[1/2-a,1/2+b]\cup[1-(\ma-a-b),1]$ for some $a>b$ (which are determined by the two equations $F_\Phi[\Mb_\ma^*,\SM(1/2-a)]=F_\Phi[\Mb_\ma^*,\SM(1/2+b)]=F_\Phi[\Mb_\ma^*,\SM(1-(\ma-a-b))]$). For $\ma\geq \ma_*$, $\xi_\ma^*=\mu/\ma$ on $[1-\ma,1]$. When $n=\lfloor \ma N\rfloor$ when $N\ra\infty$, the matrix $\Mb_{n,N}^{\IBOSS}$ obtained with IBOSS applied to the points $\fb(X_i)$ converges to $\Mb_\ma^{\IBOSS}=\Mb(\xi_\ma^{\IBOSS})$, with
$\xi_\ma^{\IBOSS}=\mu/\ma$ on $[0,\ma/2]\cup[1-\ma/2,1]$. The left panel of Figure~\ref{F:thinning-IBOSS} shows $\det(\Mb_\ma^*)$ (red solid line) and $\det(\Mb_\ma^{\IBOSS})$ (blue dotted line) as functions of $\ma\in[0,1]$. We have $\det(\Mb_\ma^{\IBOSS})=(1/960)\ma^2(\ma^4+25-40\ma+26\ma^2-8\ma^3)$, which tends to 0 as $\ma\ra 0$.
\fin

\vsp
Next examples show that IBOSS performs more comparably to Algorithm~1 for multilinear regression with intercept, where $\SM(X)=\fb(X)\fb\TT(X)$ with $\fb(X)=[1,\ X\TT]\TT$. Its performance may nevertheless be significantly poorer than that of Algorithm~1.

\paragraph{Example 8: multilinear regression with intercept, $\Phi(\Mb)=\log\det(\Mb)$}

\subparagraph{$X$ is uniformly distributed in $[-1,1]^2$.}

Direct calculation shows that, for any $\ma\in[0,1]$, the optimal measure $\xi_\ma^*$ equals $\mu/\ma$ on $[-1,1]^2 \setminus \SB_2(\0b,R_\ma)$, with $\SB_2(\0b,r)$ the open ball centered at the origin with radius $r$. Here, $R_\ma=2 \sqrt{(1-\ma)/\pi}$ when $\ma\geq 1-\pi/4$, and $R_\ma>1$ is solution of $1+\pi R^2/4 - \sqrt{R^2-1}-R^2\, \arcsin(1/R)=\ma$ when $\ma\in(1-\pi/4,1]$. The associated optimal matrix is diagonal, $\Mb_\ma^*=\diag\{1,\rho_\ma,\rho_\ma\}$, with
\bea
\rho_\ma = \left\{
\begin{array}{ll}
\frac{1}{2\ma}\, [2/3 - 2(1-\ma)^2/\pi] & \! \mbox{if } 0 \leq \ma \leq 1-\pi/4 \,, \\
\frac{1}{2\ma}\, \left[ 2/3 + \pi R_\ma^4/8 - (R_\ma^4/2)\,\arcsin(1/R_\ma)-\sqrt{R_\ma^2-1}\,(R_\ma^2+2)/6 \right] & \! \mbox{if } 1-\pi/4 < \ma \leq 1 \,.
\end{array} \right.
\eea
Extension to $d>2$ is possible but involves complicated calculations.

When $n=\lfloor \ma N\rfloor$ and $N\ra\infty$, the matrix $\Mb_{n,N}^{\IBOSS}$ obtained with IBOSS converges to $\Mb_\ma^{\IBOSS}=\Mb(\xi_\ma^{\IBOSS})$ when $n=\lfloor \ma N\rfloor$ and $N\ra\infty$, with $\xi_\ma^{\IBOSS}=\mu/\ma$ on $[-1,1]^2\setminus ([-1+a,1-a]\times[-1+b,1-b])$, with $a=\ma/2$ and $b=\ma/(1-\ma) \geq a$. The matrix $\Mb_\ma^{\IBOSS}$ is diagonal, $\Mb_\ma^{\IBOSS}=\diag\{1,D_{\ma,1},D_{\ma,2}\}$, where $\Vb_\ma^{\IBOSS}=\diag\{D_{\ma,1},D_{\ma,2}\}$ is the matrix in Theorem~\ref{P:IBOSS} with
$D_{\ma,1}=(8-5\ma+\ma^2)/12$
and $D_{\ma,2}=(8-11\ma+4\ma^2)/[3(2-\ma)^2]$. The right panel of Figure~\ref{F:thinning-IBOSS} shows $\det(\Mb_\ma^*)$ (red solid line) and $\det(\Mb_\ma^{\IBOSS})$ (blue dashed line) as functions of $\ma\in[0,1]$. Note that $\det(\Mb_\ma^*)\ra 1$ whereas $\det(\Mb_\ma^{\IBOSS})\ra 4/9$ when $\ma\ra 0$. The problem is due to selection by IBOSS of points having one coordinate in the central part of the interval.
\fin

\subparagraph{$X$ is normally distributed $\SN(\0b,\Ib_d)$.}

\begin{figure}[ht!]
\begin{center}
\includegraphics[width=.49\linewidth]{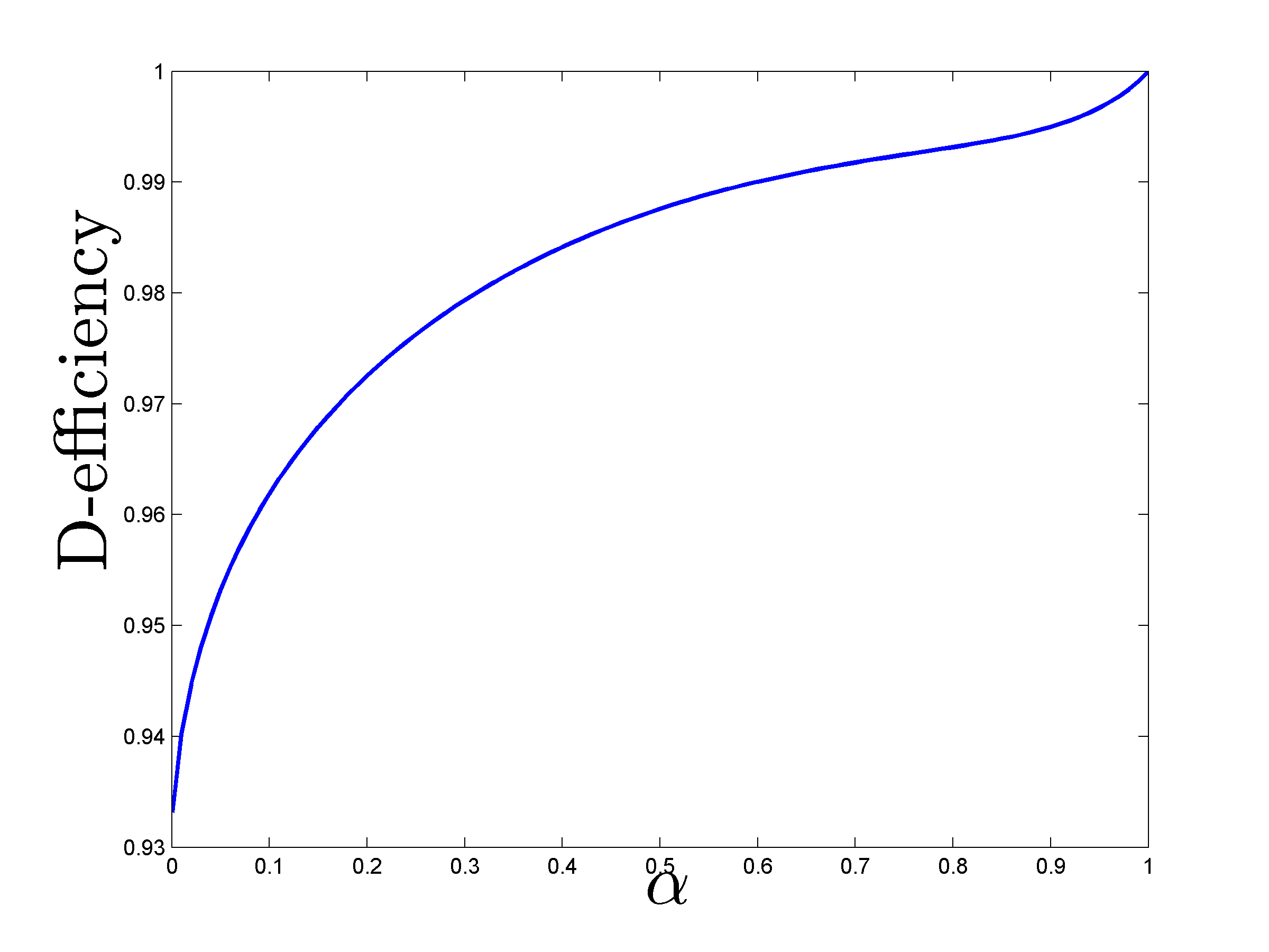} \includegraphics[width=.49\linewidth]{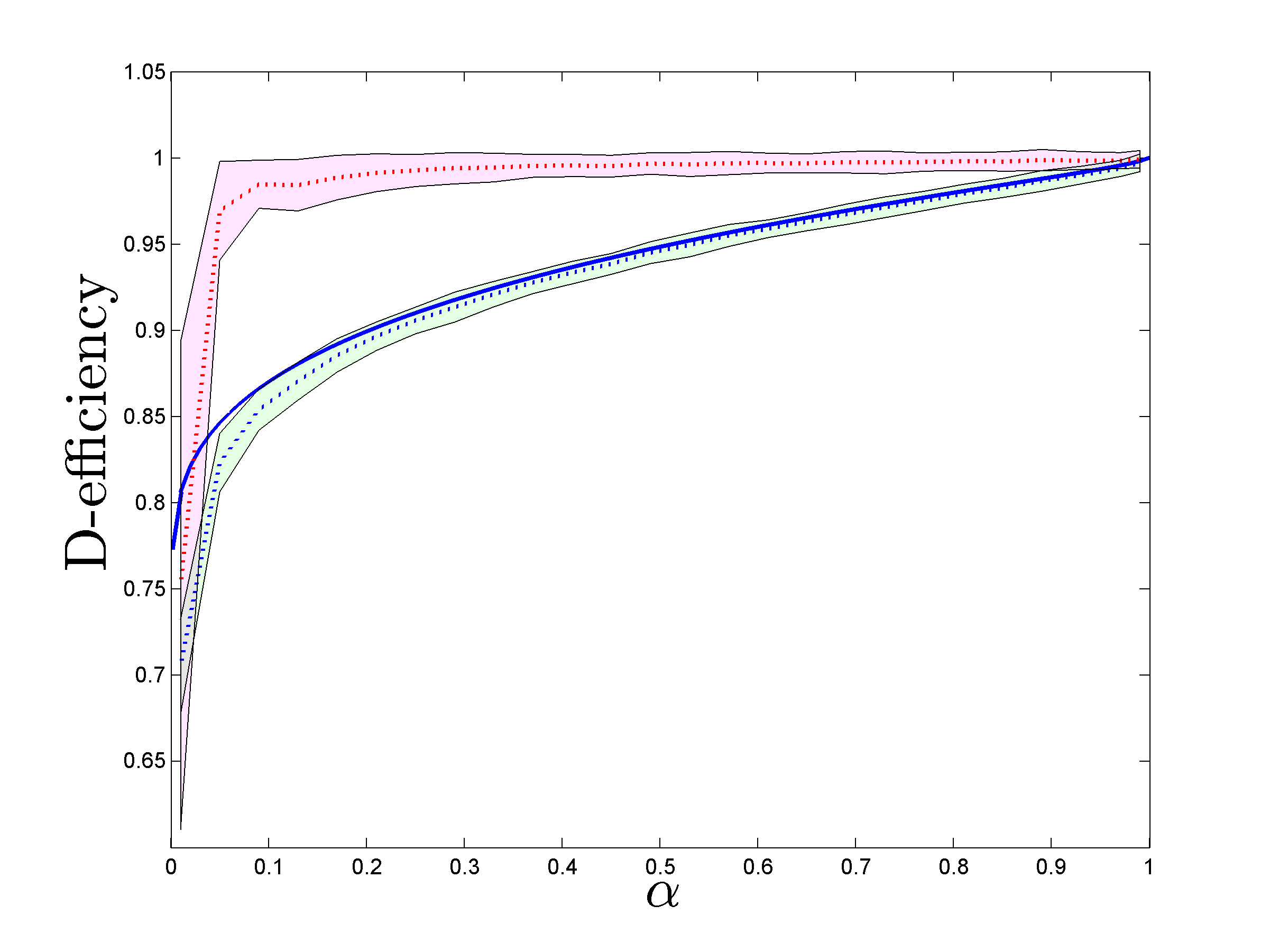}
\end{center}
\caption{\small D-efficiency of IBOSS (blue solid line) as a function of $\ma\in(0,1]$ for $d=3$ (left) and $d=25$ (right). The enveloppes on the right panel show the empirical mean efficiency $\pm$ 2 standard deviations obtained for 100 independent repetitions with $n=\lfloor \ma N\rfloor$ and $N=10\,000$ for IBOSS (green, bottom) and Algorithm~1 (magenta, top).}
\label{F:empirical-thinning-IBOSS}
\end{figure}

The expression of the optimal matrix $\Mb_\ma^*$ has been derived in Section~\ref{S:Ex2}; the asymptotic value for $N\ra\infty$ of the matrix $\Mb_{\lfloor \ma N\rfloor,N}$
is
\bea
\Mb_\ma^{\IBOSS} = \left(
                     \begin{array}{cc}
                       1 & \0b\TT \\
                       \0b & \Vb_\ma^{\IBOSS} \\
                     \end{array}
                   \right)\,,
\eea
where the expression of $\Vb_\ma^{\IBOSS}$ (here a diagonal matrix) is given in Theorem~\ref{P:IBOSS}. Figure~\ref{F:empirical-thinning-IBOSS} shows the D-efficiency $\det^{1/(d+1)}(\Mb_\ma^{\IBOSS})/\det^{1/(d+1)}(\Mb_\ma^*)$ as a function of $\ma\in(0,1]$ for $d=3$ (left) and $d=25$ (right), showing that the performance of IBOSS deteriorates as $d$ increases. We also performed series of simulations for $d=25$, with 100 independent repetitions of selections of $n=\lfloor \ma N\rfloor$ points within $\SX_N$ ($N=10\,000$) based on IBOSS and Algorithm~1. Due to the small value of $N$, we apply Algorithm~1 to replications $\SX_{mN}=\SX_N\cup\SX_N^{(2)}\cup\cdots\cup\SX_N^{(m)}$ of $\SX_N$, see Section~\ref{S:exchange}, with $m=99$ for $\ma<0.1$, $m=9$ for $0.1\leq \ma<0.5$ and $m=4$ for $\ma\geq 0.5$. The colored areas on Figure~\ref{F:empirical-thinning-IBOSS} show the variability range for efficiency, corresponding to the empirical mean $\pm$ 2 standard deviations obtained for the 100 repetitions, for IBOSS (green, bottom) and Algorithm~1 (magenta, top); note that variability decreases as $n=\lfloor \ma N\rfloor$ increases. The approximation of $\Mb_{n,N}$ obtained with IBOSS by the asymptotic matrix $\Mb_\ma^{\IBOSS}$ is quite accurate although $N$ is rather small; Algorithm~1 (incorporating $m$ repetitions of $\SX_N$) performs significantly better than IBOSS although the setting is particularly favorable to IBOSS --- it is significantly slower than IBOSS, however, when $m$ is large.

\section{Conclusions and further developments}\label{S:conclusions}

We have proposed a sequential subsampling method for experimental design (Algorithm~1) that converges to the optimal solution when the length of the sequence tends to infinity and a fixed proportion of design points is selected. Since the method only needs to keep the memory of the current information matrix associated with the design problem (or its inverse), and to update a pair of scalar variables (an estimated quantile, and an estimate of the p.d.f.\ value at the quantile), it can be applied to sequences of arbitrary length and is suitable for data streaming.

We have not tried to optimize the choice of initialization and tuning parameters in Algorithm~1. Although it does not seem critical (the same tuning has been used in all the examples presented), there is certainly an opportunity to improve, in particular concerning $\beta_0$ and $\hC_{k_0}$ (for instance, using the information that $C_{1-\ma}^*\leq 0$ whereas $\hC_{k_0}>0$ for small $\ma$ with the initialization we use).

We have only considered the case of linear models, where the information matrix does not depend on unknown parameters (equivalent to local optimum design in case of a nonlinear model), but extension to online parameter estimation in a nonlinear model with $\SM(x)=\SM(x,\mtb)$ would not require important modifications. Denote by $\hmtb^{n}$ the estimated value of the parameters after observation at the $n$ design points selected, $X_{i_1},\ldots,X_{i_n}$, say. Then, we can use $\Mb_{n_{k_0}}=(1/k_0)\, \sum_{i=1}^{k_0} \SM(X_i,\hmtb^{k_0})$ at Step~1 of Algorithm~1, and $\Mb_{n_k+1}$ given by \eqref{iter-basic-k2} can be replaced by $\Mb_{n_k+1}=[1/(n_k+1)]\, [\sum_{j=1}^{n_k} \SM(X_{i_j},\hmtb^{n_k}) + \SM(X_{k+1},\hmtb^{n_k})]$ at Step~2. Recursive estimation can be used for $k>k_0$ to reduce computational cost. For instance for maximum likelihood estimation, with the notation of Section~\ref{S:Intro}, we can update $\hmtb^{n_k}$ as
\bea
\hmtb^{n_k+1} = \hmtb^{n_k} + \frac{1}{n_k+1}\, \Mb_{n_k+1}^{-1} \frac{\mp \log\varphi_{X_{k+1},\mt}(Y_{k+1})}{\mp\mt}\bigg{|}_{\mt=\hmtb^{n_k}}
\eea
when $X_{k+1}$ is selected; see \citet{LjungS83, Tsypkin83}. A further simplification would be to update $\Mb_{n_k}$ as
$\Mb_{n_{k+1}}=\Mb_{n_k} + [1/(n_k+1)]\, [\SM(X_{k+1},\hmtb^{n_k}) - \Mb_{n_k}]$. When the $X_i$ are i.i.d.\ with $\mu$ satisfying H$_\mu$, the strong consistency of $\hmtb^{n_k}$ holds with such recursive schemes under rather general conditions when all $X_i$ are selected. Showing that this remains true when only a proportion $\ma$ is selected by Algorithm~1 requires technical developments outside the scope of this paper, but we anticipate that $\Mb_{n_k} \ra\Mb_{\ma,\bmtb}^*$ a.s., with $\Mb_{\ma,\bmtb}^*$ the optimal matrix for the true value $\bmtb$ of the model parameters.

Algorithm~1 can be viewed as an adaptive version of the treatment allocation method presented in \citep{MetelkinaP2017}: consider the selection or rejection of $X_i$ as the allocation of individual $i$ to treatment 1 (selection) or 2 (rejection), with respective contributions $\SM_1(X_i)=\SM(X_i)$ or $\SM_2(X_i)=0$ to the collection of information; count a cost of one for allocation to treatment 1 and zero for rejection. Then, the doubly-adaptive sequential allocation (4.6) of \citet{MetelkinaP2017} that optimizes a compromise between information and cost exactly coincides with Algorithm~1 where $\hC_k$ is frozen to a fixed $C$, i.e., without Step~3. In that sense, the two-time-scale stochastic approximation procedure of Algorithm~1 opens the way to the development of adaptive treatment allocation procedures where the proportion of individuals allocated to the poorest treatment could be adjusted online to a given target.

Finally, the designs obtained with the proposed thinning procedure are model-based: when the model is wrong, $\xi_\ma^*$ is no longer optimal for the true model. Model-robustness issues are not considered in the paper and would require specific developments, following for instance the approach in \citep{Wiens2005, NieWZ2018}.

\appendix

\section{Maximum of $\Phi(\Mb_{n_k})$}\label{S:max-Phi}

The property below is stated without proof in \citep{Pa05}. We provide here a formal proof based on results on conditional value-at-risk by \citet{RockafellarU2000} and \citet{Pflug2000}.

\begin{lemma}\label{L:opt-nonsequential}
Suppose that $n_k/k\ra\ma$ as $k\ra\infty$. Then, under H$_\Phi$ and H$_\SM$, for any choice of $n_k$ points $X_i$ among $k$ points i.i.d.\ with $\mu$, we have $\limsup_{k\ra\infty} \Phi(\Mb_{n_k,k}) \leq  \Phi(\Mb_\ma^*)$ a.s., where $\Mb_\ma^*$ maximizes $\Phi(\Mb)$ with respect to $\Mb\in\SMM(\ma)$.
\end{lemma}

\noindent{\em Proof.} Denote by $\Mb_{n_k,k}^*$ the matrix that corresponds to choosing $n_k$ distinct candidates that maximize $\Phi(\Mb_{n_k,k})$. The concavity of $\Phi$ implies
\be\label{opt-nonsequential-a}
\Phi(\Mb_{n_k,k}^*) \leq \Phi(\Mb_\ma^*)+ \tr[\nabla_\Phi(\Mb_\ma^*)(\Mb_{n_k,k}^*-\Mb_\ma^*)]\,.
\ee
The rest of the proof consists in deriving an upper bound on the second term on the right-hand side of \eqref{opt-nonsequential-a}.

Denote $z_i=\tr[\nabla_\Phi(\Mb_\ma^*)\SM(X_i)]$ for all $i=1,\ldots,k$ and let the $z_{i:k}$ denote the version sorted by decreasing values. Since $\Phi$ is increasing for Loewner ordering, $\Phi(\Mb) \leq \Phi(\Mb+\zb\zb\TT)$ for any $\Mb\in\SMM^\geq$ and any $\zb\in\mathds{R}^p$, and concavity implies $\Phi(\Mb+\zb\zb\TT) \leq \Phi(\Mb)+\zb\TT \nabla_\Phi(\Mb)\zb$, showing that $\nabla_\Phi(\Mb)\in\SMM^\geq$. Therefore, $z_{i:k}\geq 0$ for all $i$.

First, we may notice that $\tr[\nabla_\Phi(\Mb_\ma^*)\Mb_{n_k,k}^*] \leq (1/n_k)\, \sum_{i=1}^{n_k} z_{i:k}$ and that
\bea
\tr[\nabla_\Phi(\Mb_\ma^*)\Mb_\ma^*] = \frac{1}{\ma}\, \int_\SX \ind_{\{\tr[\nabla_\Phi(\Mb_\ma^*)\SM(x)] \geq c_{1-\ma}\}}\, \tr[\nabla_\Phi(\Mb_\ma^*)\SM(x)]\,\mu(\dd x)
\eea
with $c_{1-\ma}\geq 0$ and such that $\int_\SX \ind_{\{\tr[\nabla_\Phi(\Mb_\ma^*)\SM(x)] \geq c_{1-\ma}\}}\, \mu(\dd x)=\ma$; see \eqref{C_ma}.

Following \citet{RockafellarU2000, Pflug2000}, we then define the functions $g(x;\beta,a)=a+(1/\beta)\,[\tr[\nabla_\Phi(\Mb_\beta^*)\SM(x)-a]^+$, $x\in\SX$, $\beta\in(0,1)$, $a\in\mathds{R}$. We can then write, for any $\beta\in(0,1)$,
\be\label{Cvar1}
\tr[\nabla_\Phi(\Mb_\beta^*)\Mb_\beta^*] = \Ex\{g(X;\beta,c_{1-\beta})\} = \inf_a \Ex\{g(X;\beta,a)\} \geq c_{1-\beta}\,,
\ee
and
\bea
\frac{1}{n_k}\, \sum_{i=1}^{n_k} z_{i:k}= \Ex_{\mu_k}\{g(X;\ma_k,z_{n_k:k})\} = \inf_a \Ex_{\mu_k}\{g(X;\ma_k,a)\} \,,
\eea
where $\ma_k=n_k/k\in(\ma/2,1]$ for all $k$ larger than some $k_1$ and where $\Ex_{\mu_k}\{\cdot\}$ denotes expectation for the empirical measure $\mu_k=(1/k)\,\sum_{i=1}^k \delta_{X_i}$.

Next, we construct an upper bound on $z_{n_k:k}$. For $k>k_1$, the matrix $\Mb_k=(1/k)\,\sum_{i=1}^k \SM(X_i)$ satisfies
\be\label{UBz_{n_k:k}1}
\tr[\nabla_\Phi(\Mb_\ma^*)\Mb_k]=(1/k)\, \sum_{i=1}^k z_{i:k} \geq (n_k/k)\, z_{n_k:k} > (\ma/2)\, z_{n_k:k}  \,.
\ee
Now, $\Mb_\ma^*=\Mb(\xi_\ma^*)$ with $\xi_\ma^*=\mu/\ma$ on a set $\SX_\ma^*\subset\SX$ and $\xi_\ma^*=0$ elsewhere, and $\mu(\SX_\ma^*)=\ma\,\xi_\ma^*(\SX_\ma^*)=\ma\,\xi_\ma^*(\SX)=\ma$. H$_{\SM}$-(\textit{ii}) then implies that $\ml_{\min}(\Mb_\ma^*)=(1/\ma)\, \ml_{\min}[\int_{\SX_\ma^*} \SM(x)\,\mu(\dd x)] > \ell_\ma/\ma$, and H$_\Phi$ implies that $\|\nabla_\Phi(\Mb_\ma^*)\|<A(\ell_\ma/\ma)<\infty$. Therefore $\tr[\nabla_\Phi(\Mb_\ma^*)\Mb(\mu)]<A_\ma=A(\ell_\ma/\ma)\sqrt{pB}$ from H$_{\SM}$-(\textit{i}). Since $\tr[\nabla_\Phi(\Mb_\ma^*)\Mb_k]$ tends to $\tr[\nabla_\Phi(\Mb_\ma^*)\Mb(\mu)]$ a.s.\ as $k\ra\infty$, \eqref{UBz_{n_k:k}1} implies that there exists a.s.\ $k_2$ such that, for all $k>k_2$, $z_{n_k:k}<A_\ma/(4\ma)$.

To summarize, \eqref{opt-nonsequential-a} implies
\bea
\Phi(\Mb_{n_k,k}^*) &\leq& \Phi(\Mb_\ma^*)+ \Ex_{\mu_k}\{g(X;\ma_k,z_{n_k:k})\} - \Ex\{g(X;\ma,c_{1-\ma})\}  \\
&\leq& \Phi(\Mb_\ma^*)+ \left| \Ex_{\mu_k}\{g(X;\ma_k,z_{n_k:k})\} - \Ex\{g(X;\ma_k,c_{1-\ma_k})\} \right| \\
&& + \left|\tr[\nabla_\Phi(\Mb_{\ma_k}^*)\Mb_{\ma_k}^*] - \tr[\nabla_\Phi(\Mb_\ma^*)\Mb_\ma^*] \right| \,.
\eea
The last term tends to zero as $k$ tends to infinity, due to \eqref{Cvar1} and the continuity of conditional value-at-risk; see \cite[Prop.~13]{RockafellarU2002}. Since $c_{1-\ma_k}\leq \tr[\nabla_\Phi(\Mb_{\ma_k}^*)\Mb_{\ma_k}^*]$, see  \eqref{Cvar1}, and $\ma_k\ra\ma$, for all $k$ large enough we have $c_{1-\ma_k} \leq 2 \, \tr[\nabla_\Phi(\Mb_\ma^*)\Mb_\ma^*]$. Denote $\bar a =\max\{A_\ma/(4\ma),2\,\tr[\nabla_\Phi(\Mb_\ma^*)\Mb_\ma^*]\}$. The second term can then be rewritten as
\bea
\left| \Ex_{\mu_k}\{g(X;\ma_k,z_{n_k:k})\} - \Ex\{g(X;\ma_k,c_{1-\ma_k})\} \right| &=& \left| \inf_{a\in[0,\bar a]} \Ex_{\mu_k}\{g(X;\ma_k,a)\} - \inf_{a\in[0,\bar a]} \Ex\{g(X;\ma_k,a)\} \right|  \\
&& \leq \max_{a\in[0,\bar a]} \left| \Ex_{\mu_k}\{g(X;\ma_k,a)\} - \Ex\{g(X;\ma_k,a)\} \right| \,.
\eea
The functions $g(\cdot;t,a)$ with $t\in(\ma/2,1]$, $a\in[0,\bar a]$, form a Glivenko-Cantelli class; see \citep[p.~271]{vanderVaart98}. It implies that $\max_{a\in[0,\bar a]} \left| \Ex_{\mu_k}\{g(X;\ma_k,a)\} - \Ex\{g(X;\ma_k,a)\} \right| \ra 0$ a.s., which concludes the proof.
\carre

\vsp
The class of functions $g(\cdot;t,a)$ is in fact Donsker \citep[p.~271]{vanderVaart98}. The strict concavity of $\Phi(\cdot)$ implies that optimal matrices are unique, and in complement of Lemma~\ref{L:opt-nonsequential} we get $\|\Mb_{\lfloor \ma k\rfloor,k}^*-\Mb_\ma^*\|=\SO_p(1/\sqrt{k})$. Note that when an optimal bounded design measure $\xi_\ma^*$ is known, a selection procedure such that $n_k/k\ra\ma$ and $\Phi(\Mb_{n_k,k}) \ra  \Phi(\Mb_\ma^*)$ a.s. is straightforwardly available: simply select the points that belong to the set $\SX_\ma^*$ on which $\xi_\ma^*=\mu/\ma$.

\section{Non degeneracy of $\Mb_{n_k}$}\label{S:min-eigenvalue}

To invoke H$_{\mu,\SM}$ in order to ensure the existence of a density $\varphi_{\Mb_{n_k}}$ having the required properties for all $k$ (which is essential for the convergence of Algorithm~1, see Theorem~\ref{P:main}), we need to guarantee that $\Mb_{n_k}\in\SMM^\geq_{\ell,L}$ for all $k$, for some $\ell$ and $L$. This is the object of the following lemma.

\begin{lemma}\label{L:min-eigenvalue}
Under H$_\SM$, when $\me_1>0$ in Algorithm~1, $n_{k+1}/k >\me_1$ for all $k$ and there exists a.s.\ $\ell>0$ and $L<\infty$ such that $\Mb_{n_k}\in\SMM^\geq_{\ell,L}$ for all $k>k_0$.
\end{lemma}

\noindent{\em Proof.} Since the first $k_0$ points are selected, we have $n_k/k=1>\me_1$ for $k\leq k_0$. Let $k_*$ be the first $k$ for which $n_k/k<\me_1$. It implies that $n_{k_*}=n_{k_*-1}>(k_*-1)\,\me_1$, and \eqref{perturbation} implies that $n_{k_*+1}=n_{k_*}+1$. Therefore,
$n_{k_*+1}/k_* > \me_1+(1-\me_1)/k_*>\me_1$, and $n_k/(k-1) >\me_1$ for all $k>1$.

If the $n_k$ points were chosen randomly, $n_k>(k-1)\,\me_1$ would be enough to obtain that, from H$_\SM$, $\ml_{\min}(\Mb_{n_k}) > \ell_{\me_1}/2$ and $\ml_{\max}(\Mb_{n_k})<\sqrt{B}/2$ for all $k$ larger than some $k_1$. However, here the situation is more complicated since points are accepted or rejected according to a sequential decision rule, and a more sophisticated argumentation is required. An expedite solution is to consider the worst possible choices of $n_k$ points, that yield the smallest value of $\ml_{\min}(\Mb_{n_k})$ and largest value of $\ml_{\max}(\Mb_{n_k})$. This approach is used in Lemma~\ref{L:min-eigenvalue_aux} presented below, which permits to conclude the proof.
\carre

\begin{lemma}\label{L:min-eigenvalue_aux}
Under H$_\SM$, any matrix $\Mb_{n_k}$ obtained by choosing $n_k$ points out of $k$ independently distributed with $\mu$ and such that $n_k/k>\me>0$ satisfies $\lim\inf_{k\ra\infty} \ml_{\min}(\Mb_{n_k}) > \ell$ and $\lim\sup_{k\ra\infty} \ml_{\max}(\Mb_{n_k}) < L$ a.s.\ for some $\ell >0$ and $L<\infty$.
\end{lemma}

\noindent{\em Proof.} We first construct a lower bound on $\lim\inf_{k\ra\infty} \ml_{\min}(\Mb_{n_k})$. Consider the criterion $\Phi_{\infty}^+(\Mb)=\ml_{\min}(\Mb)$, and denote by $\Mb_{n_k,k}^*$ the $n_k$-point design matrix that minimizes $\Phi_{\infty}^+$ over the design space formed by $k$ points $X_i$ i.i.d.\ with $\mu$. We can write
$\Mb_{n_k,k}^*=(1/n_k)\, \sum_{i=1}^{n_k} \SM(X_{k_i})$, where the $k_i$ correspond to the indices of positive $u_i$ in the minimization of $f(\ub)=\Phi_{\infty}^+[\sum_{i=1}^k u_i\,\SM(X_i)]$ with respect to $\ub=(u_1,\ldots,u_k)$ under the constraints $u_i\in\{0,1\}$ for all $i$ and $\sum_i u_i=n_k$. Obviously, any matrix $\Mb_{n_k}$ obtained by choosing $n_k$ distinct points $X_i$ among $X_1,\ldots,X_k$ satisfies $\ml_{\min}(\Mb_{n_k})\geq \ml_{\min}(\Mb_{n_k,k}^*)$.

For any $\Mb\in\SMM^\geq$, denote $\SU(\Mb)=\{\ub\in\mathds{R}^p:\, \|\ub\|=1\,, \ \Mb\ub=\ml_{\min}(\Mb)\ub\}$. Then, for any $\ub\in\SU(\Mb_{n_k,k}^*)$,
$\ub\TT\Mb_{n_k,k}^*\ub = \ml_{\min}(\Mb_{n_k,k}^*)=\min_{\vb\in\mathds{R}^p:\, \|\vb\|=1} \vb\TT\Mb_{n_k,k}^*\vb =(1/n_k)\, \sum_{i=1}^{n_k} z_{i:k}(\ub)$, where the $z_{i:k}(\ub)$ correspond to the values of
$\ub\TT\SM(X_i)\ub$ sorted by increasing order for $i=1,\ldots,k$. For any $m\in\{1,\ldots,n_k-1\}$, we thus have
\bea
\ml_{\min}(\Mb_{n_k,k}^*) \geq \frac1m\, \sum_{i=1}^{m} z_{i:k}(\ub) \geq \ml_{\min}(\Mb_{m,k}^*) \,,
\eea
showing that the worst situation corresponds to the smallest admissible $n_k$; that is, we only have to consider the case when $n_k/k \ra \me$ as $k\ra\infty$.

Since $\Phi_{\infty}^+$ is concave, for any $\Mb'\in\SMM^\geq$ we have
\be\label{*1}
\ml_{\min}(\Mb') \leq \ml_{\min}(\Mb_{n_k,k}^*)+ F_{\Phi_{\infty}^+}(\Mb_{n_k,k}^*,\Mb')\,,
\ee
where $F_{\Phi_{\infty}^+}(\Mb,\Mb')=\min_{\ub\in\SU(\Mb)} \ub\TT(\Mb'-\Mb)\ub$ is the directional derivative of $\Phi_{\infty}^+$ at $\Mb$ in the direction $\Mb'$.

For any $\ma\in(0,1)$ and any $\xi_\ma\leq \mu/\ma$, there exists a set $\SX_\ma\subset\SX$ such that $\xi_\ma\geq (1-\ma)\mu$ on $\SX_\ma$ and $\mu(\SX_\ma) \geq \ma^2$. Indeed, any set $\SZ$ on which $\xi_\ma<(1-\ma)\mu$ is such that $\xi_\ma(\SZ)<(1-\ma)\,\mu(\SZ) \leq (1-\ma)$; therefore, taking $\SX_\ma=\SX\setminus\SZ$, we get $\mu(\SX_\ma)\geq \ma\,\xi_\ma(\SX_\ma)\geq \ma^2$.
Denote $\ma_k=n_k/k$, with $\ma_k>\me$ and $\ma_k\ra \me$ as $k\ra\infty$, and take any $\Mb'=\Mb(\xi_{\ma_k})\in\SMM(\ma_k)$. Applying H$_\SM$-(\textit{ii}) to the set $\SX_{\ma_k}$ defined above, we get
\bea
\ml_{\min}(\Mb') = \ml_{\min}\left( \int_{\SX} \SM(x)\,\xi_{\ma_k}(\dd x) \right) &\geq& \ml_{\min}\left( \int_{\SX_{\ma_k}} \SM(x)\,\xi_{\ma_k}(\dd x) \right) \\
&\geq& (1-\ma_k)\, \ml_{\min}\left( \int_{\SX_{\ma_k}} \SM(x)\,\mu(\dd x) \right) > (1-\ma_k)\, \ell_{\ma_k^2} \,.
\eea
For $k$ larger than some $k_1$ we have $\ma_k\in(\me,2\me)$, and therefore $\ml_{\min}(\Mb')> c_\me = (1-2\me)\,\ell_{\me^2}>0$.
The inequality \eqref{*1} thus gives, for $k>k_1$,
\be\label{LB2-1}
c_\me < \ml_{\min}(\Mb_{n_k,k}^*)+ \min_{\ub\in\SU(\Mb_{n_k,k}^*)} \min_{\Mb'\in\SMM(\ma_k)} \ub\TT(\Mb'-\Mb_{n_k,k}^*)\ub \,.
\ee
The rest of the proof follows from results on conditional value-at-risk by \citet{RockafellarU2000} and \citet{Pflug2000}. For a fixed $\ub\in\mathds{R}^p$, $\ub\neq\0b$, and $\ma\in(0,1)$, we have
\bea
\min_{\Mb'\in\SMM(\ma)} \ub\TT\Mb'\ub = \frac{1}{\ma}\, \int_\SX \ind_{\{\ub\TT\SM(x)\ub \leq a_{\ma}(\ub)\}}\,[\ub\TT\SM(x)\ub]\,\mu(\dd x) \,,
\eea
where the $\ma$-quantile $a_{\ma}(\ub)$ satisfies $\int_\SX \ind_{\{\ub\TT\SM(x)\ub \leq a_{\ma}(\ub)\}}\,\mu(\dd x)=\ma$. For any $a\in\mathds{R}$ and $\ub\in\mathds{R}^p$, denote
\bea
h(x;\ma,a,\ub)=a-\frac{1}{\ma}\,[a-\ub\TT\SM(x)\ub]^+\,, \ x\in\SX\,.
\eea
We can write
$\min_{\Mb'\in\SMM(\ma)} \ub\TT\Mb'\ub = \Ex\{h(X;\ma,a_\ma(\ub),\ub)\} = \sup_{a\in\mathds{R}} \Ex\{h(X;\ma,a,\ub)\}$,
where the expectation is with respect to $X$ distributed with $\mu$ \citep{RockafellarU2000}. Also, from \citet{Pflug2000}, for any $\ub\in\SU(\Mb_{n_k,k}^*)$ we can write
$\ub\TT\Mb_{n_k,k}^*\ub= \Ex_{\mu_k}\{h(X;\ma_k,z_{n_k:k}(\ub),\ub)\} = \sup_{a\in\mathds{R}} \Ex_{\mu_k}\{h(X;\ma_k,a,\ub)\}$,
where $\Ex_{\mu_k}\{\cdot\}$ denotes expectation for the empirical measure $\mu_k=(1/k)\,\sum_{i=1}^k \delta_{X_i}$.

Now, from H$_\SM$-(\textit{i}), for any $\ub\in\mathds{R}^p$ with $\|\ub\|=1$,
\be\label{bound-on-a-ma(u)}
(1-\ma)\, a_\ma(\ub) \leq \int_\SX \ind_{\{\ub\TT\SM(x)\ub > a_{\ma}(\ub)\}}\,[\ub\TT\SM(x)\ub]\,\mu(\dd x) < \sqrt{B} \,.
\ee
We also have
$(k-n_k)\, z_{n_k:k}(\ub) \leq \sum_{i=n_k+1}^k z_{i:k}(\ub) \leq \sum_{i=1}^k z_{i:k}(\ub) = k\, (\ub\TT\Mb_k\ub) \leq k\, \ml_{\max}(\Mb_k)$,
with $\Mb_k\ra\Mb(\mu)$ a.s.\ as $k\ra\infty$. Denote $\overline{z_\me} = 2\,\sqrt{B}/(1-2\me)$; since $\ma_k\ra \me$, from H$_\SM$2-(\textit{i}) there exists a.s.\ $k_2$ such that, for all $k>k_2$, $z_{n_k:k}(\ub)< \overline{z_\me}$ and, from \eqref{bound-on-a-ma(u)}, $a_{\ma_k}(\ub)< \overline{z_\me}$.

Therefore, for large enough $k$, for any $\ub\in\SU(\Mb_{n_k,k}^*)$,
\bea
\min_{\Mb'\in\SMM(\ma_k)} \ub\TT(\Mb'-\Mb_{n_k,k}^*)\ub &=&  \Ex\{h(X;\ma_k,a_{\ma_k}(\ub),\ub)\} - \Ex_{\mu_k}\{h(X;\ma_k,z_{n_k:k}(\ub),\ub)\} \\
&\leq& \sup_{a\in[0,\overline{z_\me}]} \left|\Ex\{h(X;\ma_k,a,\ub)\} - \Ex_{\mu_k}\{h(X;\ma_k,a,\ub)\} \right| \,.
\eea

The functions $h(\cdot;\ma,a,\ub)$ with $\ma\in(\me,2\me)$, $a\in[0,\overline{z_\me}]$ and $\ub\in\mathds{R}^p$, $\|\ub\|=1$, form a Glivenko-Cantelli class; see \citep[p.~271]{vanderVaart98}. This implies that there exists a.s.\ $k_3$ such that %
\bea
\max_{\ub\in\mathds{R}^p: \|\ub\|=1} \sup_{a\in[0,\overline{z_\me}]} \left|\Ex\{h(X;\ma_k,a,\ub)\} - \Ex_{\mu_k}\{h(X;\ma_k,a,\ub)\} \right| <c_\me/2\,, \quad \forall k>k_3 \,.
\eea
Therefore, from \eqref{LB2-1}, $\ml_{\min}(\Mb_{n_k,k}^*)>c_\me/2$ for all $k>k_3$, which concludes the first part of the proof.

\vsp
We construct now an upper bound on $\lim\sup_{k\ra\infty} \ml_{\max}(\Mb_{n_k})$ following steps similar to the above developments but exploiting now the convexity of the criterion $\Mb \ra 1/\Phi_{-\infty}^+(\Mb)=\ml_{\max}(\Mb)$. Its directional derivative is
$F_{1/\Phi_{-\infty}^+}(\Mb,\Mb')=\max_{\ub\in\SU(\Mb)} \ub\TT(\Mb'-\Mb)\ub$, with $\SU(\Mb)=\{\ub\in\mathds{R}^p:\, \|\ub\|=1\,, \ \Mb\ub=\ml_{\max}(\Mb)\ub\}$. Denote by $\Mb_{n_k,k}^*$ the $n_k$-point design matrix that maximizes $1/\Phi_{-\infty}^+$ over the design space formed by $k$ points $X_i$ i.i.d.\ with $\mu$. We can write
$\Mb_{n_k,k}^*=(1/n_k)\, \sum_{i=1}^{n_k} \SM(X_{k_i})$, where the $k_i$ correspond to the indices of positive $u_i$ in the maximization of $f(\ub)=\ml_{\max}[\sum_{i=1}^k u_i\,\SM(X_i)]$ with respect to $\ub=(u_1,\ldots,u_k)$ under the constraints $u_i\in\{0,1\}$ for all $i$ and $\sum_i u_i=n_k$. Any matrix $\Mb_{n_k}$ obtained by selecting $n_k$ distinct points $X_i$ among $X_1,\ldots,X_k$ satisfies $\ml_{\max}(\Mb_{n_k})\leq \ml_{\max}(\Mb_{n_k,k}^*)$.

For any $\ub\in\SU(\Mb_{n_k,k}^*)$ we can write
$\ub\TT\Mb_{n_k,k}^*\ub = \ml_{\max}(\Mb_{n_k,k}^*)=\max_{\vb\in\mathds{R}^p:\, \|\vb\|=1} \vb\TT\Mb_{n_k,k}^*\vb =(1/n_k)\, \sum_{i=1}^{n_k} z_{i:k}(\ub)$, where the $z_{i:k}(\ub)$ correspond to the values of
$\ub\TT\SM(X_i)\ub$ sorted by decreasing order for $i=1,\ldots,k$. For any $m\in\{1,\ldots,n_k-1\}$, we thus have
\bea
\ml_{\max}(\Mb_{n_k,k}^*) \leq \frac1m\, \sum_{i=1}^{m} z_{i:k}(\ub) \leq \ml_{\max}(\Mb_{m,k}^*) \,,
\eea
showing that the worst case corresponds to the smallest admissible $n_k$, and we can restrict our attention to the case when $\ma_k = n_k/k \ra \me$ as $k\ra\infty$.

The convexity of $1/\Phi_{-\infty}^+$ implies that, for any $\Mb'\in\SMM^\geq$,
\be\label{ml_max-1}
\ml_{\max}(\Mb') \geq \ml_{\max}(\Mb_{n_k,k}^*)+ F_{1/\Phi_{-\infty}^+}(\Mb_{n_k,k}^*,\Mb')\,.
\ee
Take $\Mb'\in\SMM(\ma_k)$, corresponding to some $\xi_{\ma_k}$. From H$_\SM$-(\textit{i}),
\bea
\ml_{\max}(\Mb')=\ml_{\max}\left[ \int_{\SX} \SM(x)\, \xi_{\ma_k}(\dd x) \right] \leq \frac{1}{\ma_k}\, \ml_{\max}[\Mb(\mu)] < \frac{\sqrt{B}}{\ma_k}\,.
\eea
Therefore, there exists some $k_1$ such that, for all $k>k_1$, $\ml_{\max}(\Mb')<2\sqrt{B}/\me$, and \eqref{ml_max-1} gives
\bea
\frac{2\,\sqrt{B}}{\me} \geq \ml_{\max}(\Mb_{n_k,k}^*)+ \max_{\ub\in\SU(\Mb_{n_k,k}^*)} \max_{\Mb'\in\SMM(\ma_k)} \ub\TT(\Mb'-\Mb_{n_k,k}^*)\ub\,.
\eea
For $a\in\mathds{R}$, $\ma\in(0,1)$ and $\ub\in\mathds{R}^p$, denote $h(x;\ma,a,\ub)=a+(1/\ma)[\ub\TT\SM(x)\ub-a]^+$, $x\in\SX$. We have
$\ml_{\max}(\Mb_{n_k,k}^*)=(1/n_k)\sum_{i=1}^{n_k} z_{i:k}(\ub)=\Ex_{\mu_k}\{h(X;\ma_k,z_{n_k:k}(\ub),\ub)\}=\inf_a \Ex_{\mu_k}\{h(X;\ma_k,a,\ub)\}$, $\ub\in\SU(\Mb_{n_k,k}^*)$, with $z_{n_k:k}(\ub)$ satisfying $0\leq n_k\, z_{n_k:k}(\ub) \leq \sum_{i=1}^{n_k} z_{i:k}(\ub) < \sum_{i=1}^k z_{i:k}(\ub)=k\,\ml_{max}(\Mb_k)$. Also, for any $\ma\in(0,1)$ and $\ub\in\mathds{R}^p$, $\ub\neq \0b$,
$\max_{\Mb'\in\SMM(\ma)} \ub\TT\Mb'\ub = \Ex\{h(X;\ma,a_\ma(\ub),\ub)\} = \inf_a \Ex\{h(X;\ma,a,\ub)\}$, where $a_\ma(\ub)$ satisfies
$\int_\SX \ind_{\{\ub\TT\SM(x)\ub \geq a_{\ma}(\ub)\}}\, \mu(\dd x)= \ma$, and H$_\SM$-(\textit{i}) implies that $0\leq a_\ma(\ub)\leq (1/\ma) \int_\SX \ind_{\{\ub\TT\SM(x)\ub \geq a_{\ma}(\ub)\}}\,\ub\TT\SM(x)\ub\, \mu(\dd x) < \sqrt{B}/\ma$. Since $\ma_k=n_k/k\ra\me$ and $\Mb_k\ra\Mb(\mu)$ a.s., there exists a.s.\ $k_2$ such that, for all $k>k_2$, $0\leq a_{\ma_k}(\ub) < 2\sqrt{B}/\me$ and $0\leq z_{n_k:k}(\ub) < 2\sqrt{B}/\me$. This implies that, for $\ub\in\SU(\Mb_{n_k,k}^*)$ and $k>k_2$,
\bea
\max_{\Mb'\in\SMM(\ma_k)} \ub\TT(\Mb'-\Mb_{n_k,k}^*)\ub &=&  \Ex\{h(X;\ma_k,a_{\ma_k}(\ub),\ub)\} - \Ex_{\mu_k}\{h(X;\ma_k,z_{n_k:k}(\ub),\ub)\} \\
&\leq& \sup_{a\in[0,2\sqrt{B}/\me]} \left|\Ex\{h(X;\ma_k,a,\ub)\} - \Ex_{\mu_k}\{h(X;\ma_k,a,\ub)\} \right| \,.
\eea
The rest of the proof is similar to the case above for $\ml_{\min}$, using the fact that the functions $h(\cdot;\ma,a,\ub)$ with $\ma\in(\me,2\me)$, $a\in[0,2\sqrt{B}/\me]$ and $\ub\in\mathds{R}^p$, $\|\ub\|=1$, form a Glivenko-Cantelli class.
\carre

\section{Convergence of $\hC_k$}\label{S:fk}

We consider the convergence properties of \eqref{SAQuantile} when the matrix $\Mb_k$ is fixed, that is,
\be\label{C_k-appendix}
\hC_{k+1} = \hC_k + \frac{\beta_k}{(k+1)^q}\, \left(\ind_{\{Z_{k+1} \geq \hC_k\}} - \ma\right)\,,
\ee
where the $Z_k$ have a fixed distribution with uniformly bounded density $f$ such that $f(C_{1-\ma})>0$. We follow the arguments of \citet{Tierney83}. The construction of $\beta_k$ is like in Algorithm~1, with $\beta_k= \max\{\min(1/\hf_k,\beta_0\,k^\mg)\}$ and $\hf_k$ following the recursion
\be\label{f_k-appendix}
 \hf_{k+1}= \hf_k + \frac{1}{(k+1)^q}\, \left[ \frac{1}{2\,h_{k+1}}\, \ind_{\{|Z_{k+1}-\hC_k|\leq h_{k+1}\}} - \hf_k \right]
\ee
with $h_k=h/k^\mg$.

\begin{theo}\label{P:CV-of-Ck}
Let $\ma\in(0,1)$, $\beta_0>0$, $h>0$, $1/2<q \leq 1$, $0<\mg<q-1/2$. Let $F$ be a distribution function such that $f(t)=\dd F(t)/\dd t$ exists for all $t$, is uniformly bounded, and is strictly positive in a neighborhood of $C_{1-\ma}$, the unique value of $C$ such that $F(C)=1-\ma$. Let $(X_i)$ be an i.i.d.\ sequence distributed with $F$ and define $\hC_k$ and $\hf_k$ by \eqref{C_k-appendix} and \eqref{f_k-appendix} respectively, with $\beta_k= \min\{1/\hf_k,\beta_0\,k^\mg\}$ and $h_k=h/k^\mg$. Then, $\hC_k \ra C_{1-\ma}$ a.s.\ when $k\ra\infty$.
\end{theo}

\noindent{\em Proof.}
We first show that $\hf_k$ is a.s.\ bounded.
From the mean-value theorem, there exists a $t_k$ in $[\hC_k-h_{k+1},\hC_k+h_{k+1}]$ such that $\Pr\{|Z_{k+1}-\hC_k|\leq h_{k+1}\}=F(\hC_k+h_{k+1})-F(\hC_k-h_{k+1})=2\,h_{k+1}\,f(t_k)$. Denote $\mo_{k+1}=\ind_{\{|Z_{k+1}-\hC_k|\leq h_{k+1}\}}-2\,h_{k+1}\,f(t_k)$.
We can write
\bea
\hf_{k+1}= (1-B_k)\,\hf_k + A_k + A'_k
\eea
where $B_k=1/[(k+1)^q]$, $A_k=\mo_{k+1}/[2\,h_{k+1}\,(k+1)^q]$ and $A'_k=B_k\,f(t_k)$. Therefore,
\bea
\hf_{k+1} = \hf_1\, \prod_{i=1}^k (1-B_i) + \sum_{j=1}^k (A_j+A'_j)\, \prod_{i=j+1}^k (1-B_i) \,.
\eea
We have $\prod_{i=1}^k (1-B_i)<\exp(-\sum_{i=1}^k B_i) \ra 0$ as $k\ra\infty$ since $q\leq 1$. Next, for $h_k=h/k^\mg$ and $0<\mg<q-1/2$, $\sum_k 1/[h_k\,k^q]^2 < \infty$, $\sum_{j=1}^k A_j$ forms an $\SL^2$-bounded martingale and therefore converges a.s.\ to some limit. Lemma~2 of \citet[p.~190]{AlbertG67} then implies that $\sum_{j=1}^k A_j\, \prod_{i=j+1}^k (1-B_i) \ra 0$ a.s.\ as $k\ra\infty$. Consider now the term $T_k=\sum_{j=1}^k A'_j\, \prod_{i=j+1}^k (1-B_i)$. Since $f$ is bounded, $A'_j< \bar f\, B_j$ for some $\bar f<\infty$ and
\bea
T_k < \bar f\, \sum_{j=1}^k B_j\, \prod_{i=j+1}^k (1-B_i) = \bar f\, \left[1- \prod_{i=1}^k (1-B_i) \right] < \bar f\,,
\eea
where the equality follows from \citet[Lemma~1, p.~189]{AlbertG67}. This shows that $\hf_k$ is a.s.\ bounded. Therefore, $\beta_k= \min\{1/\hf_k,\beta_0\,k^\mg\}$ is a.s.\ bounded away from zero.

We consider now the convergence of \eqref{C_k-appendix}. Following \citet{Tierney83}, define
\bea
D_k= \frac{\beta_k}{(k+1)^q} \, \left\{\ind_{\{Z_{k+1} \geq \hC_k\}} - [1-F(\hC_k)]\right\} \mbox{ and } E_k = \frac{\beta_k}{(k+1)^q} \, \frac{F(\hC_k)-(1-\ma)}{\hC_k - C_{1-\ma}} \,.
\eea
Denote by $\SF_k$ the increasing sequence of $\ms$-fields generated by the $X_i$; we have $\Ex\{D_k|\SF_k\}=0$ and $\Ex\{D_k^2|\SF_k\}= \beta_k^2\,F(\hC_k)\,[1-F(\hC_k)]/(k+1)^{2q}$.
We can rewrite \eqref{C_k-appendix} as $\hC_{k+1} - C_{1-\ma} = (\hC_k - C_{1-\ma} )\,(1-E_k) + D_k$, which gives
\bea
\Ex\{(\hC_{k+1} - C_{1-\ma})^2|\SF_k\} = (\hC_k - C_{1-\ma} )^2\,(1-E_k)^2 + \frac{\beta_k^2}{(k+1)^{2q}} \,F(\hC_k)\,[1-F(\hC_k)] \,.
\eea
$E_k\geq 0$ for all $k$, $[F(\hC_k)-(1-\ma)]/(\hC_k - C_{1-\ma})$ is bounded since $f$ is bounded, and therefore $E_k \ra 0$.
Since $\beta_k \leq \beta_0\, k^\mg$ and $0<\mg<q-1/2$, $\sum_k \beta_k^2/(k+1)^{2q} < \infty$. Robbins-Siegmund Theorem (1971)\nocite{RobbinsS71} then implies that $\hC_k$ converges a.s.\ to some limit %
and that $\sum_k (\hC_k - C_{1-\ma} )^2\,[1-(1-E_k)^2] <\infty$ a.s.; since $E_k\ra 0$, we obtain $\sum_k (\hC_k - C_{1-\ma} )^2\,E_k <\infty$ a.s.
Since $q\leq 1$, $\beta_k$ is a.s.\ bounded away from zero, and $f$ is strictly positive in a neighborhood of $C_{1-\ma}$, we obtain that $\sum_k E_k = \infty$,
implying that $\hC_k \ra C_{1-\ma}$ a.s., which concludes the proof.
\carre

\vsp
\citet{Tierney83} uses $q=1$; the continuity of $f$ at $C_{1-\ma}$ then implies that $f_k\ra f(C_{1-\ma})$ a.s., and his construction also achieves the optimal rate of convergence of $\hC_k$ to $C_{1-\ma}$, with $\sqrt{k}(\hC_k-C_{1-\ma}) \rad \SN(0,\ma(1-\ma)/f^2(C_{1-\ma})$ as $k\ra\infty$.

\section{Lipschitz continuity of $C_{1-\ma}(\Mb)$}\label{S:Lipschitz}

\begin{lemma}\label{L:Lipschitz}
Under H$_\Phi$ and H$_{\mu,\SM}$, the $(1-\ma)$-quantile $C_{1-\ma}(\Mb)$ of the distribution $F_\Mb$ of $Z_\Mb(X)=F_\Phi[\Mb,\SM(X)]$ is a Lipschitz continuous function of $\Mb\in\SMM_{\ell,L}^\geq$.
\end{lemma}

\noindent{\em Proof.} For any $\Ab\in\SMM^>$, define the random variable $T_\Ab(X)=\tr[\Ab\SM(X)]$ and denote $G_\Ab$ its distribution function and $Q_{1-\ma}(\Ab)$ the associated $(1-\ma)$-quantile. We have $Z_\Mb(X)=T_{\nabla_\Phi(\Mb)}(X)-\tr[\nabla_\Phi(\Mb)\Mb]$, and therefore
\be\label{Quantile-Lipschitz-main}
C_{1-\ma}(\Mb)=Q_{1-\ma}[\nabla_\Phi(\Mb)]-\tr[\nabla_\Phi(\Mb)\Mb] \,.
\ee

We fist show that $\tr[\nabla_\Phi(\Mb)\Mb]$ is Lipschitz continuous in $\Mb$. Indeed, for any $\Mb$, $\Mb'$ in $\SMM_{\ell,L}^\geq$, we have
\be
\left| \tr[\nabla_\Phi(\Mb')\Mb'] - \tr[\nabla_\Phi(\Mb)\Mb]\right| &\leq& \|\Mb'\| \, \|\nabla_\Phi(\Mb')-\nabla_\Phi(\Mb)\|+ \|\nabla_\Phi(\Mb)\|\, \|\Mb'-\Mb\| \nonumber \\
&<& [L\sqrt{p}\,K_\ell+A(\ell)]\, \|\Mb'-\Mb\| \,, \label{Quantile-Lipschitz0}
\ee
where we used H$_\Phi$ and the fact that $\Mb, \Mb' \in\SMM_{\ell,L}^\geq$.

Consider now $G_\Ab$ and $G_{\Ab'}$ for two matrices $\Ab$ and $\Ab'$ in $\SMM^>$. We have
\bea
G_{\Ab'}(t)-G_{\Ab}(t) = \int_\SX \left( \ind_{\{\tr[\Ab'\SM(x)]\leq t \}} - \ind_{\{\tr[\Ab\SM(x)]\leq t \}} \right)\,\mu(\dd x) \,,
\eea
and therefore
\bea
\left| G_{\Ab'}(t)-G_{\Ab}(t) \right| &\leq& \Prob\left\{\min\{ \tr[\Ab'\SM(X)],\tr[\Ab\SM(X)]\} \leq t \right. \\
 && \hspace{2cm} \left. \leq \max\{ \tr[\Ab'\SM(X)],\tr[\Ab\SM(X)]\} \right\} \\
&& \hspace{-1cm} \leq \Prob\left\{ \tr[(\Ab-\|\Ab-\Ab'\|\Ib_p)\SM(X)] \leq t \leq \tr[(\Ab+\|\Ab-\Ab'\|\Ib_p)\SM(X)] \right\} \,,
\eea
with $\Ib_p$ the $p\times p$ identity matrix. Since $\Ab- \ml_{\min}(\Ab)\,\Ib_b\in\SMM^\geq$, denoting $b_1=1-\|\Ab-\Ab'\|/\ml_{\min}(\Ab)$ and $b_2=1+\|\Ab-\Ab'\|/\ml_{\min}(\Ab)$, we obtain
\be
\left| G_{\Ab'}(t)-G_{\Ab}(t) \right| &\leq& \Prob\left\{ b_1\,\tr[\Ab\SM(X)] \leq t \leq b_2\,\tr[\Ab\SM(X)] \right\} \nonumber \\
&& = \Prob\left\{ \tr[\Ab\SM(X)] \leq \frac{t}{b_1} \bigwedge \tr[\Ab\SM(X)]\geq \frac{t}{b_2} \right\} \nonumber \\
&& = G_{\Ab}(t/b_1)-G_{\Ab}(t/b_2) \,. \label{Quantile-Lipschitz1}
\ee
In the rest of the proof we show that $Q_{1-\ma}[\nabla_\Phi(\Mb)]$ is Lipschitz continuous in $\Mb$. Take two matrices $\Mb,\Mb'\in\SMM_{\ell,L}^\geq$, and consider the associated $(1-\ma)$-quantiles $Q_{1-\ma}[\nabla_\Phi(\Mb)]$ and $Q_{1-\ma}[\nabla_\Phi(\Mb')]$, which we shall respectively denote $Q_{1-\ma}$ and $Q'_{1-\ma}$ to simplify notation. From H$_{\mu,\SM}$, the p.d.f.\ $\psi_\Mb$ associated with $G_{\nabla_\Phi(\Mb)}$ is continuous at $Q_{1-\ma}$ and satisfies $\psi_\Mb(Q_{1-\ma})>\me_{\ell,L}$. From the identities
\bea
\int_{-\infty}^{Q_{1-\ma}} \psi_\Mb(z)\, \dd z = \int_{-\infty}^{Q'_{1-\ma}} \psi_{\Mb'}(z)\, \dd z = 1-\ma \,,
\eea
we deduce
\be\label{Quantile-Lipschitz2}
\hspace{-0.5cm} \left| \int_{Q_{1-\ma}}^{Q'_{1-\ma}} \psi_\Mb(z)\, \dd z \right| = \left| \int_{-\infty}^{Q'_{1-\ma}} [\psi_{\Mb'}(z)-\psi_\Mb(z)]\, \dd z \right| = \left| G_{\nabla_\Phi(\Mb')}(Q'_{1-\ma}) - G_{\nabla_\Phi(\Mb)}(Q'_{1-\ma}) \right|\,.
\ee
From H$_\Phi$, when substituting ${\nabla_\Phi(\Mb)}$ for $\Ab$ and ${\nabla_\Phi(\Mb')}$ for $\Ab'$ in $b_1$ and $b_2$, we get
$b_1>B_1=1-K_\ell\|\Mb'-\Mb\|/a(L)$ and $b_2<B_2=1+K_\ell\|\Mb'-\Mb\|/a(L)$, showing that $Q'_{1-\ma}\ra Q_{1-\ma}$ as $\|\Mb'-\Mb\|\ra 0$. Therefore, there exists some $\beta_1$ such that, for $\|\Mb'-\Mb\|<\beta_1$ we have
\be\label{Quantile-Lipschitz3}
\left| \int_{Q_{1-\ma}}^{Q'_{1-\ma}} \psi_\Mb(z)\, \dd z \right| > \frac12\, \left| Q'_{1-\ma} - Q_{1-\ma} \right|\, \me_{\ell,L} \,.
\ee
Using \eqref{Quantile-Lipschitz1}, we also obtain for $\|\Mb'-\Mb\|$ smaller than some $\beta_2$
\bea
\left| G_{\nabla_\Phi(\Mb')}(Q'_{1-\ma}) - G_{\nabla_\Phi(\Mb)}(Q'_{1-\ma}) \right| &\leq&
G_{\nabla_\Phi(\Mb)}(Q'_{1-\ma}/B_1)-G_{\nabla_\Phi(\Mb)}(Q'_{1-\ma}/B_2) \\
&<& 2 \psi_\Mb(Q'_{1-\ma})\, \left( \frac{Q'_{1-\ma}}{B_1}-\frac{Q'_{1-\ma}}{B_2} \right) \\
&& \hspace{-1.5cm} < \ 4\|\Mb'-\Mb\|\,\psi_\Mb(Q'_{1-\ma})\,Q'_{1-\ma}\,\frac{a(L)}{K_\ell\left(a^2(L)/K_\ell^2-\|\Mb'-\Mb\|^2\right)}\,.
\eea
Therefore, when $\|\Mb'-\Mb\|< a(L)/(K_\ell\sqrt{2})$,
\bea
\left| G_{\nabla_\Phi(\Mb')}(Q'_{1-\ma}) - G_{\nabla_\Phi(\Mb)}(Q'_{1-\ma}) \right| < \kappa\,\|\Mb-\Mb'\|
\eea
with $\kappa = 8\bar\varphi_\Mb\,Q'_{1-\ma}\,K_\ell/a(L)$, where $\bar\varphi_\Mb$ is the upper bound on $\varphi_\Mb$ in H$_{\mu,\SM}$. Using \eqref{Quantile-Lipschitz2} and \eqref{Quantile-Lipschitz3} we thus obtain, for $\|\Mb'-\Mb\|$ small enough,
\bea
\left| Q_{1-\ma}[\nabla_\Phi(\Mb')] - Q_{1-\ma}[\nabla_\Phi(\Mb)] \right| < 2\kappa/\me_{\ell,L}\,\|\Mb-\Mb'\|\,,
\eea
which, combined with \eqref{Quantile-Lipschitz0} and \eqref{Quantile-Lipschitz-main}, completes the proof.
\carre

\section*{Acknowledgements}
The work of the first author was partly supported by project INDEX (INcremental Design of EXperiments) ANR-18-CE91-0007 of the French National Research Agency (ANR).

\end{document}